\documentclass[a4paper,11pt]{article}
\usepackage{cite}
\pdfoutput=1 

\usepackage{jcappub} 

\usepackage[T1]{fontenc} 

\usepackage{caption}
\usepackage{hyperref}
\usepackage{subcaption}
\usepackage{graphicx}
\usepackage{multirow}
\usepackage{comment}
\usepackage{makecell}
\usepackage{amsmath}
\usepackage{booktabs}
\usepackage[dvipsnames]{xcolor}
\usepackage{tikz}
\usepackage{mathrsfs}
\usepackage[margin=1in]{geometry}
\usepackage{adjustbox}   

\newcommand{\cosmogrid}{\texttt{CosmoGridV1}}
\newcommand{\pkdgrav}{\texttt{PkDGraV3}}
\newcommand{\ufalcon}{\texttt{UFalconv2}}
\newcommand{\healpix}{\texttt{HEALPix} }

\newcommand{\hillipop}{\texttt{HiLLiPoP}}
\newcommand{\lollipop}{\texttt{LoLLiPoP}}

\author[a]{Alexander Reeves,}
\author[b, c]{Andrina Nicola,}
\author[a]{Alexandre Refregier,}

\title{Tuning the cosmic instrument: robust cosmology through combined probes}

\affiliation[a]{Institute for Particle Physics and Astrophysics, ETH Zürich, Wolfgang-Pauli-Strasse 27, CH-8093 Zürich, Switzerland}

\affiliation[b]{Jodrell Bank Centre for Astrophysics, Department of Physics and Astronomy, The University of Manchester, Manchester M13 9PL, UK}
\affiliation[c]{Argelander Institut f\"ur Astronomie, Universit\"at Bonn, Auf dem H\"ugel 71, 53121 Bonn, Germany}

\emailAdd{areeves@phys.ethz.ch}

\abstract{As wide-field surveys yield increasingly precise data, multiprobe analyses offer significant advantages. In this work, we use our previously developed framework for jointly analyzing cosmic microwave background (CMB) and large-scale structure data. We analyze combinations of three CMB (\textit{Planck} PR3, \textit{Planck} PR4, and ACT+WMAP) datasets, DESI Y1 Baryon Acoustic Oscillation (BAO) data, and a $9\times 2$pt low-$z$ dataset comprising KiDS-1000, BOSS DR12, and \textit{Planck} CMB lensing/Integrated Sachs Wolfe (including all cross-correlations). We first assess internal consistency, finding a mild ($<2\sigma$) tension between CMB and low-$z$ datasets in the full parameter space and hints of systematics in \textit{Planck} PR3 and KiDS-1000. We then derive constraints in $\Lambda\mathrm{CDM}$ and, motivated by recent DESI results, dynamical dark energy ($w_0w_a\mathrm{CDM}$) and free neutrino mass extensions. In $\Lambda \mathrm{CDM}$, we derive a novel $9\times2$pt constraint of $S8=0.777^{+0.017}_{-0.017}$ and find strong consistency among CMB datasets. In $w_0w_a\mathrm{CDM}$, adding low-$z$ to CMB+BAO tightens $(w_0,w_a)$ constraints by 50\% (in figure-of-merit terms) in our baseline combination of \textit{Planck} PR4 + low-$z$ + BAO. The posterior accommodates a cosmological constant ($w_0 = -1, w_a = 0$) within $1\sigma$, in contrast to the $\sim2\sigma$ preference for evolving dark energy from CMB+BAO alone. For neutrino masses, our baseline dataset yields a systematics-robust constraint of $M_\nu<0.12\mathrm{eV}$ in $\nu\Lambda\mathrm{CDM}$. Allowing dynamical dark energy and free neutrino mass ($\nu w_0w_a\mathrm{CDM}$) broadens and shifts the neutrino mass posterior higher, yielding a $1.8\sigma$ constraint ($M_\nu=0.16^{+0.09}_{-0.09}\mathrm{eV}$) in our baseline. Our analysis demonstrates the power of multiprobe analyses for assessing tensions, identifying systematics and providing robust constraints.
}

\begin{document}
\maketitle
\flushbottom

\section{Introduction}

As cosmological datasets become increasingly precise, so too do the demands on our understanding and control of potential systematic effects and tensions between different measurements. Over the last few years, several low-redshift probes have shown mild to moderate discrepancies with Cosmic Microwave Background (CMB) constraints under the standard $\Lambda \mathrm{CDM}$ framework~\cite{Abdalla:2022yfr}. One prominent example is the so-called `Hubble tension,' which refers to the mismatch between CMB-inferred measurements of the Hubble constant ($H_0$) and those determined from Type-Ia supernovae or time delays in strongly lensed quasars~\cite{Riess:2016jrr, Riess:2019cxk, Riess:2021jrx, H0LiCOW:2019pvv}. Although the significance of this tension is debated---a recent analysis based on James Webb Space Telescope data finds an $H_0$ value consistent with CMB analyses~\cite{Freedman:2024eph}---it remains a major focus of current research. At the perturbation level, a second milder discrepancy exists in the parameter $S8=\sigma_8 \sqrt{\Omega_m/0.3}$, with weak-lensing surveys either alone or in combination with galaxy clustering data often favoring a value systematically smaller than that inferred from the CMB by $\sim 2$--$3\sigma$~\cite{DES:2021bvc, KiDS:2020suj, HSC:2018mrq} (see Ref.~\cite{Chen:2024vvk} for a counterexample finding a value of $S8$ consistent with CMB measurements). Whilst, increased baryonic feedback~\cite{Amon:2022azi, Preston:2023uup, DES:2024iny} has been proposed as a potential explanation, other works question whether such effects alone are sufficient~\cite{Salcido:2024qrt, McCarthy:2023ism}. These tensions have led to a variety of proposals for physics beyond the standard $\Lambda \mathrm{CDM}$ model (see Refs.~\cite{Schoneberg:2021qvd, DiValentino:2021izs} for recent reviews), although no consensus on an extended model has yet emerged.

The recently released Dark Energy Spectroscopic Instrument (DESI) Y1 data~\cite{DESI:2024uvr, DESI:2024mwx} has further sharpened attention on two specific extensions: (1) dynamical dark energy, wherein the cosmological constant $\Lambda$ is replaced by a time-evolving equation of state, and (2) $\nu \Lambda \mathrm{CDM}$, where the sum of neutrino masses ($M_\nu$) is varied. Intriguingly, DESI finds a mild preference (at the $2$--$4\sigma$ level) for an evolving dark energy equation of state when its BAO measurements are combined with \textit{Planck}~\cite{Planck:2018vyg} data and supernovae from Pantheon+~\cite{Brout:2022vxf} or the Dark Energy Survey Year 5 (DESY5) sample~\cite{DES:2024jxu}\footnote{Note earlier work from Ref.~\cite{Zhao:2017cud} also found a mild preference for dynamical dark energy when combining SDSS BAO and Union3 supernovae data.}. While these indications are not definitive---and may stem from unaccounted systematic errors in some of the data~\cite{Colgain:2024mtg, Efstathiou:2024xcq, Gialamas:2024lyw}---they motivate further analysis using complementary datasets. In parallel, constraints on the total neutrino mass have tightened considerably, with the DESI collaboration reporting a limit of $M_\nu < 0.072\,\mathrm{eV}$ (95\% CL) when combining DESI Y1 BAO with \textit{Planck} PR3 (\texttt{plik}-based) CMB data~\cite{DESI:2024mwx}, representing a significantly more stringent upper bound compared previous results~\cite{Tanseri:2022zfe, Vagnozzi:2017ovm, DiValentino:2021hoh, Wang:2024hen, Craig:2024tky, DiValentino:2024xsv, Jimenez:2022dkn, Gariazzo:2022ahe, Lattanzi:2020iik, RoyChoudhury:2018gay, RoyChoudhury:2019hls} and pushing the $M_\nu$ upper limit towards the minimum allowed value from neutrino oscillation experiments ($M_\nu>0.06 \mathrm{eV}$ under normal ordering (NO)). Furthermore, several recent works have extended the neutrino mass parameter space to (unphysical) negative values finding that combinations of CMB and DESI data posteriors have peaks at $M_\nu<0$, putting pressure on the standard model interpretation of neutrinos \cite{Craig:2024tky}. Taken together, these findings demonstrate both the opportunities and the challenges presented by the latest generation of cosmological surveys, emphasizing the need for careful systematics control and consistency tests with other complementary datasets. 

In this context, the combined analysis of multiple cosmological probes is strongly motivated. Such analyses offer significant advantages over single-probe measurements. Chief among these are the ability to identify and mitigate systematic effects uncorrelated across different datasets, the facilitation of an assessment of discrepancies between data placed on equal footing, and the breaking of parameter degeneracies for tighter constraints on cosmological models. By incorporating measurements spanning a wide range of scales and redshifts, multiprobe techniques provide a robust end-to-end test of cosmological models and their possible extensions. In recent years, numerous combined probe studies have been performed using data from contemporary surveys. A prominent example in large-scale structure (LSS) is the $3\times2$pt methodology, which combines weak lensing (WL) and galaxy clustering at the two-point correlation level; this approach is standard for photometric galaxy surveys such as the Dark Energy Survey (DES)~\cite{DES:2017myr}, the Hyper Suprime-Cam (HSC) Survey~\cite{HSC:2018mrq}, and the Kilo-Degree Survey (KiDS)~\cite{Heymans:2020gsg, Hildebrandt:2018yau}. Beyond $3\times2$pt analyses, considerable work has focused on cross-correlating CMB secondary signals---including CMB lensing~\cite{Lewis:2006fu}, the Integrated Sachs-Wolfe (ISW) effect~\cite{SachsWolfe1967}, and the thermal Sunyaev-Zeldovich (tSZ) effect---with low-redshift tracers. Notable examples include the cross-correlation of CMB lensing with galaxy clustering~\cite{Piccirilli:2022myi, Singh:2016xey, White:2021yvw, Doux:2017tsv, Krolewski:2019yrv, HerschelATLAS:2014txv, Sailer:2024coh, Farren:2024rla, ACT:2024nrz}, ISW cross-correlations~\cite{Planck:2015fcm, Krolewski:2021znk}, the cross-correlation of CMB lensing with galaxy weak lensing~\cite{Robertson:2020xom, ACT:2023skz, Marques:2020dsb}, and tSZ/kSZ--weak lensing cross-correlations~\cite{Troster:2021gsz, DES:2024iny}.

A key objective within multiprobe cosmology has been to build a unified framework capable of analyzing data from several surveys simultaneously. A seminal development in this direction was the \texttt{cosmolike} framework~\cite{Eifler:2013fit, Krause:2016jvl}, which enabled a $6\times2$pt exploration combining cosmic shear, galaxy-galaxy lensing, galaxy clustering, photometric baryon acoustic oscillations (BAO), cluster number counts, and cluster weak lensing. More recent efforts extend this to include CMB secondary information. Notably, Ref.~\cite{Sgier:2021bzf} applies a map-based formalism~\cite{Nicola:2016eua, Nicola:2016qrc} to jointly analyze cosmic shear, galaxy clustering, ISW, and CMB lensing. Additional extended multiprobe frameworks have been proposed in e.g. Refs.~\cite{Xu:2023qmp, Fang:2023efj, Eifler:2024jai, Garcia-Garcia:2021unp, Ruiz-Zapatero:2022zpx, Johnston:2024wiz, Ruiz-Zapatero:2023hdf}.

In our previous work~\cite{Reeves:2023kjx} (hereafter Paper I), we introduced a multiprobe pipeline for joint cosmological analysis. The pipeline was validated using mock datasets, combining primary CMB, CMB lensing, projected galaxy clustering, and galaxy weak lensing measurements—including all cross-correlations between these probes (yielding 12 distinct spherical harmonic power spectra, $C_\ell$). Here, we present the first application of that pipeline to real survey data. We assess the internal consistency of the dataset and identify outlier spectra before ultimately deriving robust cosmological constraints. Finally, we perform a model comparison to determine if there is a statistical preference over $\Lambda\mathrm{CDM}$ for any of the models considered in this study. 

The rest of this paper is structured as follows: first,
in Section~\ref{sec:datasets} we describe the various datasets used in this analysis. In Section~\ref{sec:pipeline}, we outline our methodology to derive the masks, covariance matrix, and data vector for these datasets required for our later analysis. In Sections~\ref{sec:modeling} and ~\ref{sec:inference} we detail respectively the theoretical modeling including the use of \texttt{JAX}-based emulators and the inference set-up used in this framework. We present our results in Section~\ref{sec:results}, beginning with an assessment of internal consistency and goodness of fit before a comprehensive analysis of cosmological constraints in $\Lambda\mathrm{CDM}$ and extensions.  Finally, in Section~\ref{sec:conclusion} we draw our conclusions. Additional material supporting these results is presented in the Appendices, including justification for some of the assumptions made in the likelihood modeling and extra parameter tables.

\section{Datasets\label{sec:datasets}}

In this work, we analyze data from 5 different cosmological surveys. These are KiDS-1000\footnote{\url{https://kids.strw.leidenuniv.nl/DR4/lensing.php}}, BOSS-DR12\footnote{\url{https://data.sdss.org/sas/dr12/boss/}}, \textit{Planck} (PR3 and PR4)\footnote{\url{http://pla.esac.esa.int/pla/\#home}}, Atacama Cosmology Telescope (ACT) DR4\footnote{\url{https://lambda.gsfc.nasa.gov/product/act/act_dr4_maps_info.html}} and DESI\footnote{\url{https://data.desi.lbl.gov/doc/}}. We split these data into two main categories: CMB data comprising CMB primary auto- and cross- temperature and polarization correlations (TTTEEE) and low-$z$ data which consists of CMB lensing\footnote{Note here even though the CMB weak lensing kernel has support over the redshift range to recombination, $z*$, we consider this to be low-$z$ since the kernel peaks at $z\sim2$.}, projected galaxy clustering and galaxy shear data. We treat the BAO data as a separate addition and it is \emph{not} included in our baseline low-$z$ collection of data. 

\paragraph{KiDS-1000}  
We use data from the fourth release of the Kilo-Degree Survey (KiDS)~\cite{Kuijken:2019gsa, KiDS:2020suj}, covering 1006 square degrees and comprising $\sim 2\times10^7$ galaxies observed across nine photometric bands, including \textit{ugri} and near-infrared contributed by VIKING~\cite{Giblin:2020quj, KiDS:2020suj}. We analyze angular power spectra ($C_\ell$) using the pseudo-$C_\ell$ formalism, as in Refs.~\cite{Troster:2021gsz, KiDS:2021opn}.

\paragraph{BOSS DR12} 
We analyze the final public data release of BOSS, which is part of the Sloan Digital Sky Survey (SDSS) III project. This release (BOSS DR12) includes two primary spectroscopic galaxy catalogs: LOWZ and CMASS.  The galaxies within these samples are targeted by applying color-color and color-magnitude cuts to the SDSS photometric catalog~\cite{Reid:2015gra,SDSS-III:2015hof}.

\paragraph{\textit{Planck} 2018 and PR4 Power Spectra}
The \textit{Planck} satellite measured nearly full-sky observations of the anisotropies in the polarization and temperature of the CMB between 2009 and 2013~\cite{Planck:2018vyg}. We use data from both the third data release (PR3) and the final legacy release (PR4) of \textit{Planck}, including temperature and polarization anisotropies~\cite{Planck:2018vyg, Planck:2018nkj, Tristram:2023haj}. PR4 uses the \texttt{NPIPE} pipeline with improved calibration and inclusion of repointing data.  

\paragraph{\textit{Planck} 2018 Lensing}
We use gravitational lensing maps from the PR3 \textit{Planck} data release~\cite{Planck:2018lbu}. While more recent lensing maps from PR4~\cite{Carron:2022eyg} and ACT DR6~\cite{ACT:2023kun} offer higher signal-to-noise, we leave the incorporation of these datasets for future work.

\paragraph{ACT DR4 Primary Power Spectra}
ACT is a 6-meter telescope in Chile that observes $\sim 18,000$ deg$^2$ of the sky. Temperature and polarization data from the fourth data release (DR4) of ACT are used in this analysis~\cite{Mallaby-Kay:2021tuk, ACT:2020gnv}.  

\paragraph{DESI Y1 BAO}
Finally, we include data from the first public release of DESI~\cite{DESI:2024uvr, DESI:2024mwx}, which comprises BAO measurements in seven redshift bins spanning $0.1 < z < 4.2$. These measurements are based on $\sim 6$ million galaxies and quasars, providing constraints on the transverse comoving distance and Hubble parameter relative to the sound horizon.

\section{Masks, covariance matrix and data vector\label{sec:pipeline}} 
Our analysis distinguishes between modeling CMB and low-$z$ data. The low-$z$ data are modeled using a simulation-based covariance matrix, with the pseudo-$C_\ell$ data vector measured from projected data maps. In contrast, the CMB and BAO data are treated at the likelihood level. Details of the likelihood implementation for CMB and BAO data are given in Section~\ref{subsec:likelihood}. This section describes our methodology for the low-$z$ data, including mask creation, data-vector production, and covariance matrix generation. We provide a high-level summary here and refer readers to Paper I~\cite{Reeves:2023kjx} for further details. We use a fiducial resolution corresponding to \healpix $\mathrm{NSIDE}=2048$ for all maps.

\subsection{Map-level mock simulations} \label{subsec:simualtions}
\begin{table}[h]
\centering
\begin{tabular}{cc}
    \hline
    Parameter & Value \\
    \hline
    $\Omega_{m}$ & 0.26 \\
    $\sigma_{8}$ & 0.84 \\
    $n_{s}$ & 0.9649 \\
    $\Omega_{b}$ & 0.0493 \\
    $H_{0}$(Km/s/Mpc) & 67.3 \\
    $M_\nu$(eV) & 0.06 \\
    \hline
\end{tabular}
\caption{Fiducial Cosmological parameters used in the \cosmogrid.}
\label{tab:fidcosmo}
\end{table}

For the low-$z$ probes, we create map-level realizations of each probe using the \ufalcon{}~\cite{Reeves:2023kjx, Sgier:2018soj, Sgier:2020das} code. \ufalcon{} is a publicly available software\footnote{\url{https://cosmology.ethz.ch/research/software-lab/UFalcon.html}} designed to generate signal realizations of several LSS probes including galaxy weak lensing, CMB lensing, projected galaxy clustering and ISW maps given an input N-body lightcone. To create our covariance matrix we use the 200 fiducial simulations of the \cosmogrid{} simulation suite\footnote{\url{http://www.cosmogrid.ai/}} ~\cite{Kacprzak:2022pww}. \cosmogrid{} is a large suite of simulations produced with the GPU-accelerated N-Body code \pkdgrav\ \cite{Potter:2016ttn}. The fiducial cosmology is shown in Table~\ref{tab:fidcosmo}. We add 10 random survey noise realizations to each of the \ufalcon{} generated signal realizations to get 2000 pseudo-random mock maps from which the low-$z$ multiprobe covariance matrix can be estimated. Finally, we must include the appropriate survey mask in all of our simulated maps. We detail below how noise realizations, tomographic $n(z)$ distributions, and masks are generated for each of the probes considered:

\paragraph{Galaxy clustering} The BOSS-DR12 mask is constructed from acceptance and veto masks provided by the BOSS collaboration in the \texttt{MANGLE} format~\cite{Swanson:2007aj, Hamilton:2003ea}. The acceptance mask represents the ratio of galaxies with measured redshifts to those targeted:
\begin{equation}
C = \frac{N_{\text{observed}}}{N_{\text{targeted}}}.
\end{equation}
Following Ref.~\cite{Loureiro:2018qva}, we convert the acceptance mask to a high-resolution \healpix map ($\mathrm{NSIDE}=16384$), apply the veto masks (which remove regions affected by observational systematics~\cite{Reid:2015gra}), and then downgrade the combined map to our analysis resolution ($\mathrm{NSIDE}=2048$). A binary mask is then produced by assigning 1 to pixels with values greater than 0.75, and 0 otherwise, following the methodology of Ref.~\cite{Doux:2017tsv}. We use two broad redshift bins for our BOSS DR12 modeling which cover the following ranges: $0<z<0.4$ (LOWZ) and $0.4<z<0.8$ (CMASS) and are taken from the BOSS data release. Finally, Poisson noise realizations are created by randomizing galaxy positions within the BOSS-DR12 survey mask. We model the shot noise as following a Poisson distribution as this is accurate on the large linear scales that we analyze in this work (see e.g. Ref.~\cite{Schmittfull:2018yuk} for a discussion of non-Poisson shot noise effects that become important at small scales.). 

\paragraph{Cosmic Shear} We create a survey mask using the `lensing weights' derived from \texttt{lensfit}, following Ref.~\cite{Troster:2021gsz}\footnote{\url{https://github.com/tilmantroester/KiDS-1000xtSZ.}}. The \texttt{lensfit} weights are binned into pixels of an $\mathrm{NSIDE}=2048$ \healpix map to produce a mask that weights shear pixel values according to shape measurement uncertainty. The survey galaxies are subdivided into five distinct redshift bins with associated photometric redshift distributions provided in the KiDS-1000 data release covering a total redshift range between $0.1<z_{\mathrm{phot}}<1.2$~\cite{Hildebrandt:2020rno}. Noise realizations for KiDS-1000 are generated by applying random rotations to galaxy ellipticities in the KiDS-1000 catalog, effectively removing lensing correlations while preserving shape noise. The rotated galaxies are then projected into tomographic maps and added to the pure signal realizations from \ufalcon{}. 

\paragraph{CMB Lensing} For \textit{Planck} PR3 lensing, we use the masks provided by the \textit{Planck} collaboration at $\mathrm{NSIDE}=2048$. These masks are apodized using a Gaussian smoothing kernel of FWHM $1.0$ deg to mitigate power leakage from noise-dominated small-scale modes~\cite{Krolewski:2021yqy, Garcia-Garcia:2021unp, Krolewski:2021znk}. CMB lensing noise realizations are generated by subtracting the cosmological signal and mean field bias from the spherical harmonic coefficients ($a_{\ell m}$s) of the \textit{Planck} Full Focal Plane (FFP10) simulations\footnote{Available from the Planck Legacy Archive: \url{https://pla.esac.esa.int/home}.}. These are then projected to maps at \healpix $\mathrm{NSIDE}=2048$ using \texttt{healpy}'s \texttt{alm2cl} function. This approach makes the approximation that the noise can be treated as a linear addition to the signal (i.e. there is no noise-signal coupling). This was checked in Ref.~\cite{Nicola:2016qrc} to be an accurate assumption for the noise power spectrum at the 3.5\% level which is sufficient for covariance matrix estimation.

\paragraph{ISW Temperature} We use the \textit{Planck} PR3 temperature mask once again apodized with a Gaussian smoothing kernel of FWHM $1.0$ deg following Ref.~\cite{Krolewski:2021znk}. To add noise realizations for the \textit{Planck} ISW map, we add an FFP10 end-to-end simulation containing both CMB signal and noise components to the ISW signal maps. We are therefore double-counting the ISW signal component in these simulations, though the ISW contribution from the FFP10 simulations is uncorrelated with the other signal maps. We expect this to have a negligible impact on the final covariance matrix as the ISW component is subdominant to the other contributions to the temperature signal map and the noise map. We validated this approximation in Paper I by comparing the diagonal covariance elements to an analytically derived Gaussian ISW covariance, finding deviations below 5\%. 

\subsection{Measuring pseudo-$C_\ell$s \label{subsec:measuring_pseudo}}
We employ the \texttt{NaMaster} framework~\cite{Alonso:2018jzx} to compute pseudo-$C_\ell$ power spectra from projected maps of both simulations and observational data. This approach corrects for the mode coupling induced by the survey mask, ensuring accurate power spectrum estimates in the presence of partial sky coverage. The ensemble average of the cut-sky $C_\ell$ measurements, $\langle\tilde{C}^{XY}_\ell\rangle$, is related to the full-sky power spectrum $C_\ell$ through a mode coupling matrix $M_{\ell \ell'}$
\begin{equation} \label{equation}
\begin{split}
\langle\tilde{C}_{\ell}^{XY}\rangle = \sum_{\ell '} M_{\ell \ell'} C_{\ell '}.
\end{split}
\end{equation}
Recovering the full-sky $C_\ell$ from cut-sky data is generally not possible due to the information loss caused by the survey mask. In this pipeline we apply the MASTER algorithm~\cite{Hivon:2001jp}, implemented in \texttt{NaMaster}~\cite{Alonso:2018jzx}, which by treating the true angular power spectrum as piecewise constant over specified multipole bins can overcome this problem. In particular, this approximation enables the inversion of the binned mode coupling matrix, yielding an unbiased estimate of the full-sky bandpower $C_\ell$s in each bin. 

Additionally, to address pixelization effects, we incorporate the window function obtained from \texttt{healpy}'s \texttt{pixwin} function at $\mathrm{NSIDE}=2048$. This window function is treated as an effective ``beam'' and included in the \texttt{NaMaster} \texttt{NmtField} object for each map. We note that this method of correcting for pixelization effects at high $\ell/\mathrm{NSIDE}$ is only valid in the limit of a high-source pixel density for the given tracer (see appendix A of Ref.~\cite{Nicola:2020lhi}). See also Refs.~\cite{Wolz:2024dro, BaleatoLizancos:2023jbr} for an alternative method to bypass these issues by avoiding pixelization. 

\subsection{Simulation-based covariance matrix} \label{subsec:covariance_matrix}
Once we have made signal-correlated noisy mock realizations for each probe following the procedure outlined in Section~\ref{subsec:simualtions}, we apply the appropriate survey mask to each projected map and measure the auto- and cross-pseudo-$C_\ell$s using the \texttt{NaMaster} framework outlined in Section \ref{subsec:measuring_pseudo}. From these 2000 mock data vector realizations we estimate the multiprobe covariance matrix for the low-$z$ probes. We ignore the cosmological dependence of the covariance matrix which is sufficiently accurate for the datasets used in this work~\cite{Kodwani:2018uaf}. We note here that whilst we use our CMB lensing simulations to estimate the cross-covariance terms for the CMB lensing autocorrelation (i.e. the $<C_\ell^{\kappa \kappa}C_\ell^{xy}>$ where $x$, $y$ are two other probes in the pipeline), we insert the official \textit{Planck} covariance matrix block for $<C_\ell^{\kappa \kappa}C_\ell^{\kappa \kappa}>$ (see further description in Section~\ref{subsec:likelihood}). We checked, in Paper I, that this covariance matrix is converged and agrees within 10\% along the diagonal with an independent analytical calculation for the same matrix.

\subsection{Data vector}
\begin{figure}
\centering
\includegraphics[scale=0.26]{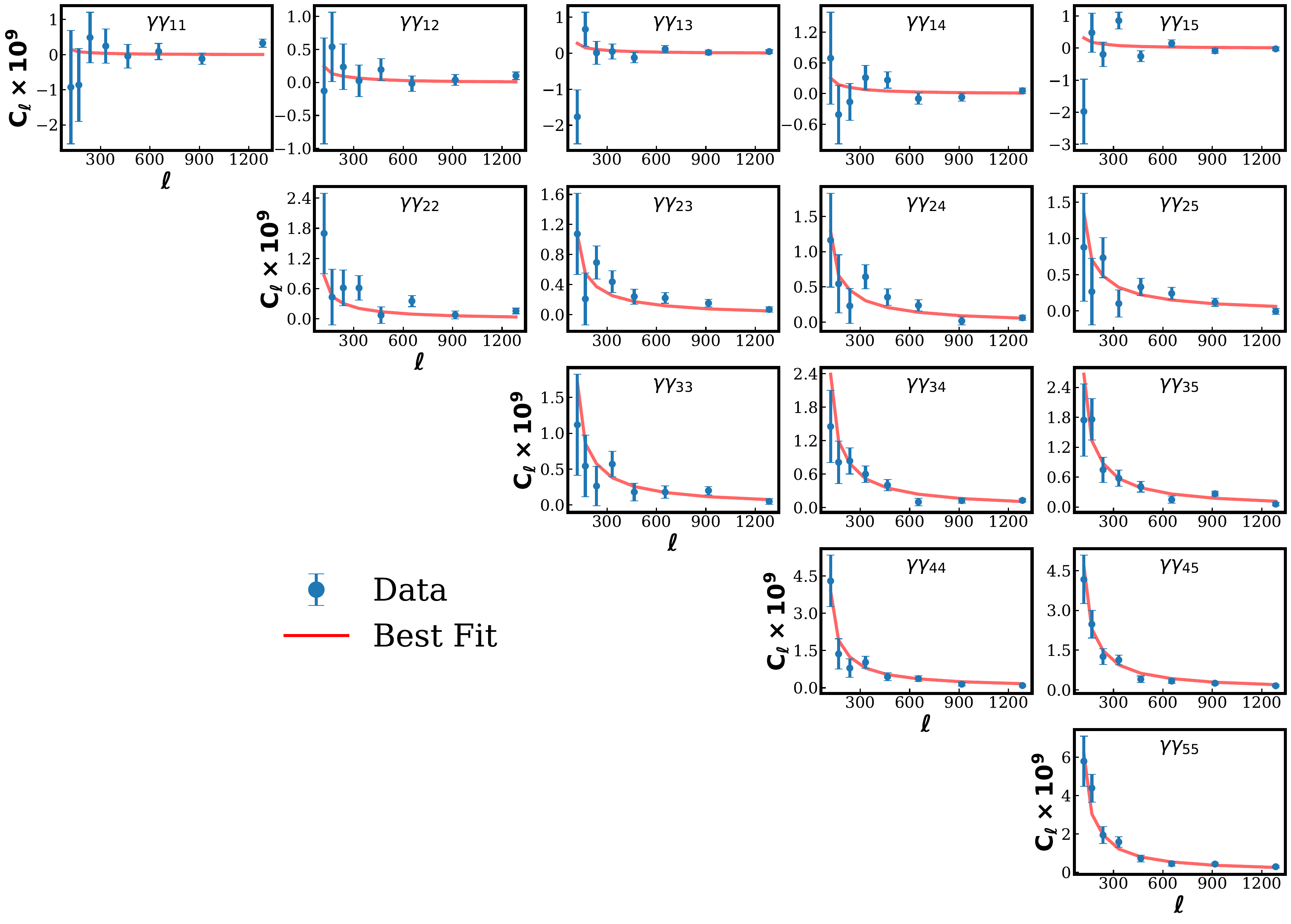}
\caption{Measured Pseudo-$C_\ell$s vs bestfit theory predictions for the KiDS-1000 shear spectra. We use the following acronym system: $\gamma_{1-5}$ refers to the 5 tomographic redshift bins of KiDS-1000. \label{fig:shear_bestfit_vs_data}}
\end{figure}

\begin{figure}
\centering
\includegraphics[scale=0.32]{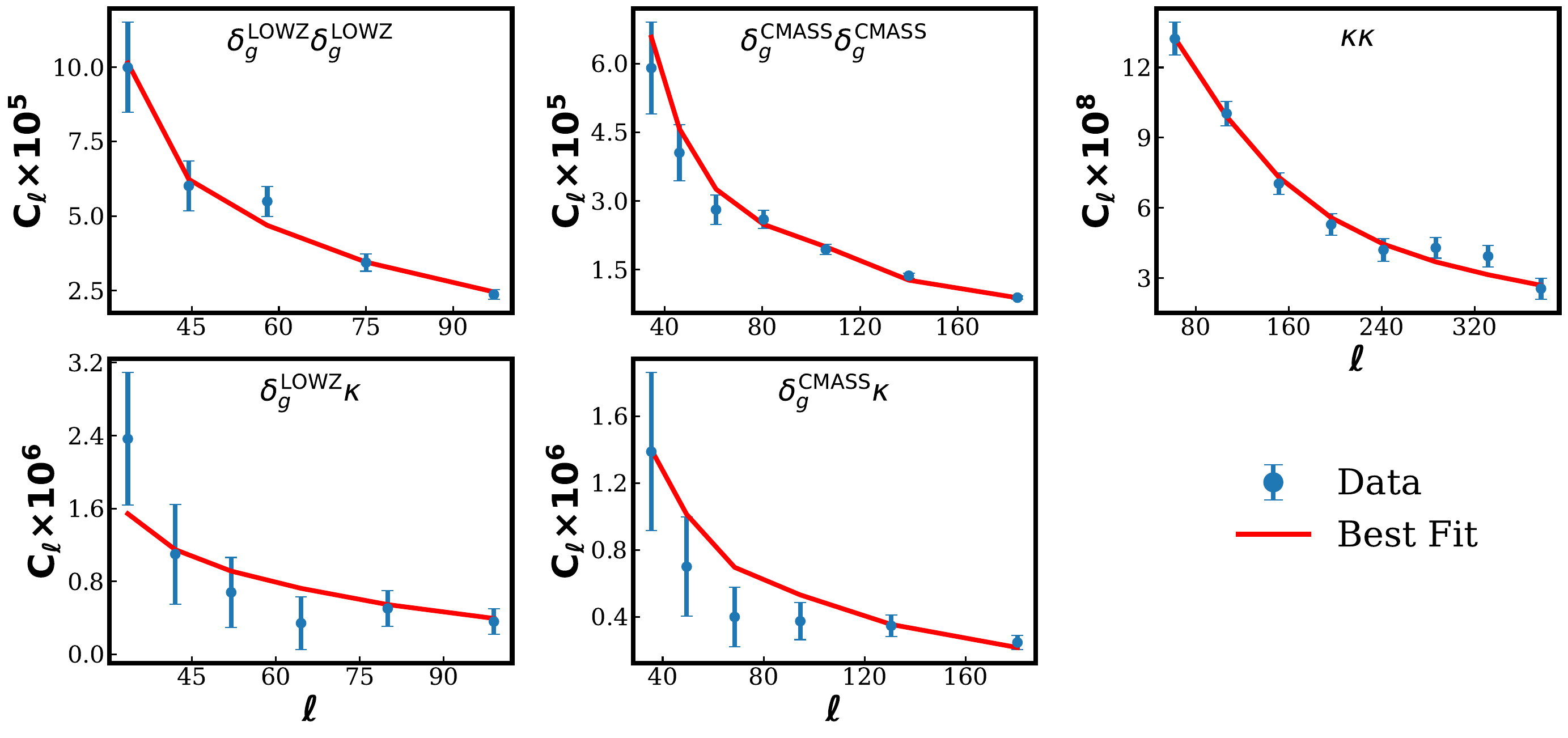}
\caption{Measured Pseudo-$C_\ell$s vs bestfit theory predictions for the BOSS DR12 and \textit{Planck} CMB lensing auto-spectra. We use the following acronym system: $\kappa$ refers to the CMB lensing spectra whilst $\delta_g^{\mathrm{LOWZ}}$ and $\delta_g^{\mathrm{CMASS}}$ are the BOSS DR12 LOWZ and CMASS spectra respectively. \label{fig:auto_bestfit_vs_data}}
\end{figure}

\begin{figure}
\centering
\includegraphics[width=\textwidth]{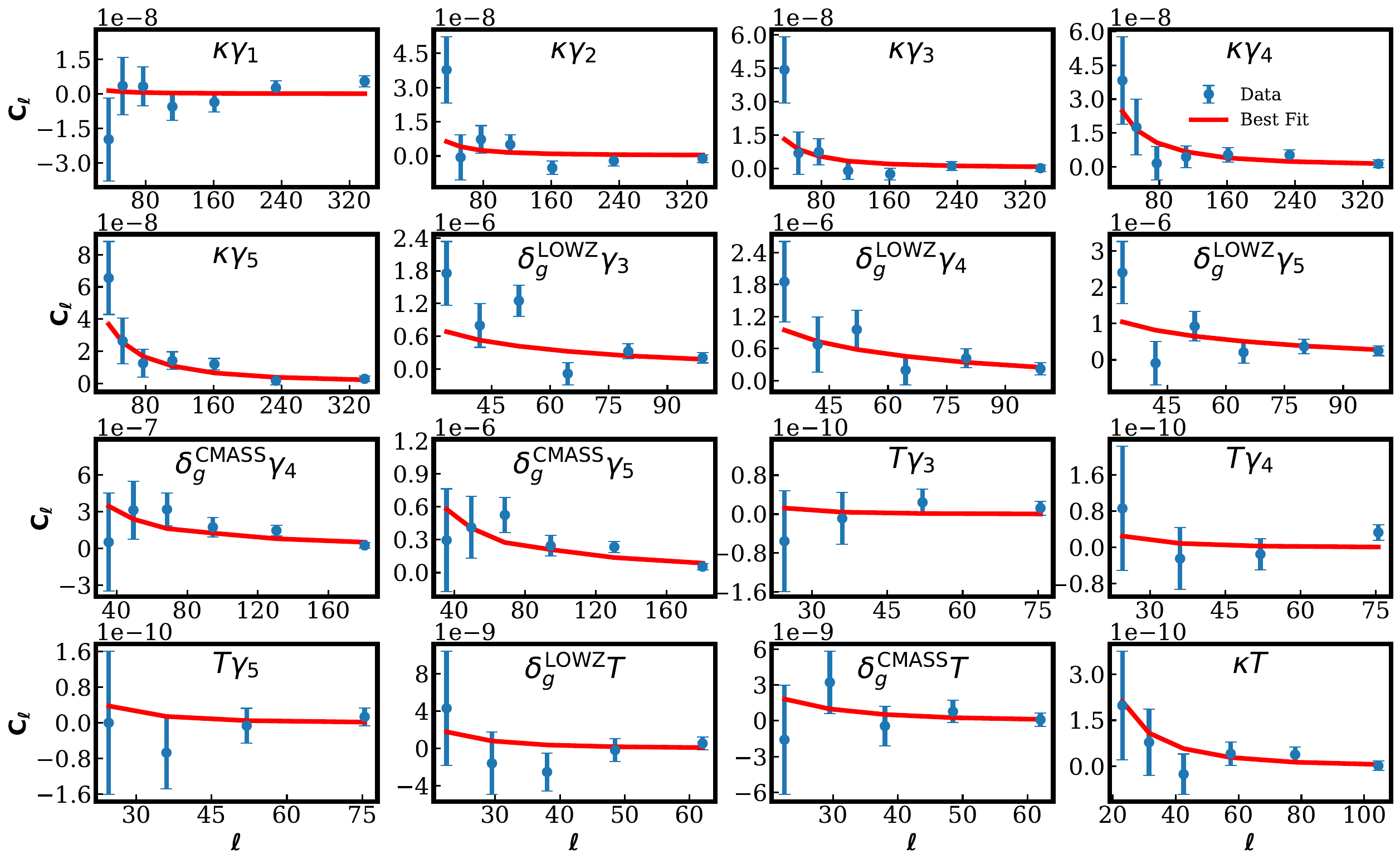}
\caption{Measured Pseudo-$C_\ell$s vs bestfit theory predictions for the measured cross-spectra in the pipeline. We use the following acronym system: $\kappa$ refers to the CMB lensing spectra whilst $\delta_g^{\mathrm{LOWZ}}$ and $\delta_g^{\mathrm{CMASS}}$ are the BOSS DR12 LOWZ and CMASS spectra respectively and $\gamma_{1-5}$ refers to the 5 tomographic redshift bins of KiDS-1000. \label{fig:cross_bestfit_vs_data}}
\end{figure}

\begin{table}[ht]
    \centering
    \caption{Redshift bins, number of objects, sky fraction, and source pixel density in the BOSS DR12 and KiDS-1000 datasets. In the case of KiDS-1000 we use the lensing weights to derive an effective sky coverage which explains the slightly lower footprint in the first two redshift bins.}
    \label{tab:lowz_data}
    \begin{tabular}{llccc}
    \toprule
    \textbf{Survey} & \textbf{Bin} & \textbf{Number of objects} & \textbf{Sky fraction} & \textbf{Source pixel density} \\
    \midrule
    BOSS DR12 & LOWZ  & 390\,200  & 0.20 & 0.04 \\
    BOSS DR12 & CMASS & 823\,193  & 0.22 & 0.07 \\
    \midrule
    KiDS-1000 & 1     & 1\,792\,136 & 0.016 & 2.2 \\
    KiDS-1000 & 2     & 3\,681\,319 & 0.019 & 3.8 \\
    KiDS-1000 & 3     & 6\,148\,102 & 0.020 & 5.9 \\
    KiDS-1000 & 4     & 4\,544\,395 & 0.020 & 4.5 \\
    KiDS-1000 & 5     & 5\,096\,059 & 0.020 & 5.0 \\
    \bottomrule
    \end{tabular}
\end{table}

To produce the data vector for the pipeline, we again distinguish between the CMB, BAO, and low-$z$ datasets. The CMB and BAO data vectors are modeled using the official data products from their respective likelihoods, as described in Section~\ref{subsec:likelihood}. In contrast, the low-$z$ data vector is constructed by projecting relevant quantities onto \healpix maps with a resolution of $\mathrm{NSIDE}=2048$ and measuring the associated auto- and cross-correlation pseudo-$C_\ell$s using the \texttt{NaMaster} framework outlined in Section~\ref{subsec:measuring_pseudo}. This results in 7 distinct auto- and cross- correlations for the low-$z$ dataset. These data are shown along with the bestfit $\Lambda\mathrm{CDM}$ prediction in our baseline dataset (\textit{Planck} PR4 + low-$z$) in Figs.~\ref{fig:shear_bestfit_vs_data}, ~\ref{fig:auto_bestfit_vs_data} and \ref{fig:cross_bestfit_vs_data}. Below, we detail the projection process for each probe map required to measure the low-$z$ data vector: 

\paragraph{KiDS-1000}
We pixelise the shear catalogues into \healpix maps $\gamma$ with
\begin{equation}
    \gamma(\mathbf{n_i}) = \frac{1}{w(\mathbf{n_i})} \sum_{j \in \mathbf{n_i}} w_j \epsilon_j,
\end{equation}
where $\mathbf{n_i}$ denotes the position of the i-th pixel and the sum runs over all galaxies in that pixel, which are characterized by their weight $w_j$ and the two ellipticity components, which we denote here with $\mathbf{\epsilon_j}$. In our case, the weights $w_j$ are given by the \texttt{lensfit} weights and account for uncertainty in the shape measurement.
The weight map $w(\mathbf{n_i})$ is defined as $w(\mathbf{n_i})=\sum_{j \in \mathbf{n_i}} w_j$. The multiplicative shear bias is accounted for by scaling all ellipticity estimates in a tomographic bin $\epsilon \to \frac{1}{1+m_z} \epsilon$ where we use the multiplicative bias values $m_z$ from Ref.~\cite{Giblin:2020quj}. We correct for the additive shear bias by subtracting the weighted ellipticity mean in each tomographic bin, following Ref.~\cite{Giblin:2020quj}. The auto-power spectra for galaxy shear are subject to a noise bias, which we estimate analytically following ~\cite{Nicola:2020lhi}
\begin{equation}
    N_\ell = \Omega_{pix} \langle \sum_{j \in \mathbf{n_i}} w_j^2 \sigma^2_{\gamma, j} \rangle_{pix},
\end{equation}
where $\Omega_{pix}$ is the pixel area, $\sigma^2_{\gamma, j} = \frac{1}{2}|\epsilon_j|^2$ with $|\epsilon_j|^2= \epsilon_1^2 + \epsilon_2^2$ and the average is taken over all pixels. 

Our decision to upgrade the resolution to \healpix $\mathrm{NSIDE}=2048$ from $\mathrm{NSIDE}=1024$ in Paper I~\cite{Reeves:2023kjx} is chiefly due to avoiding biases from pixel window effects in the KiDS-1000 cosmic shear dataset. We estimate that our method of dealing with the pixelization effect in our pseudo-$C_\ell$ measurement biases our power spectrum recovery by at most 0.5\% according to Figure 10 of Appendix A in Ref.~\cite{Nicola:2020lhi} given $\ell_{max}=1500$ and the pixel density values reported in Table~\ref{tab:lowz_data}.

\paragraph{BOSS-DR12}
Based on the LOWZ and CMASS galaxy catalogs, we construct galaxy overdensity maps. First, we select galaxies from the data catalogs with redshifts within $z \in [0.0, 0.4]$ and $z \in [0.4, 0.8]$ for LOWZ and CMASS, respectively using both the South and North Galactic cap selections. For both samples, we then compute the number of objects in each pixel
$\mathbf{n_j}$ weighted by the total galaxy weight which is computed as follows: 
\begin{equation}\label{eqt:weight}
    w_{tot} = w_{star}w_{see}(w_{cp}+w_{noz}-1), 
\end{equation}
where $w_{tot}$ is the total galaxy weight, $w_{star}$ accounts for the stellar contamination, $w_{see}$ gives a weight based on atmospheric seeing conditions, $w_{cp}$ is a fiber collision weight and $w_{noz}$ accounts for redshift failures. The LOWZ and CMASS data maps $\delta_g$ are then produced as: 
\begin{equation}
    \delta_g(\mathbf{n_i}) = \sum_{j \in \mathbf{n_i}} w_{tot_j} / \Bar{n},
\end{equation}
where $\mathbf{n_i}$ denotes the position of the i-th pixel and the sum runs over all galaxies in that pixel, $\Bar{n}$ is the average weighted pixel density of galaxies $\Bar{n} = \frac{1}{N_{pix}} \sum_{n_i}  \sum_{j \in \mathbf{n_i}} w_{tot_j}$ where $N_{pix}$ is the number of \healpix pixels within the survey mask. 

\paragraph{Planck 2018 temperature maps for ISW measurements}
In this work, we utilize the \textit{Planck} PR3 temperature maps for our ISW analysis. Although the more recent \texttt{NPIPE} maps are available, we expect the differences introduced by using these updated maps to be negligible given the low signal-to-noise ratio of this part of our data vector. The PR3 foreground-reduced CMB anisotropy maps have been derived from the nine frequency channel maps using four component-separation algorithms: \texttt{Commander}, \texttt{NILC}, \texttt{SEVEM} and \texttt{SMICA}~\cite{Planck:2018yye}. All maps are provided in Galactic coordinates at a \texttt{Healpix} resolution of $\mathrm{NSIDE}=2048$. For the present analysis we use the \texttt{Commander} maps as these are the recommended maps for measurements on large angular scales\footnote{Available at the Planck Legacy Archive, \url{http:/pla.esac.esa.int}}. 

\paragraph{Planck 2018 lensing map}
We use the publicly available \textit{Planck} PR3 lensing maps to measure CMB lensing cross-correlations. The Planck PR3 lensing measurement reconstructs the lensing convergence from a linear combination of minimum variance quadratic estimators whose inputs are Wiener-filtered and inverse-variance-weighted \texttt{SMICA} temperature and polarization multipoles with $100 \leq \ell \leq 2048$. When computing the CMB lensing cross-correlations with the other LSS tracers we also account for the normalization correction that depends on the overlap mask between the two tracers (see Appendix~\ref{appendix:mc_norm} for further details). 

\subsection{Scale cuts \label{subsec:scale_cuts}}
The scale cuts applied in this analysis are chosen to limit the analysis to ranges where the data can be adequately described using relatively simple models with a limited number of nuisance parameters. Specifically, as detailed in the following subsections, we exclude high-$\ell$ regions where small-scale processes such as baryonic effects (for weak lensing) or non-linear clustering (for galaxy clustering) become significant and where the ISW signal-to-noise is too low. We also avoid very low-$\ell$ regions, where box size limitations in simulations (i.e. under-sampling of larger-than-box size modes), large-scale systematic effects, inaccuracy of the Limber approximation~\cite{LoVerde:2008re} and redshift-space distortions~\cite{SDSS:2006egz} complicate the modeling.

\paragraph{KiDS-1000 auto-correlations} For KiDS-1000 auto-correlations, we adopt the scale range $100<\ell<1500$, consistent with the analysis of Ref.~\cite{Troster:2021gsz}. Higher $\ell$ values are excluded due to the difficulty of accurately modeling non-linear power spectra and baryonic feedback at these scales.

\paragraph{BOSS DR12} For galaxy clustering, both in auto- and cross-correlation, we use scales where the linear bias model is applicable. The maximum multipole $\ell_{\text{max}}$ is determined such that the associated Fourier modes satisfy $k<0.1 \mathrm{Mpc}^{-1}$ to avoid the regime where non-linear clustering becomes important at small scales~\cite{Doux:2017tsv}. The minimum multipole is set conservatively at $\ell_{\text{min}}=30$ for autocorrelations to avoid scales where redshift-space distortions dominate~\cite{SDSS:2006egz} and to avoid box size effects in the simulation-based covariance matrix. 

\paragraph{KiDS-1000 cross-correlations with BOSS DR12} For cross-correlations between KiDS-1000 and BOSS DR12, the scale cuts are dictated by the stricter of the cuts applied to the respective auto-correlations. Additionally, we follow Ref.~\cite{Heymans:2020gsg}, in excluding cross-correlation bins where there is significant overlap between KiDS-1000 source galaxies and BOSS DR12 galaxies, as in this regime the signal is dominated by intrinsic alignments (IA) and the simple non-linear alignment (NLA) model used in this analysis may not be sufficiently accurate (see further details on modeling in Section~\ref{sec:modeling}). We hence include only the correlations $C_\ell^{\gamma_3 \delta_{\text{lowz}}}, C_\ell^{\gamma_4 \delta_{\text{lowz}}}, C_\ell^{\gamma_5 \delta_{\text{lowz}}}, C_\ell^{\gamma_4 \delta_{\text{cmass}}}$, and $C_\ell^{\gamma_5 \delta_{\text{cmass}}}$.

\paragraph{\textit{Planck} CMB lensing cross-correlations} For correlations involving \textit{Planck} CMB lensing, we use the range $30<\ell<400$. The lower limit avoids scales where box size effects and the Limber approximation may introduce inaccuracies~\cite{LoVerde:2008re}, while the upper limit follows the conservative \textit{Planck} 2018 lensing analysis cuts~\cite{Planck:2018lbu}, where lensing bias uncertainties become significant. 

\paragraph{\textit{Planck} ISW cross-correlations} For ISW cross-correlations, we use a slightly extended range with $\ell_{\text{min}}=20$ to improve the signal-to-noise ratio (S/N), which peaks on larger scales. While this lower limit could marginally affect the accuracy of the recovered covariance matrix due to low-$\ell$ box size effects in the simulation-based covariance matrix, the impact is expected to be minimal given the inherently low S/N of ISW measurements relative to other probes. We exclude the $T\gamma_1$ and $T\gamma_2$ cross-correlations, as we find their low S/N makes simulation-based estimates of their covariance unreliable.

\begin{table}[ht]
\centering
\caption{Multipole ranges and binning for all auto- and cross-correlations. 
We inherit the \textit{Planck} (primary and lensing auto-correlation), ACT and WMAP scale cuts and binning from the official analyses, hence no explicit log-binning is shown 
for these spectra.}
\label{tab:scale_cuts}
\begin{tabular}{|lcc|}
\hline
\textbf{Dataset / Correlation} & \boldmath$\ell$\textbf{-range} & \textbf{\#\,log-bins} \\
\hline
\multicolumn{3}{|c|}{\textbf{CMB Datasets}} \\
\hline
Planck PR3/PR4 TT, EE (low-$\ell$) & 2--30 & --- \\
Planck PR3/PR4 TTTEEE (high-$\ell$) & 30--2508 & --- \\
WMAP TT (low-$\ell$) & 2--1200 & --- \\
WMAP TE (low-$\ell$) & 24--800 & --- \\
ACT DR4 TTTEEE (high-$\ell$) & 350--4125 & --- \\
$\kappa \,\kappa$     & 44--400 & --- \\
\hline
\multicolumn{3}{|c|}{\textbf{LSS Auto-Correlations}} \\
\hline
$\delta_l \,\delta_l$ & 30--110 & 5 \\
$\delta_c \,\delta_c$ & 30--210 & 7 \\
$\gamma \,\gamma$     & 100--1500 & 8 \\
\hline
\multicolumn{3}{|c|}{\textbf{LSS Cross-Correlations}} \\
\hline
$\delta_l \,\gamma$   & 30--110 & 6 \\
$\delta_c \,\gamma$   & 30--210 & 6 \\
$\kappa \,\gamma$     & 30--400 & 7 \\
$\delta_l \,\kappa$   & 30--110 & 6 \\
$\delta_c \,\kappa$   & 30--210 & 6 \\
$T \,\kappa$          & 20--120 & 6 \\
$T \,\gamma$          & 20--90  & 4 \\
$T \,\delta_l$        & 20--70  & 5 \\
$T \,\delta_c$        & 20--70  & 5 \\
\hline 
\end{tabular}
\end{table}

\section{Modeling and emulator\label{sec:modeling}}
\subsection{Theoretical modeling and code implementation \label{subsec:theory_modeling}} 

\paragraph{Low-$z$ spectra} As in Paper~I, we use the Limber approximation~\cite{Limber:1954zz} to compute the projected auto- and cross-correlations in our low-$z$ dataset. The theoretical modeling adopts a linear bias prescription for galaxy clustering and a non-linear alignment (NLA) model for intrinsic alignments. We do not include magnification bias in our galaxy modeling, though we expect this to have a negligible impact on our constraints (see Ref.~\cite{Heymans:2020gsg} who found this to be negligible in the context of $3\times2$pt analysis with KiDS-1000 and BOSS). In contrast to Paper I, we employ the CCL library~\cite{LSSTDarkEnergyScience:2018yem} with a \texttt{camb}~\cite{Lewis:2002ah} backend instead of the previous \texttt{classyobs} code. We made this change as we found that the \texttt{classyobs} implementation of the Limber integrals was sensitive to the sampling of the integrand in redshift space whereas CCL was found to be more stable. In addition to this, the use of CCL allows us to update our baryonic feedback modeling from \texttt{HMcode-2016} to \texttt{HMcode-2020}~\cite{Mead:2020vgs}, using the single-parameter variant with $\log_{10}(T_{\rm AGN}/k)$ characterizing the feedback strength.

Corrections must be applied to ensure consistency between the binned theoretical predictions and data bandpowers. The \texttt{Master} algorithm assumes that $C_\ell$s are piece-wise constant. In practice, theoretical $C_\ell$s are not constant within a given $\ell$ bin, which introduces discrepancies. To account for this, we first couple the theoretical $C_\ell$ using the mode coupling matrix $M_{\ell \ell'}$, then bin these coupled values to match the format of the data bandpowers, and finally decouple them using the inverted binned mode coupling matrix thereby correcting for the piecewise-constant assumption~\cite{Alonso:2018jzx}.

\paragraph{CMB} For the CMB primary power spectra, we use the Boltzmann solver \texttt{class}~\cite{Blas:2011rf} to directly make training data for the emulators. We increased the default precision settings of \texttt{class} using: 
\begin{itemize}
    \item \texttt{accurate\_lensing}=1
    \item \texttt{k\_max\_tau0\_over\_l\_max}=15
    \item \texttt{perturbations\_sampling\_stepsize} = 0.15
\end{itemize}
these settings were found in Ref.~\cite{Bolliet:2023sst} to be an optimal trade-off between computation speed and accuracy. The first two parameters here ensure accuracy for high-$\ell$ lensed CMB primary power spectra which is required for the ACT DR4 likelihood.

\paragraph{BAO spectra} Finally, for the BAO data we employ the Boltzmann solver \texttt{class} to generate theoretical predictions for the relevant BAO observables at the redshifts probed by the DESI Y1 BAO dataset~\cite{DESI:2024mwx, DESI:2024uvr, DESI:2024lzq}. Specifically, for each redshift bin in the DESI Y1 sample, we compute the comoving distance $D_\mathrm{M}(z)$, the Hubble distance $D_\mathrm{H}(z) = 1/H(z)$, or the volume-averaged distance $D_\mathrm{V}(z) = \bigl[D_\mathrm{M}(z)^2 \, z \, D_\mathrm{H}(z)\bigr]^{1/3}$, depending on which observable is measured in that bin in the DESI data release. Each of these distances is then divided by the sound horizon at the drag epoch, $r_d$, so that our predictions match the form of the DESI Y1 BAO data.  

\paragraph{Neutrino masses and excess lensing parameter} We model neutrino masses using three degenerate mass states, which is sufficiently accurate given the precision of current observations~\cite{Herold:2024nvk, Vagnozzi:2017ovm, Giusarma:2016phn, Archidiacono:2016lnv}. In certain analyses, we also consider a free lensing amplitude rescaling parameter, $A_L$, which we refer to as \textbf{$A_{\mathrm{lens}}$} in this text. Introduced in~\cite{Calabrese:2008rt}, this phenomenological parameter scales the amplitude of the lensing potential used to calculate the lensed CMB primary anisotropies (i.e., the TTTEEE spectra):
\begin{equation}\label{eqn:a_l_definition}
    C_\ell^{\phi \phi} \to A_L C_\ell^{\phi \phi}.
\end{equation}
Importantly, the lensing potential is \emph{not} rescaled when computing the reconstructed CMB lensing auto-correlation (our definition is therefore equivalent to the $A_{\mathrm{smear}}$ of Ref.~\cite{Loverde:2024nfi}). This provides an internal consistency check of the CMB data, with the expectation that $A_L = 1$ in $\Lambda \mathrm{CDM}$. 

\subsection{\texttt{JAX}-based emulators}
We use neural networks to accelerate the inference stage of our framework, following the approach introduced in Paper I~\cite{Reeves:2023kjx}. In brief, we first create training data for each spectrum in our pipeline by sampling cosmological parameters on a Latin Hypercube (LHC) grid. At every sampled point, we use our theory code to generate the target spectra. To reduce their dynamic range, we take the logarithm of each dataset, except for $C_\ell^{TE}$, where the presence of negative values precludes a simple log transform. For $C_\ell^{TE}$, we apply the strategy from Ref.~\cite{SpurioMancini:2021ppk}, performing a Principal Component Analysis (PCA) and retaining the 512 highest-variance modes to mitigate large variance in the training data and thereby improve model accuracy. 

We adopt the neural-network design from Ref.~\cite{SpurioMancini:2021ppk}\footnote{Available at \url{https://github.com/alessiospuriomancini/cosmopower}}, implemented in \texttt{TensorFlow}~\cite{tensorflow2015-whitepaper}. Each model consists of four hidden layers with 512 neurons per layer. The activation function in each layer is given by
\begin{equation}
\label{eqt:activation}
    f(\vec{x}) \;=\; \bigg(\vec{\gamma} \;+\;\big(1 + \exp(- \vec{\beta}\cdot\vec{x})\big)^{-1}\,(1-\vec{\gamma})\bigg)\,\cdot\,\vec{x},
\end{equation}
where $\vec{\beta}$ and $\vec{\gamma}$ are trained alongside the usual network weights. We use the \texttt{ADAM} optimizer~\cite{adamxyz} with default parameters, a batch size of 1024, and train each emulator for 1000 epochs. Across 20,000 test points, the median relative absolute deviation between emulated and true spectra remains below 0.1\% for all models (see Paper I~\cite{Reeves:2023kjx} for further validation plots of the emulator accuracy).

The main update in our pipeline is that after training the emulators in \texttt{TensorFlow}, we load the optimized weights and biases into a \texttt{JAX} framework using \texttt{flax}~\cite{flax2020github}, following an approach similar to Ref.~\cite{Piras:2023aub}. We also now incorporate a newly developed emulator for the BAO quantities described in Section~\ref{subsec:theory_modeling}, enabling us to include this additional dataset in our framework. 

\section{Likelihood and inference \label{sec:inference}}
\subsection{Likelihood} \label{subsec:likelihood}
In this section, we describe how we bring the covariance matrix and data vector together with our emulators to compute the likelihood $\mathscr{L}(\theta)$. We also describe how the likelihoods between different probes are combined. All likelihoods used in this pipeline have been (re-)written in \texttt{JAX} enabling a rapid computation that benefits from \texttt{JAX}'s just-in-time (JIT) compilation and GPU acceleration.

\paragraph{low-$z$ probes} We assume a Gaussian likelihood with a fixed covariance matrix for the low-$z$ probes and hence our likelihood takes the form: 
\begin{equation}
    \log(\mathscr{L}^{LSS}) = -0.5* (\vec{T}(\theta)-\vec{D})^TC^{-1}(\vec{T}(\theta)-\vec{D}),
\end{equation}
where $T(\theta)$ is the theory prediction at the parameters $\theta$, $\vec{D}$ is the concatenated data vector for each of the LSS probes measured from the projected maps and $C$ is the LSS covariance matrix estimated from simulations. The naïve estimate of the inverse covariance matrix from a finite number of simulations is biased to be too large giving rise to erroneously small confidence regions~\cite{Hartlap:2006kj}. We hence de-bias our estimate using the `Hartlap' factor: 
\begin{equation}\label{eqt:hartlap}
    C^{-1} \to \frac{n-p-2}{n-1} C^{-1}, 
\end{equation}
where $n=2000$ is the number of independent realizations used to estimate the covariance and $p=245$ is the number of concatenated bandpower data points.

\paragraph{CMB lensing auto-correlation} \label{subsubsec:cmb_lensing}
As in Paper I, we use the \textit{Planck} CMB-marginalized likelihood for the $C_\ell^{\kappa \kappa}$ auto-correlation~\cite{Planck:2018lbu}. In principle, combining CMB lensing and primary CMB data in a single inference requires consistently re-evaluating the normalization and $N_1$ bias from the sampled CMB power spectra. The CMB-marginalized approach integrates out the dependence on the primary CMB spectra, leaving a likelihood sensitive only to the lensing potential power spectrum, $C_L^{\phi \phi}$. We showed in Paper I, that using the CMB-marginalized likelihood alongside the primary CMB likelihood yields constraints indistinguishable from the fully consistent treatment. To implement this likelihood we take the official CMB marginalized covariance matrix from the \textit{Planck} PR3 release and insert this into the appropriate sub-block of the low-$z$ covariance matrix. We also use the \textit{Planck} team's estimates of the CMB lensing bandpowers and the appropriate bandpower binning matrices. We exclude the lowest multipole bin of the lensing auto-correlation likelihood, as we find that our simulation-based covariance estimates of the cross-covariance (i.e. $\mathrm{COV}(C_\ell^{\kappa \kappa}C_\ell^{x,y})$ where $x$, $y$ are two other spectra in the pipeline) at large scales suffer from box size effects. This is because this autocorrelation bin includes regions where box size effects become important for the \cosmogrid{} simulations used here ($\ell_{min} \lesssim 10$, see Ref.~\cite{Sgier:2018soj} for an in-depth discussion of box size effects in \texttt{UFalcon}). 

Collectively, the joint $9 \times 2$pt dataset consisting of KiDS-1000, BOSS-DR12, \textit{Planck} PR3 lensing and ISW and all associated cross-correlations will be referred to as  ``\textbf{low-z}''. 

\paragraph{CMB primary (TTTEEE)} The CMB primary power spectra likelihoods are computed following the exact procedures outlined by the respective collaborations. We use the collaboration-produced likelihoods, including covariance matrices, binning matrices, and non-Gaussian low-$\ell$ likelihoods. We neglect the cross-covariance between LSS tracers and the CMB primary probes. This simplification is justified by the fact that the amplitude of the cross-correlation with LSS tracers is negligible compared to the dominant CMB primary power spectra (see Ref.~\cite{Schmittfull:2013uea}, though this statement may need to be re-addressed for future surveys~\cite{Peloton:2016kbw}).

\paragraph{Planck PR3}
For our \textit{Planck} PR3 modeling, we follow the approach from Paper I, relying on official data products for the CMB primary power spectra. We re-implement the Python-based high-$\ell$ \texttt{Plik\_lite} likelihood, \texttt{planck-lite-py}\footnote{\url{https://github.com/heatherprince/planck-lite-py}}~\cite{Prince:2019hse} in the \texttt{JAX} framework.  This likelihood has already been marginalized over the foreground parameters of the \textit{Planck} sky model following Ref.~\cite{Dunkley:2013vu}. For $\ell < 30$ in TT and EE, we use a \texttt{JAX} implementation of the the log-normal binned likelihood \texttt{planck-low-py} from Ref.~\cite{Prince:2021fdv}\footnote{\url{https://github.com/heatherprince/planck-low-py}}. There, the binned data in $D_\ell^{TT/EE} = \ell(\ell+1)C_\ell^{TT/EE}/(2\pi)$ is well described by an offset-lognormal distribution
\begin{equation}
    \mathscr{L}(x) \;=\; \frac{1}{\sqrt{2\pi}\,\sigma\,(x - x_0)}\,\exp\!\Bigl[-\,\frac{\bigl(\ln\,(x - x_0)\;-\;\mu\bigr)^2}{2\,\sigma^2}\Bigr],
\end{equation}
where $x_0$ is an offset parameter and $(\mu,\,\sigma)$ specify the mean and variance in log-space. We adopt the low-$\ell$ parameter values given by Ref.~\cite{Prince:2021fdv} for \textit{Planck} PR3. There is just one nuisance parameter, shared by the high-$\ell$ and low-$\ell$ likelihood, $A_{\mathrm{Planck}}$, which rescales all primary spectra by $C_\ell^{TT/TE/EE} \to A_{\mathrm{Planck}}^2\,C_\ell^{TT/TE/EE}$. Following Ref.~\cite{Planck:2018vyg}, we impose a Gaussian prior $A_{\mathrm{Planck}} \sim \mathcal{N}(1,0.025)$.  We term this likelihood ``\textbf{\textit{Planck} PR3}''.

\paragraph{Planck PR4}
We re-implement both the \hillipop\ high-$\ell$ likelihood~\cite{Tristram:2023haj}\footnote{\url{https://github.com/planck-npipe/hillipop}} and the low-$\ell$ ($\ell<30$) \lollipop\ likelihood~\cite{Tristram:2020wbi,Tristram:2021tvh}\footnote{\url{https://github.com/planck-npipe/lollipop}} in \texttt{JAX}. The \hillipop\ likelihood includes a foreground model that accounts for:
\begin{itemize}
    \item Galactic dust
    \item Cosmic infrared background (CIB)
    \item Thermal and kinetic Sunyaev--Zel'dovich (tSZ and kSZ) effects
    \item Point sources from radio and infrared star-forming galaxies
    \item The CIB\,$\times$\,tSZ cross-correlation
\end{itemize}
These contributions are modeled at the power-spectrum level by fitting templates (with free amplitudes and spectral indices) to the data. Additional nuisance parameters capture the instrument calibration across \textit{Planck}'s three frequency channels (see Ref.~\cite{Tristram:2023haj} for details). We follow the exact priors and computational steps of Ref.~\cite{Tristram:2023haj}. We will refer to this likelihood as ``\textbf{\textit{Planck} PR4}''.

\paragraph{ACT DR4}  In this work, we make use of the Python implementation of the ACT DR4 likelihood \texttt{pyactlike}\footnote{\url{https://github.com/ACTCollaboration/pyactlike}}~\cite{ACT:2020gnv, ACT:2020frw}. Given that the ACT data does not contain a measurement of the low-$\ell$ polarization power spectrum it is not possible to directly constrain the $\tau$ parameter as at high-$\ell$ this parameter is highly degenerate with an overall rescaling of the amplitude of the spectra. The ACT collaboration imposes a relatively conservative Gaussian prior on the $\tau$ parameter of $\tau=0.065 \pm 0.015$ based on \textit{Planck} PR3 and WMAP estimates. Here we instead use a slightly stronger prior based on the updated \texttt{LoLLiPoP} constraint of $\tau = 0.0590 \pm 0.0061$ when using this dataset. There is a single nuisance parameter for this likelihood, $y_p$ which marginalizes the polarization efficiency and rescales the spectra $C_\ell^{EE}\to y_p^2C_\ell^{EE}$, $C_\ell^{TE}\to y_pC_\ell^{EE}$. We follow the ACT team in placing a flat symmetric prior on this parameter centered on $y_p=1$ ($y_p\in[0.9,1.1]$)~\cite{ACT:2020gnv}. 

We supplement the high-$\ell$ ACT-DR4 likelihood with low-$\ell$ information from WMAP using the final 9-year data release~\cite{Hinshaw_2013}, making this dataset \textit{Planck}-independent. Following the prescription in \cite{ACT:2020gnv}, we combine with the low-$\ell$ WMAP TT/TE likelihood which covers $2 \leq \ell \leq 1200$ in TT and $24 \leq \ell \leq 800$ in TE. We do not use the low-$\ell$ polarization (EE) likelihood from WMAP\footnote{There is evidence of dust contamination in this likelihood~\cite{Planck:2013win}}. We follow \cite{ACT:2020gnv} in ignoring correlations between the datasets across the $\ell$-range where they overlap which is justified by the fact that at least one of the experiments is noise-dominated across the region of $\ell$-overlap. We therefore sum the individual log-likelihoods: 
\begin{equation}
    -2\log(\mathscr{L}_{tot}) = -2\log(\mathscr{L}_{ACT}) -2\log(\mathscr{L}_{WMAP}) 
\end{equation}
We have re-written each of these likelihoods in \texttt{JAX} and will later refer to their combination as ``\textbf{ACT + WMAP}''.

\paragraph{DESI Y1 BAO} In this work, we employ the latest BAO scale measurements from the DESI Data Release 1 (DR1) \citep{DESI:2024mwx}. The DESI DR1 data includes:
\begin{itemize}
    \item The Bright Galaxy Sample (BGS) covering $0.1 < z < 0.4$,
    \item The Luminous Red Galaxy Sample (LRG) in two redshift bins: $0.4 < z < 0.6$ and $0.6 < z < 0.8$,
    \item The Emission Line Galaxy Sample (ELG) spanning $1.1 < z < 1.6$,
    \item The combined LRG+ELG sample covering $0.8 < z < 1.1$,
    \item The Quasar Sample (QSO) in the redshift range $0.8 < z < 2.1$,
    \item The Lyman-$\alpha$ Forest Sample (Ly$\alpha$) with $1.77 < z < 4.16$.
\end{itemize}
We take the BAO measurements from the \texttt{cobaya}~\cite{Torrado:2020dgo} repository, which maintains a collection of publicly available data.\footnote{\url{https://github.com/CobayaSampler/bao_data}} In particular, we implement a simple Gaussian likelihood using the covariance provided by the DESI team and our BAO emulator for theoretical predictions. As for the CMB primary spectra, we neglect the cross-covariance of this likelihood with the low-$z$ data vector where because we use projected LSS quantities the BAO information content is negligible compared to the DESI data (see Appendix~\ref{appendix:bao_information} where we explore and justify this statement). We refer to this likelihood as ``\textbf{DESI Y1 BAO}''. 

\subsection{Inference pipeline} 

\begin{table}[h]
\centering
\begin{tabular}{|ccc|}
    \hline
    parameter & prior min & prior max \\
    \hline 
    \multicolumn{3}{|c|}{cosmology} \\
    \hline
    $\omega_b$ & 0.018 & 0.026 \\
    $\Omega_{m}$ & 0.15 & 0.8 \\
    $n_s$ & 0.85 & 1.05\\
    $h$ & 0.6 & 0.9 \\
    $\sigma_8$ & 0.4 & 1.3 \\
    $\tau_{\text{reio}}$ & 0.02 & 0.12\\    
    \hline 
    \multicolumn{3}{|c|}{nuisance} \\
    \hline
    $A_\mathrm{IA}$ & -6 & 6 \\
    $b_{\mathrm{LOWZ}}$ & 0.1 & 5 \\
    $b_{\mathrm{CMASS}}$ & 0.1 & 5 \\
    $\log T_{AGN}$ & 7.1 & 8.5 \\ 
    $\Delta_{z1}-\Delta_{z5}$ & \multicolumn{2}{c|}{$\mathcal{N}(\mu_z, \textbf{C}_z)$} \\
    \hline
    \multicolumn{3}{|c|}{extensions} \\
    \hline
    $M_\nu$ & 0.0 & 0.4 \\
    $A_L$ & 0 & 5\\
    $w_0$ & -3 & 0.5 \\
    $w_a$ & -3 & 2 \\
    \hline
\end{tabular}
\caption{The parameters varied in the analysis and the associated priors. Note there are extra nuisance parameters associated with the ACT DR4, WMAP, and \hillipop\ likelihoods which we exclude here but note that we follow the same procedure as the respective collaborations in placing priors on these. We show the uniform prior on $\tau$ used for the \textit{Planck} likelihoods, when instead using the ACT likelihood we impose a Gaussian prior of $\tau= 0.0590 \pm 0.0061$ as explained in the text.\label{tab:vary_params}}
\end{table}

To perform the main cosmological inference in this pipeline we use the \texttt{emcee} package~\cite{Foreman-Mackey:2012any}. The \texttt{emcee} algorithm uses an ensemble of walkers in parallel and capitalizes on an affine-invariant ensemble sampling technique to efficiently explore the parameter space. Writing the likelihood module in \texttt{JAX}, enables a rapid inference process as we leverage \texttt{JAX}'s GPU acceleration and the vectorization afforded by the \texttt{vmap} function to compute the likelihood for each walker in parallel. For the analysis setups presented in this work, a fully converged MCMC run takes at most 25 minutes on a single GPU core of an Nvidia Tesla A100 node and usually less than 5 minutes. We tested the convergence of our MCMC by checking that the estimated autocorrelation time ($\tau$)~\cite{Foreman-Mackey:2012any} converged to a stable value and that the number of steps for each walker in the ensembles satisfied $N_{steps}>100*\tau$ for all parameters in the model. We use a common set of priors for each of our analyses which are specified in Table~\ref{tab:vary_params}. We have excluded the various CMB likelihood nuisance parameters which are described in the relevant subsections in Section~\ref{subsec:likelihood} and generally follow the set-up used by the collaborations associated with the given likelihood. 

\section{Results \label{sec:results}} 

\subsection{Internal consistency \label{subsec:internal_consistency}}

When combining different datasets, it is important to examine their consistency. In this work, we carry out several assessments of consistency both before and after combining data. We begin by separating the observations into CMB, CMB + DESI Y1 BAO, and our low-$z$ data vector. We make this split as we want to test the consistency of the new $9\times 2$pt dataset with the CMB and optionally the BAO data that we wish to combine with. Our first test is a simple one-dimensional Gaussian tension in the $S8$ parameter, defined by 
\begin{equation}\label{eqt:sigma_s8}
\sigma S8 = \frac{\Delta S8}{\sqrt{\sigma^2_{\mathrm{CMB}} + \sigma^2_{\mathrm{low-z}}}},
\end{equation}
where $\Delta S8$ is the difference in the mean values, and $\sigma_{\mathrm{CMB}}$ and $\sigma_{\mathrm{low-z}}$ are the corresponding standard deviations of the CMB- and low-$z$-derived $S8$ posteriors. These results are shown in the first column of Table~\ref{tab:tensions}. 

\begin{table}
\centering
\caption{Comparison of the one-dimensional Gaussian $S8$ discrepancy and the tension across the full parameter space ($n_\sigma$) relative to the low-$z$ dataset, for various CMB datasets under different cosmological models. The lower halves of each model section show results with DESI Y1 BAO data included. The error bar on the full parameter space tension comes from the scatter in the estimate from 10 independent normalizing flows (see text).}
\label{tab:tensions}
\begin{tabular}{llcc}
\hline
\textbf{Model} & \textbf{Experiment} & \boldmath{$\sigma S8$ (Gaussian)} & \boldmath{$n_\sigma$ (All Params)}\\
\hline
\multirow{3}{*}{$\Lambda\mathrm{CDM}$}
  & \textit{Planck} PR3             & 2.45 & $1.964 \pm 0.006$ \\
  & \textit{Planck} PR4             & 2.15 & $1.703 \pm 0.005$ \\
  & ACT (DR4) + WMAP                & 2.04 & $1.595 \pm 0.004$ \\
\cline{2-4}
  & \textit{Planck} PR3 + DESI Y1   & 1.76 & $1.271 \pm 0.004$ \\
  & \textit{Planck} PR4 + DESI Y1   & 1.68 & $1.301 \pm 0.004$ \\
  & ACT (DR4) + WMAP + DESI Y1      & 1.60 & $1.252 \pm 0.004$ \\
\hline
\multirow{3}{*}{$\nu \Lambda\mathrm{CDM}$}
  & \textit{Planck} PR3             & 2.54 & $1.655 \pm 0.005$ \\
  & \textit{Planck} PR4             & 2.12 & $1.469 \pm 0.004$ \\
  & ACT (DR4) + WMAP                & 1.62 & $1.699 \pm 0.005$  \\
\cline{2-4}
  & \textit{Planck} PR3 + DESI Y1   & 1.66 & $1.320 \pm 0.004$ \\
  & \textit{Planck} PR4 + DESI Y1   & 1.62 & $1.263 \pm 0.004$ \\
  & ACT (DR4) + WMAP + DESI Y1      & 1.41 & $1.172 \pm 0.005$ \\
\hline
\multirow{3}{*}{$w_0w_a\mathrm{CDM}$}
  & \textit{Planck} PR3             & 0.00 & $1.042 \pm 0.005$ \\
  & \textit{Planck} PR4             & 0.28 & $0.725 \pm 0.004$ \\
  & ACT (DR4) + WMAP                & 0.46 & $0.970 \pm 0.005$ \\
\cline{2-4}
  & \textit{Planck} PR3 + DESI Y1   & 0.80 & $1.734 \pm 0.006$ \\
  & \textit{Planck} PR4 + DESI Y1   & 1.05 & $1.282 \pm 0.004$ \\
  & ACT (DR4) + WMAP + DESI Y1      & 0.87 & $0.923  \pm 0.001$\\
\hline
\end{tabular}
\end{table}

Within $\Lambda \mathrm{CDM}$, we find a weaker $S8$ tension than the $3.4\sigma$ reported by the KiDS-1000 team~\cite{KiDS:2020suj}, primarily because including additional components in the low-$z$ data vector shifts the $S8$ constraint towards higher values. Furthermore, the updated \textit{Planck} PR4 likelihood exhibits moderately lower tension with low-$z$ data compared to PR3, and the ACT+WMAP constraints reduce it still further, albeit due to larger error bars rather than a shift in the mean $S8$. When DESI Y1 BAO data are added to the CMB, the tension drops below $2\sigma$ in all cases, reflecting the slightly lower $\Omega_m$ preferred by BAO measurements. 

We observe that extended cosmological models tend to reduce the tension: broader uncertainties in $S8$ for both the CMB and low-$z$ datasets and shifts in the CMB central values both act to reconcile $S8$ across probes. In the $\nu\Lambda \mathrm{CDM}$ scenario, for instance, a larger neutrino mass leads to an increased $\Omega_m$ (due to the geometric degeneracy in CMB data~\cite{Loverde:2024nfi}) but decreased $\sigma_8$ (due to small-scale power suppression from free streaming neutrinos) when considering CMB primary data alone, the net result is a small decrease in the mean value of $S8$ and a broadening of the posterior. Under a dynamical dark energy model, the CMB primary data constraints on $\sigma_8$ and $\Omega_m$ correlate strongly with $w_0$ and $w_a$, driving $S8$ toward values that are fully consistent with the low-$z$ data vector. In contrast to $\Lambda\mathrm{CDM}$, the inclusion of DESI Y1 BAO data in the dynamical dark energy model pushes $S8$ slightly higher by breaking parameter degeneracies, but the overall tension remains $\lesssim 1\sigma$ for all three CMB likelihoods. 

While a one-dimensional comparison of $S8$ is a useful starting point, these datasets constrain a larger volume in multidimensional parameter space. Consequently, a more robust tension metric must account for discrepancies among all shared parameters. Here, we adopt the parameter-difference distribution approach recommended by Ref.~\cite{DES:2020hen}. Concretely, we define
\begin{equation}
    \mathcal{P}(\Delta \theta) \;=\;
    \int_{\Pi}\,\mathcal{P}_A(\theta)\,\mathcal{P}_B(\theta - \Delta \theta)\,\mathrm{d}\theta,
\end{equation}
where $\theta$ represents the parameters common to two datasets, $\Pi$ is the support of the prior, and $\mathcal{P}_A$, $\mathcal{P}_B$ are the respective posteriors for experiments $A$ and $B$, which are assumed to be uncorrelated. Tension is then quantified by the fraction of the parameter-difference posterior above its value at $\Delta \theta = 0$:
\begin{equation}
    \label{eqt:tension_metric}
    \Delta \;=\; 
    \int_{\mathcal{P}(\Delta \theta) \;>\;\mathcal{P}(0)}\,\mathcal{P}(\Delta \theta)\,\mathrm{d}\Delta \theta.
\end{equation}
The probability $\Delta$ can be cast as an effective number of standard deviations for easier comparison to the 1D-$S8$ tension metric following:
\begin{equation}
    n_\sigma \;=\; \sqrt{2} \,\mathrm{Erf}^{-1}\!\bigl(\Delta\bigr).
\end{equation}

We compute these metrics using \texttt{tensiometer}\footnote{\url{https://github.com/mraveri/tensiometer}}~\cite{Raveri:2021wfz} which fits a normalizing flow to the parameter-difference posterior, which in turn allows us to evaluate the integral in Eq.~\eqref{eqt:tension_metric} with high accuracy. We train 10 independent normalizing flows to get an estimate of the scatter of this quantity.\footnote{We checked that using an alternative kernel density estimator technique, which is also implemented in the \texttt{tentiometer} package (see Ref~\cite{Raveri:2021wfz} for details), to perform this integral yields nearly identical results.} 

In the full parameter space (second column of Table~\ref{tab:tensions}), the $\Lambda \mathrm{CDM}$ measured tension is lower compared to the one-dimensional $S8$ comparisons with all CMB datasets having $n_\sigma < 2$ in $\Lambda \mathrm{CDM}$, reflecting the fact that the parameter-space average tension is slightly less than the 1D $S8$ projection. The $\nu \Lambda \mathrm{CDM}$ results display a similar trend albeit with the tension reduced compared to $\Lambda \mathrm{CDM}$ due to the broader uncertainties and suppression of small-scale structure growth via neutrinos. Interestingly, we find that for the dynamical dark energy model, there is a larger tension in the full parameter space compared to the $S8$ parameter projection suggesting that there is increased tension in other parts of the parameter space, which we will return to when deriving constraints on this model. We note that despite this increase, the full parameter space tension for this model is at a level $n_\sigma<2$ for all combinations of data. 

\begin{figure}
\centering
\includegraphics[scale=0.31]{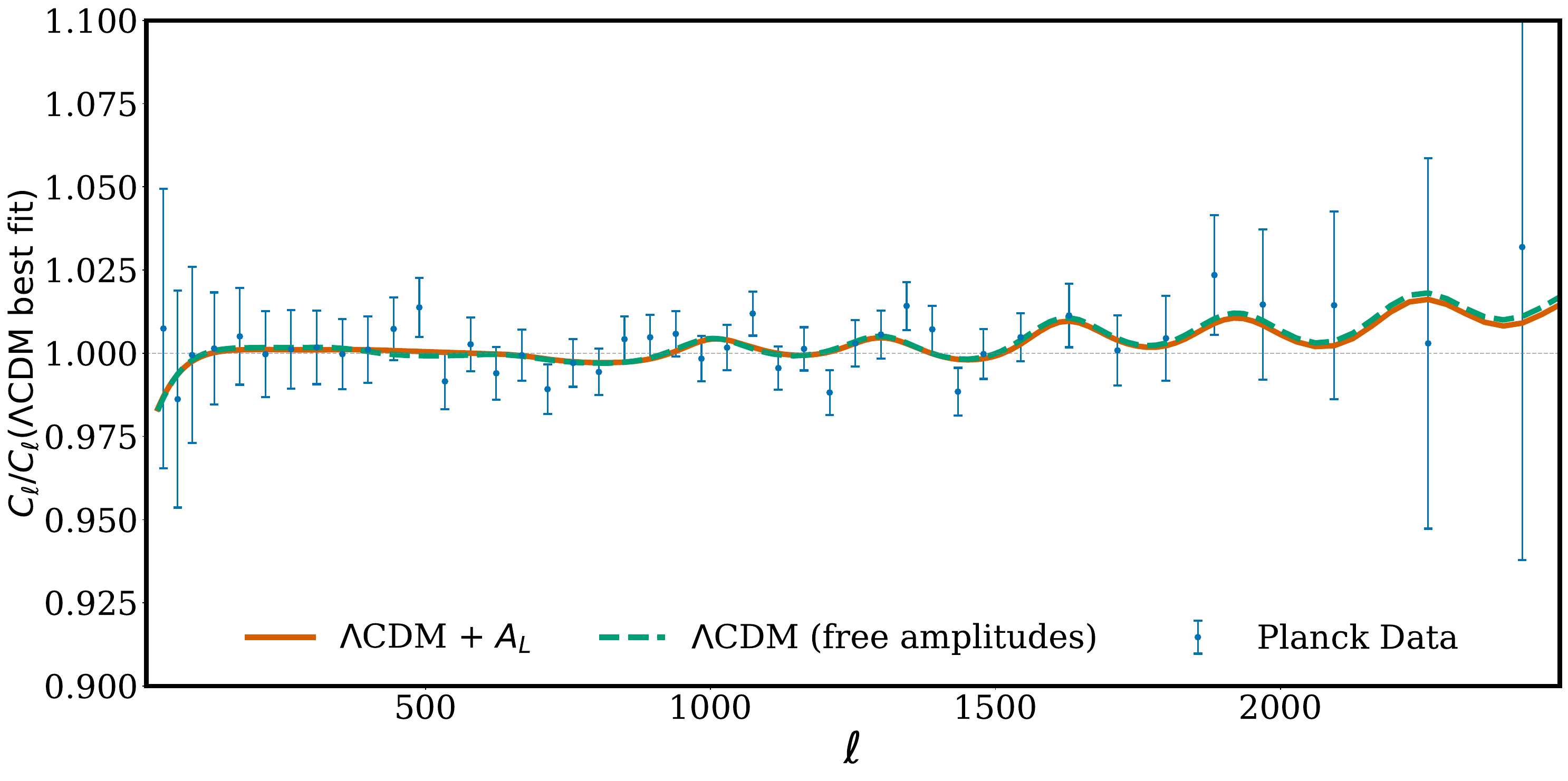}
\caption{Planck (PR3) TT data (re-binned to highlight features), bestfit to $\Lambda \mathrm{CDM}$+$A_\mathrm{lens}$ model, bestfit to free amplitudes model each normalized by bestfit in $\Lambda \mathrm{CDM}$ model. As explained in the text this demonstrates that the preference for $m_t^{\mathrm{planck}}/m_e^{\mathrm{planck}}<1$ is due to the same residual in $\Lambda \mathrm{CDM}$ that gives rise to the preference for $A_L>1$ in this likelihood. \label{fig:planck_residuals}} 
\end{figure}

\begin{figure}[ht]
    \centering
    \begin{subfigure}[t]{\textwidth}
        \centering
        \includegraphics[width=\textwidth]{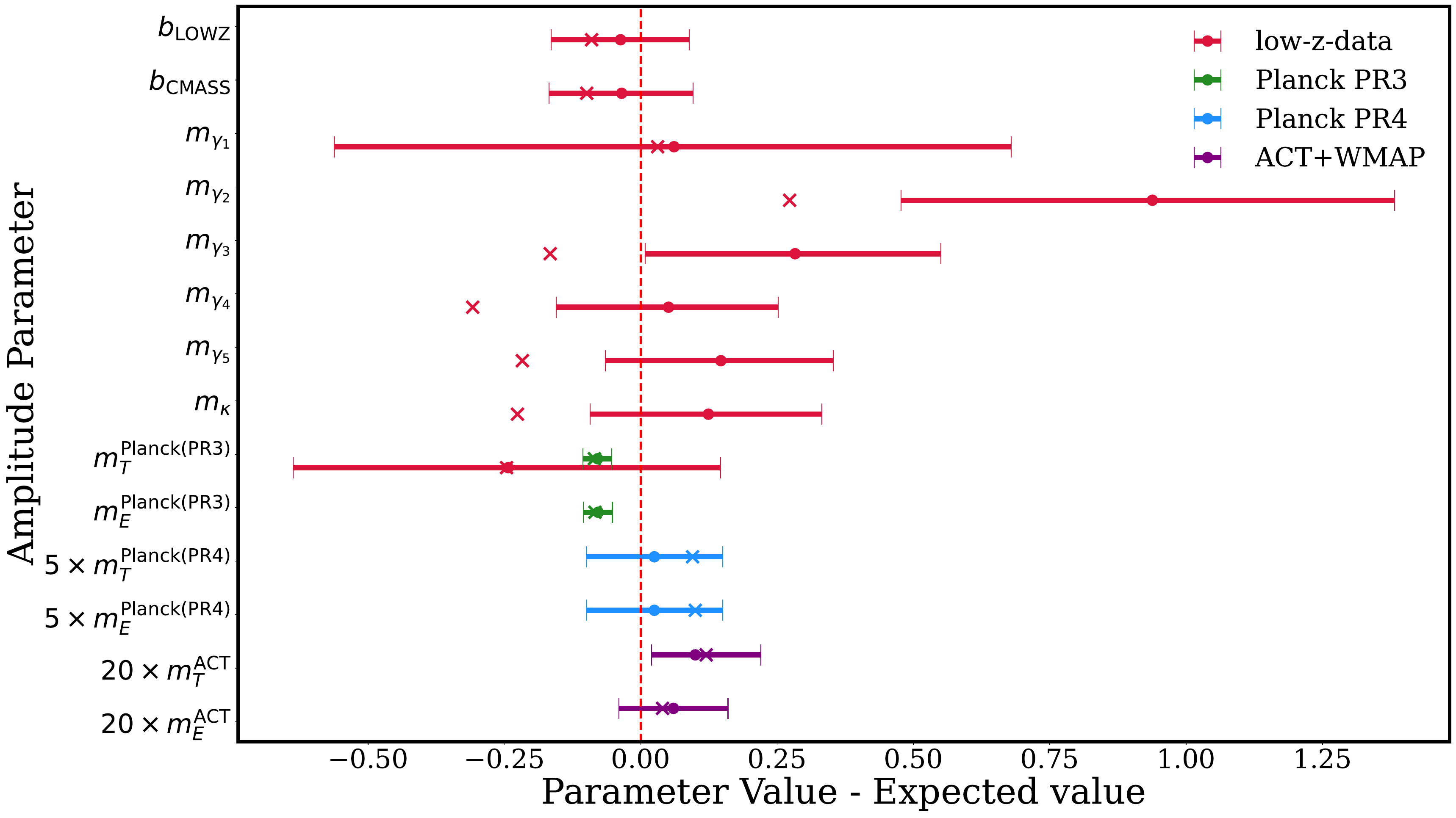}
        \caption{Single likelihood constraints.\label{fig:multiplicative_single}}
        \label{fig:single_constraints}
    \end{subfigure}
    \vspace{0.5cm}
    \begin{subfigure}[t]{\textwidth}
        \centering
        \includegraphics[width=\textwidth]{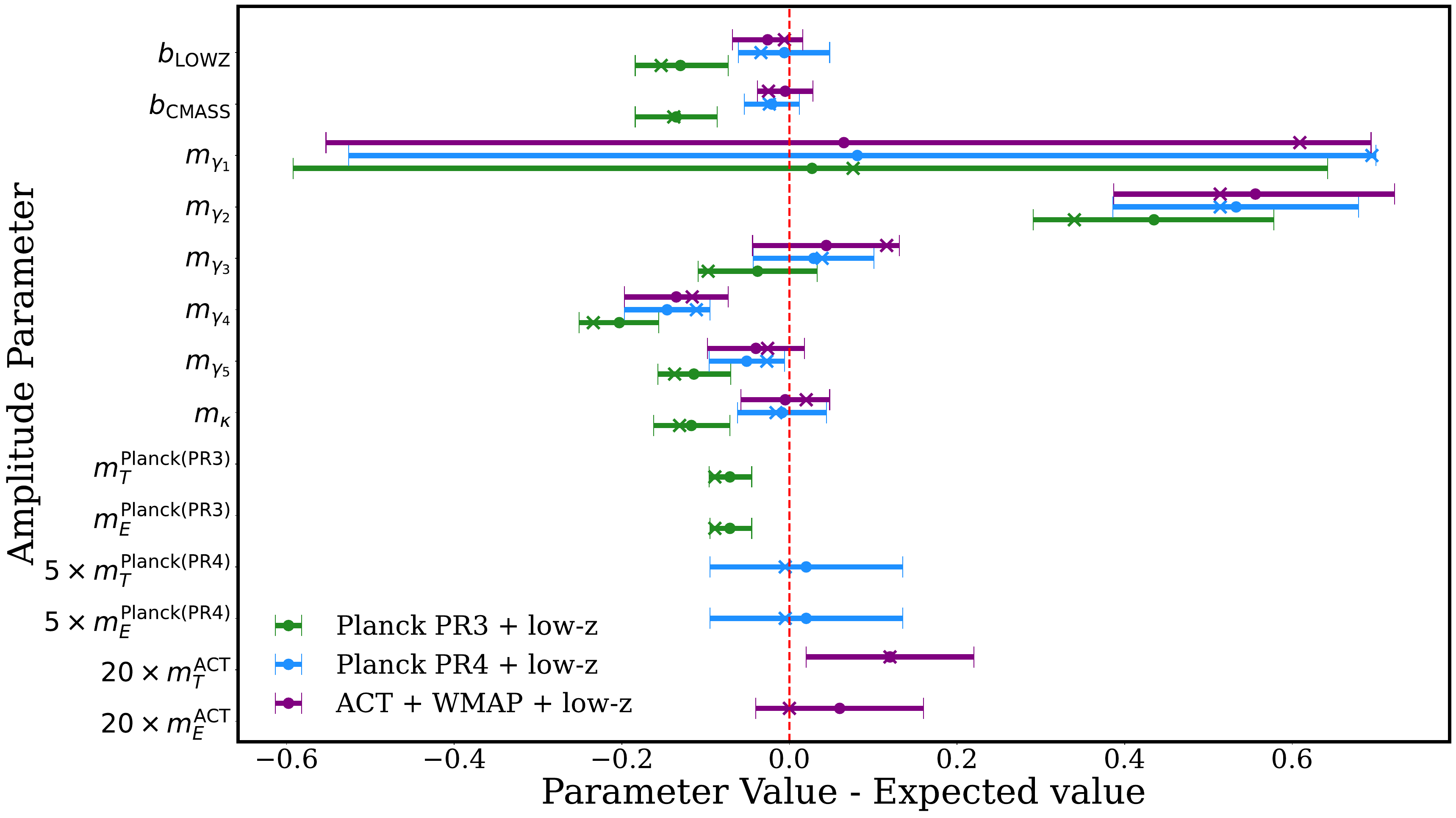}
        \caption{Combined data constraints.\label{fig:multiplicative_multi}}
        \label{fig:multi_constraints}
    \end{subfigure}
    \caption{1D constraints on the multiplicative rescaling parameters for various combinations of data in the $\Lambda \mathrm{CDM}$ model with fixed neutrino mass $M_\nu=0.06\mathrm{eV}$. The dots represent the Bayesian mean whilst the crosses are the best-fit points found in the chain which can deviate from the mean due to projected effects as explained in the text. The expected values for the parameters are typically zero except for the linear bias parameters for which we take the bestfit values found in ~\cite{Doux:2017tsv}, and the shear multiplicative bias parameters for which we take the mean values reported in Table I of ~\cite{Giblin:2020quj}. We have rescaled the amplitude parameters associated with the ACT and \textit{Planck} PR4 likelihoods for visualization purposes.\label{fig:multiplicative constraints}}
    
\end{figure}

In order to better understand the effects driving the mild tensions observed between the different data sets considered in our analysis, we allow for a free amplitude for each of the probes: 
\begin{equation}
    C_\ell^{XY} \to (1+m_{X})(1+m_{Y}) C_\ell^{XY}
\end{equation}
and let these amplitude parameters vary with a wide uniform prior $m_{i} \in [-1,1]$ (a similar methodology to Ref.~\cite{Sgier:2021bzf})\footnote{For the multiplicative factor $m_{\gamma2}$ associated with the second redshift bin of KiDS-1000 we had to widen this prior to $m_{\gamma2} \in [-2,2]$ in order to fully resolve the posterior.}. Note that we do not ascribe a free amplitude to the galaxy clustering $C_\ell$s as these power spectra are already modeled in an amplitude-agnostic manner with a linear bias. We remove the $A_{\mathrm{planck}}$ and $y_p$ nuisance parameters associated with the \textit{Planck} and ACT likelihoods respectively and replace these with the appropriate amplitude parameters defined above. The idea of this test is to see how far the free amplitude parameters shift when all of the datasets are combined: a statistically significant shift indicates an amplitude discrepancy in a given spectrum. 

Fig.~\ref{fig:multiplicative_single} displays the resulting free amplitude constraints, found when allowing both the $\Lambda \mathrm{CDM}$ parameters and the free amplitudes for the relevant spectra to vary, for each individual CMB likelihood and separately our low-$z$ data under $\Lambda \mathrm{CDM}$. The majority of amplitude parameters are consistent with the expected value within $1\sigma$, but several notable shifts emerge. First, the $m_{\gamma 2}$ parameter, associated with the amplitude rescaling of the data in the second KiDS-1000 redshift bin, exhibits a mild positive shift at the $2\sigma$ level. The KiDS Collaboration also found that this bin’s cosmology differs from the other four (Appendix B.2 of~\cite{KiDS:2020suj}), which they argue could be attributed to a statistical fluctuation or high-$z$ contaminants in the second redshift bin. In the latter scenario, increased lensing amplitude would be derived from interloper objects at high-$z$ producing an anomalously high $C_\ell$ amplitude. This preference is also present at a slightly higher significance when analyzing the KiDS-1000 data by itself without the other components of the low-$z$ data vector ($m_{\gamma 2}=1.33^{+0.51}_{-0.51}$ representing a $2.6\sigma$ shift from the expected value). We note that we observe relatively strong projection effects for the KiDS-1000 and CMB lensing multiplicative parameters when analysing the low-$z$ data alone due to the relatively low constraining power of this combination, though this does not change our interpretation. 

We find that the data prefer a lower-than-expected value for the PR3 rescaling parameters $m_T^{planck}$ and $m_E^{planck}$ at a $\sim 3 \sigma$ level each. When inspecting the MCMC chains for this analysis we also found that there was a preference for an increased $\sigma_8$ parameter compared to the $\Lambda \mathrm{CDM}$ constraints. We can understand these preferences as a manifestation of the `$A_{\mathrm{lens}}$' tension which is known to be present in this likelihood~\cite{Renzi:2017cbg}. By allowing for a high $\sigma_8$ whilst maintaining the correct amplitude for the CMB power spectra, having $m^{\textit{Planck} \, (\text{PR3})}_T, m^{\textit{Planck} \, (\text{PR3})}_E < 0$ produces a cosmology with extra lensing at low redshifts which fits the PR3 residual in $\Lambda CDM$. This can be seen in Figure~\ref{fig:planck_residuals} where the residuals of the best-fit model with free amplitudes compared to the baseline $\Lambda \mathrm{CDM}$ model are almost identical to the best fit $A_{\mathrm{lens}}$ model i.e. they are fitting to the same lensing residual in the $C_\ell^{TT}$ PR3 data. For the other CMB experiments (ACT DR4 + WMAP and \textit{Planck} PR4) we do not see such a preference which is consistent with this interpretation as the collaborations find $A_L$ consistent with the expected value for these data. 

In Fig.~\ref{fig:multiplicative_multi} we show the constraints on the multiplicative parameters when we combine the CMB with the low-$z$ data. Combining with CMB information substantially tightens constraints on the low-$z$ amplitude parameters relative to the low-$z$ data alone scenario. We find that even with the addition of the low-$z$ data the preference for $m^{\textit{Planck} \, (\text{PR3})}_T, m^{\textit{Planck} \, (\text{PR3})}_E < 0$ remains with a similar level of significance and the low-$z$ multiplicative parameters are systematically shifted to lower than their expected values. Similar to the \textit{Planck} PR3 alone case, we find a preference for an increased $\sigma_8$ in these chains (which enables fitting the residual associated with the $A_{\mathrm{lens}}$ systematic in \textit{Planck} PR3 as discussed above), this in turn requires the low-$z$ multiplicative parameters to shift lower to fit the data. For the combinations with the other CMB datasets (\textit{Planck} PR4 and ACT+WMAP), we see a general preference for $m_{\gamma X}<0$ for the low-$z$ multiplicative parameters (aside from $m_{\gamma2}$ which remains $>0$), and especially for $m_\gamma^4<0$ which prefers a value $\sim 10\%$ lower than the expected value (at a significance of $2.86\sigma$ for the combination of \textit{Planck} PR4 + low-$z$). This is a manifestation of the $S8$ or amplitude discrepancy between CMB measurements and weak lensing measurements. With this test we can identify that KiDS-1000 redshift bin 4 shows the largest amplitude discrepancy within the low-$z$ data vector. Given the increased constraining power, we now find evidence for the previously observed amplitude excess in KiDS redshift bin 2 at $3.64\sigma$ with the combination of \textit{Planck} PR4 + low-$z$. We also only very small shifts between the mean and bestfit values of the chains indicating negligible projection effects. 

Overall, whilst most of the data have internally consistent amplitudes, there is evidence for two notable systematic discrepancies: a mild excess in the second KiDS-1000 redshift bin amplitude and a $\sim3\sigma$ deficit in the \textit{Planck} PR3 temperature and polarization amplitudes linked to the known $A_{\mathrm{lens}}$ anomaly. We will explore the impact of both of these discrepancies on our cosmological parameter inference in the following sections. We also identify KiDS-1000 redshift bin 4 as being the most influential in the amplitude discrepancy between the CMB and low redshift data. We note that we did not assess the internal consistency of the DESI Y1 BAO data as we could not fit this neatly into the amplitude scaling approach taken here. Several recent works have identified a mild internal tension in the DESI LRG data at $z_{\mathrm{eff}}=0.51$, which can impact cosmological constraints in dynamical dark energy models~\cite{Wang:2024pui, Colgain:2024xqj}, we leave exploration of this to future work.

\subsection{Goodness of fit}
\begin{table}[ht]
\centering
\scriptsize
\resizebox{0.8 \textwidth}{!}{%
\begin{tabular}{|lc|cc|}
\hline
\textbf{Experiment} & $\mathbf{n_{data}}$ & \multicolumn{2}{c|}{$\mathbf{\Lambda \mathrm{CDM}}$} \\
 & & $\chi^2_{\text{red}}$ & PTE  \\
\hline
\textit{Planck} PR3 (high-$\ell$)       & 613   & 0.96  & 0.72  \\
\textit{Planck} PR4 (high-$\ell$)       & 29758 & 1.03  & 0.00094   \\
ACT (DR4) (high-$\ell$)                 & 260   & 1.07  & 0.18  \\
low-$z$ data                           & 245   & 1.11  & 0.091 \\ 
low-$z$ (no KiDS bin 2)               & 200   & 1.02  & 0.31   \\
\textit{Planck} PR3 (high-$\ell$) + low-$z$ & 858   & 1.01  & 0.34  \\
\textit{Planck} PR4 (high-$\ell$) + low-$z$ & 30003 & 1.03  & 0.00059  \\
ACT (DR4) (high-$\ell$) + low-$z$     & 505   & 1.08  & 0.090  \\
\hline
\end{tabular}%
}
\caption{Bayesian goodness-of-fit comparison for the $\Lambda \mathrm{CDM}$ model across different datasets with both the reduced $\chi^2$ metric and the Bayesian PTE. \label{tab:goodness_of_fit}}
\end{table}

To assess goodness of fit, we compute the Bayesian probability to exceed (PTE) in $\Lambda \mathrm{CDM}$ using the posterior predictive distribution (PPD). This quantifies how extreme the observed data are compared to synthetic datasets generated from the posterior~\cite{Gelman1996, Meng1994}, providing a test of how well the $\Lambda \mathrm{CDM}$ model fits our datasets. Furthermore, by examining how the PTE changes when combining datasets, we can further evaluate whether the data are well described by a single set of model parameters. Assuming the likelihood can be well approximated as a Gaussian this can be determined as (following the same methodology as outlined in Appendix D of Ref.~\cite{Sailer:2024coh}): 
\begin{equation}\label{eqt:pte}
    \mathrm{PTE} \;=\;
    \int \mathrm{d}\theta \,
    \left[
       1 \;-\;
       \frac{\gamma\!\bigl( \tfrac{N_d}{2},\, \tfrac{\chi^2(\theta)}{2}\bigr)}
            {\Gamma\!\bigl(\tfrac{N_d}{2}\bigr)}
    \right]
    P(\theta),
\end{equation}
where $N_d$ is the number of data points, $P(\theta)$ is the posterior, and $\gamma$ and $\Gamma$ are the incomplete and complete Gamma functions, respectively. We compute the PTE only for the high-$\ell$ CMB primary likelihoods as the low-$\ell$ likelihoods are typically non-Gaussian and hence the assumption behind equation~\ref{eqt:pte} breaks down. We compute this quantity for the various combinations of data in Table~\ref{tab:goodness_of_fit}. 

Table~\ref{tab:goodness_of_fit} summarizes these PTE values for individual and combined datasets. The key findings are:

\begin{itemize}
    \item \textbf{\textit{Planck} PR4 (\hillipop\ Likelihood).} 
    As reported by Ref.~\cite{Tristram:2023haj}, the high-$\ell$ \hillipop\ likelihood yields a low PTE suggesting a poor fit to the data. This may stem from slightly underestimated instrumental noise in the $\ell$-range $500<\ell<1000$ leading to excess (i.e. larger than error bar) scatter in the data as argued in Refs.~\cite{Tristram:2023haj,Rosenberg:2022sdy}. The \textit{Planck} \texttt{NPIPE} processing uses a common mask for each frequency channel which, by virtue of cosmic variance cancellation in some frequency combinations, makes the likelihood sensitive to noise modeling~\cite{Rosenberg:2022sdy}. Refs.~\cite{Rosenberg:2022sdy, Tristram:2023haj} argue that this excess scatter does not affect cosmological inference as this residual cannot be fit by an improved foreground modeling or different cosmology. Similar excesses are not observed in the \texttt{Plik\_lite} likelihood, which has increased effective errors on high-$\ell$ TT by up to a factor of two due to the foreground marginalization~\cite{Planck:2018vyg} and uses different masks for the different frequency channels thus reducing the sensitivity to noise and foreground modeling uncertainties.

    \item \textbf{KiDS Redshift Bin 2.} 
    We find a relatively low PTE (0.091) for our low-$z$ dataset. However, we show that removing  KiDS-1000 bin 2 from the data vector gives a substantively improved PTE (0.31) and represents a good fit under the $\Lambda \mathrm{CDM}$ model. This is consistent with the amplitude test (Sec.~\ref{subsec:internal_consistency}) where we found this bin to exhibit internal tension. Still, we find that in most cases removing this bin does not appreciably shift cosmological constraints, so we follow the KiDS team in retaining it in our later analysis but carefully testing each time that this has no impact on our conclusions (see Appendix~\ref{appendix:kids_bin_2} for exploration of this bin's impact on our results). 

    \item \textbf{Combination of CMB and low-$z$ Data.}
    Adding the low-$z$ data to any of the CMB likelihoods does not significantly alter the reduced $\chi^2$, and the Bayesian PTE shows a mild decrease reflecting the $\sim2\sigma$ tension between these data.
\end{itemize}

Based on the results of this section, we choose the \textit{Planck} PR4 + low-$z$ combination as our baseline configuration as it achieves competitive parameter constraints, avoids strong evidence for systematics, and does not rely on external priors. We find at most (for the \textit{Planck} PR3 likelihood) a $2\sigma$ tension between the low-$z$ dataset and the CMB in the full parameter space, which is reduced when combining with DESI Y1 BAO data and further still when considering extended models. We also demonstrate that the amplitudes of the spectra are broadly self-consistent within $\Lambda \mathrm{CDM}$ but identify two hints of systematics in the \textit{Planck} PR3 likelihood and the KiDS-1000 redshift bin 2. Finally, we find the goodness of fit is not drastically changed when combining our data compared to the uncombined fits. The results of these tests taken together provide us with confidence in jointly analyzing these datasets, provided that we carefully interpret our results in the context of the hints of tensions and systematics identified in this section. 

\subsection{$\Lambda CDM$ constraints}

\subsubsection{Comparison of low-redshift inference of $S8$ with other works}
\begin{figure}[ht!]
    \centering
    \includegraphics[width=1.\textwidth]{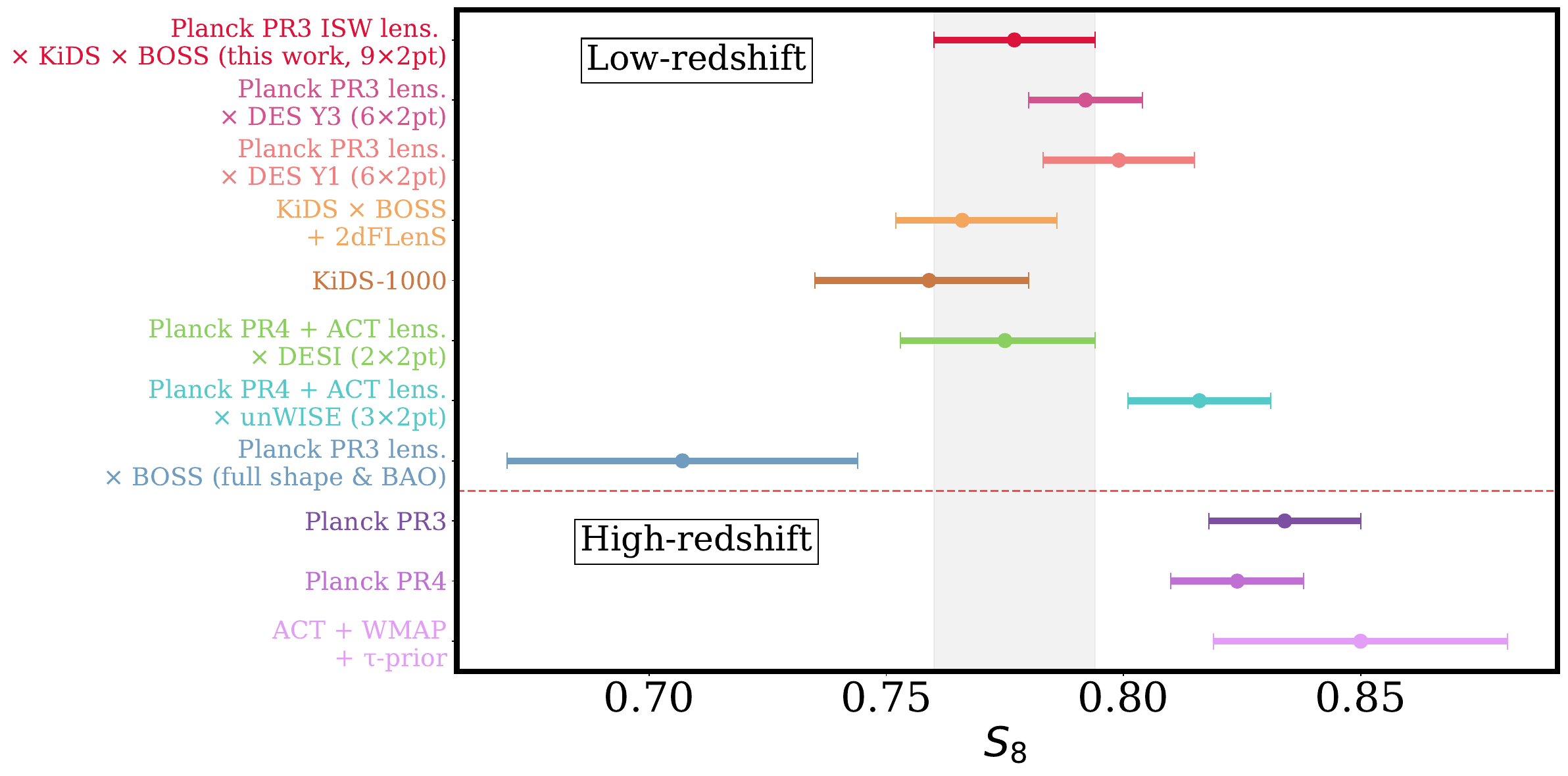}
    \caption{\textbf{A comparison of various $S8$ measurements.} We include:
    (1) \textit{Planck} PR3 ISW + lensing $\times$ KiDS $\times$ BOSS (\emph{this work}), 
    (2) \cite{DES:2022urg} (\emph{\textit{Planck} PR3 lensing $\times$ DES Y3 ($6 \times 2$pt)})
    (3) \cite{Xu:2023qmp} (\emph{\textit{Planck} PR3 lensing $\times$ DES Y1 ($6 \times 2$pt)}),
    (4) \cite{Heymans:2020gsg} (\emph{KiDS $\times$ BOSS + 2dFLenS}),
    (5) \cite{KiDS:2020suj} (\emph{KiDS-1000}),
    (6) \cite{Sailer:2024coh} (\emph{\textit{Planck} + ACT lensing $\times$ DESI}),
    (7) \cite{Farren:2024rla} (\emph{\textit{Planck} + ACT lensing $\times$ unWISE ($3 \times 2$pt)}),
    (8) \cite{Chen:2024vvk} (\emph{\textit{Planck} CMB lensing $\times$ BOSS, including full shape + BAO constraints}),
    (9) \cite{Planck:2018vyg} (\emph{\textit{Planck} PR3}),
    (10) \cite{Tristram:2020wbi} (\emph{\textit{Planck} PR4}),
    (11) \cite{ACT:2020gnv} (\emph{ACT + WMAP}). The CMB constraints shown here are for the CMB primary TTTEEE spectra only (no CMB lensing is added). The gray shaded region marks the $1\sigma$ confidence interval of this work.
    }
    \label{fig:comparison_lowz}
\end{figure}

In Fig.~\ref{fig:comparison_lowz}, we show a comparison of our $S8$ measurement:
\begin{equation}
    S8 = \sigma_8\sqrt{\Omega_m/0.3} = 0.777^{+0.017}_{-0.017},
\end{equation}
from the low-$z$ data ($9\times2$pt combination of \textit{Planck} PR3 CMB lensing and ISW, KiDS-1000 weak lensing and BOSS-DR12 galaxy clustering considering all possible cross-correlations) compared to a selection of other contemporary measurements. The closest related measurement to ours in terms of the number and type of cross-correlations is that of Refs.~\cite{DES:2022urg, Xu:2023qmp}, shown in the second and third line of Fig.~\ref{fig:comparison_lowz}. This is a combination of \textit{Planck} PR3 lensing with DES galaxy weak lensing and clustering including the CMB lensing auto-correlation ($6 \times 2$pt analysis). Our $S8$ constraint is compatible with the DES Y3 constraint of Ref.~\cite{DES:2022urg} at the $0.7 \sigma$ level (where the $\sigma$ shift is computing according to Equation~\ref{eqt:sigma_s8}). Our result is also compatible at the $\sim 1 \sigma$ level with the DES Y1 constraint in Ref.~\cite{Xu:2023qmp} which uses a PCA-based approach to marginalize over Baryonic feedback. Further down in Fig.~\ref{fig:comparison_lowz}, our measurement remains consistent within $1\sigma$ of the KiDS $3\times2$pt analysis with BOSS and 2dFLenS~\cite{Heymans:2020gsg}. However, it is approximately $1\sigma$ higher than the KiDS-1000 weak lensing-only result~\cite{KiDS:2020suj} as discussed in Section~\ref{subsec:internal_consistency}. We also show the results of recent studies using updated CMB lensing likelihoods from \textit{Planck} PR4~\cite{Carron:2022eyg} and ACT DR6~\cite{ACT:2023kun}, alongside galaxy clustering data from DESI Y1 LRGs and unWISE~\cite{Sailer:2024coh,Farren:2024rla}. Our $S8$ value agrees within $1\sigma$ with the result of Ref.~\cite{Sailer:2024coh}, though it is $1.7\sigma$ lower than the $3\times2$pt constraint reported in Ref.~\cite{Farren:2024rla}.
Finally, for the low-redshift data, Ref.~\cite{Chen:2022jzq}, employs an effective Lagrangian perturbation theory to combine 3D galaxy clustering, reconstructed BAO, and CMB lensing measurements. Their approach yields an $S8$ value that lies more than $1\sigma$ below our result. However, we note that this analysis does not incorporate a normalization correction to the CMB lensing cross-correlation and may be subject to volume-related biases~\cite{Sailer:2024coh}. In the bottom panel of Fig.~\ref{fig:comparison_lowz}, we show the constraints from the three CMB likelihoods used in this analysis finding a tension at the $\sim 2\sigma$ level as discussed in Section~\ref{subsec:internal_consistency}. 

\subsubsection{$\Lambda \mathrm{CDM}$ parameter inference}
\begin{figure}[!htb]
\centering
\begin{subfigure}[t]{0.48\textwidth}
    \centering
    \includegraphics[width=\textwidth]{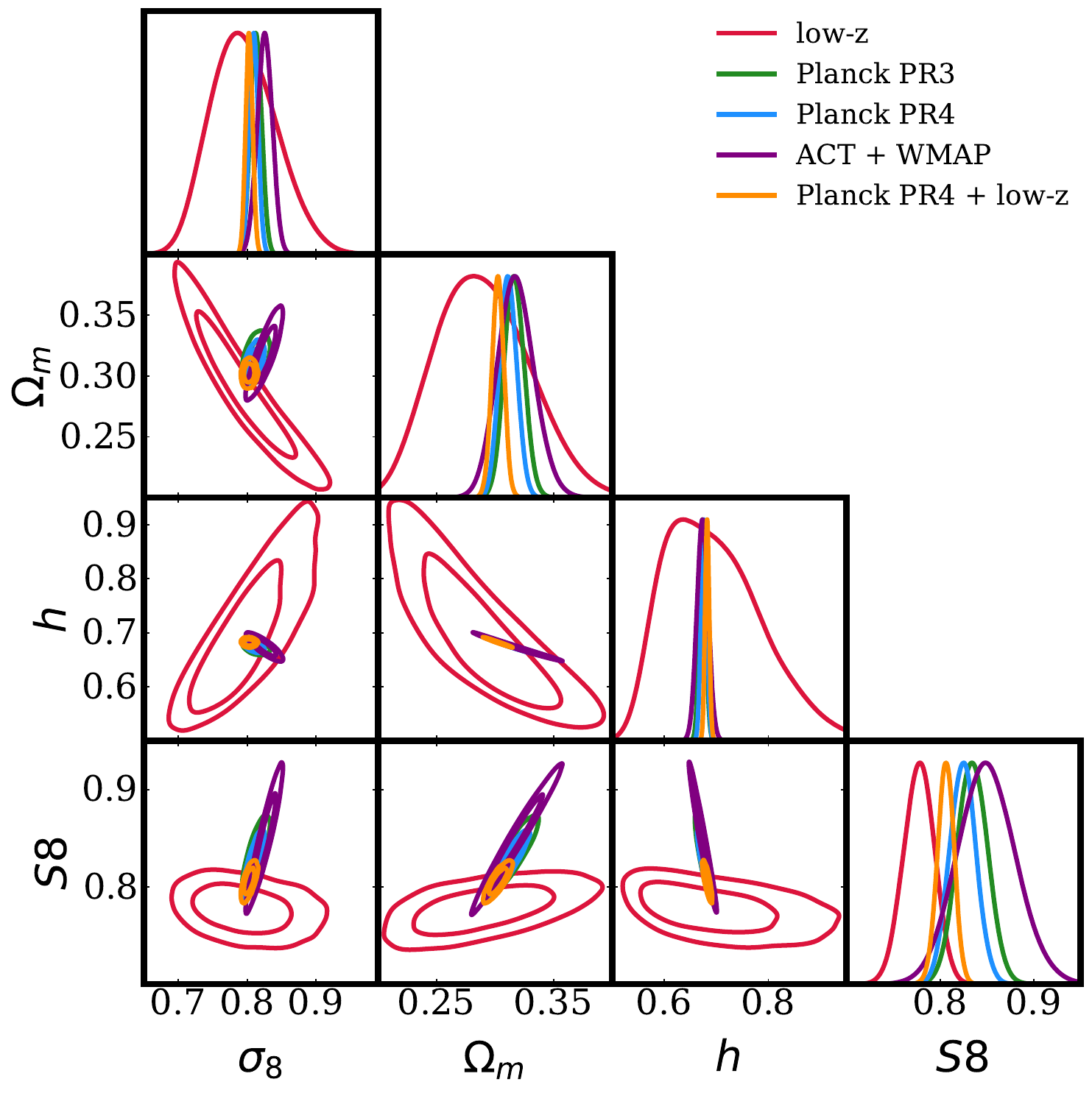}
    \caption{The constraints in the $\Lambda \mathrm{CDM}$ model considering the low-$z$ data and the various CMB primary likelihoods analyzed in this work. We additionally show the location of the combined constraint in our baseline configuration (\textit{Planck} PR4 + low-$z$, orange contour).\label{fig:baseline_single}}
\end{subfigure}
\hfill
\begin{subfigure}[t]{0.48\textwidth}
    \centering
    \includegraphics[width=\textwidth]{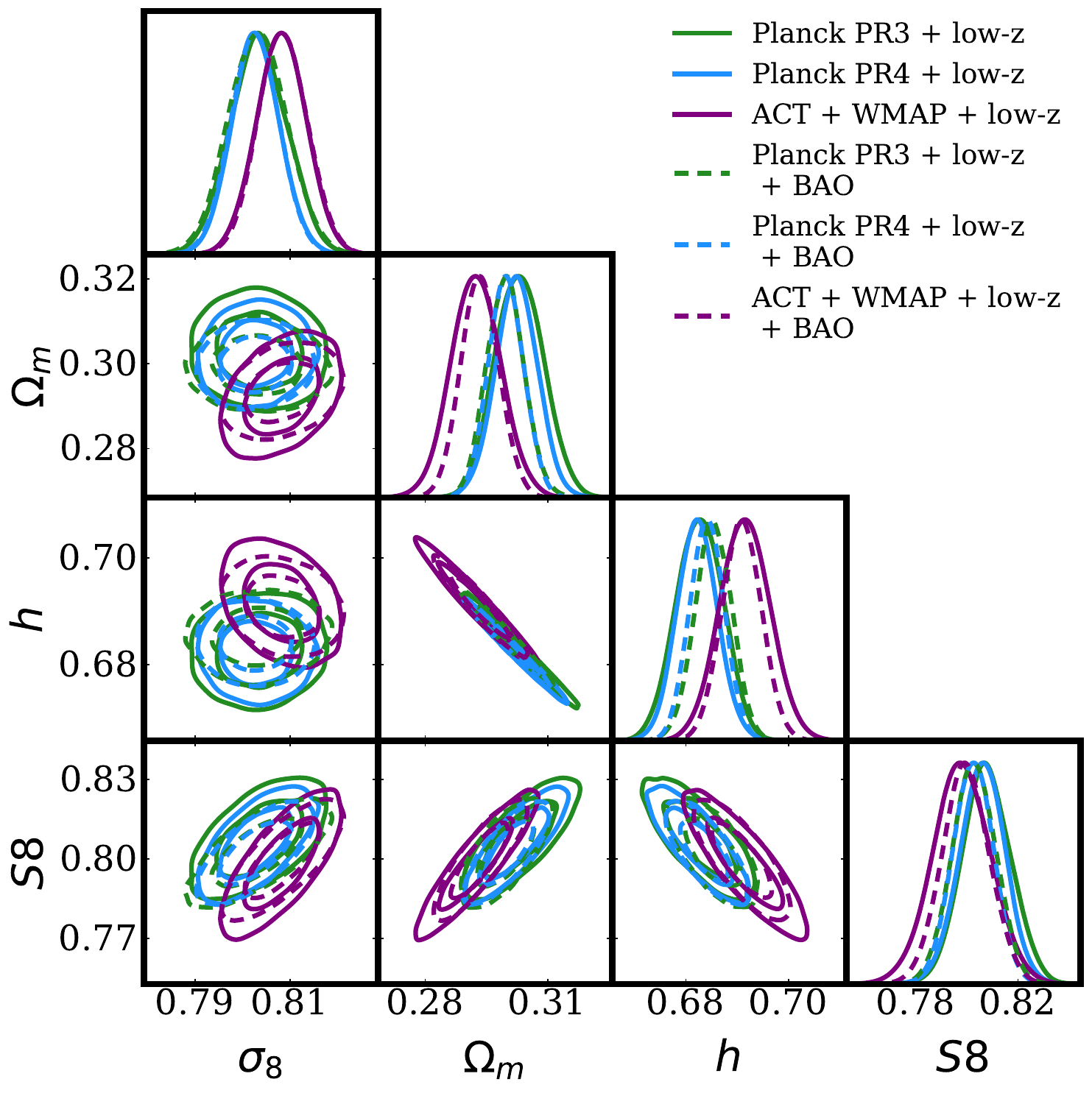}
    \caption{The constraints in the $\Lambda \mathrm{CDM}$ model considering the combination of the CMB likelihoods and the low-$z$ dataset and optionally DESI Y1 BAO (dashed contours).\label{fig:baseline_combined}}
    
\end{subfigure}
\caption{Constraints from various combinations of datasets in the $\Lambda \mathrm{CDM}$ model.}

\end{figure}

\begin{table}[ht]
\centering
\caption{Cosmological and selected nuisance parameter constraints for the $\Lambda \mathrm{CDM}$ model using single and combined datasets. \label{tab:lcdm_constraints}}
\resizebox{\textwidth}{!}{
\setlength{\tabcolsep}{2.5pt} 
\renewcommand{\arraystretch}{1.} 
\begin{tabular}{lccccccc}
\hline
\hline
\textbf{Parameter} & low-$z$ & \makecell{\textit{Planck} \\ PR3} & \makecell{\textit{Planck} \\ PR4} & \makecell{ACT \\ +WMAP} & \makecell{\textit{Planck} \\ PR3+low-$z$} & \makecell{\textit{Planck} \\ PR4+low-$z$} & \makecell{ACT+WMAP \\ +low-$z$} \\
\hline
\multicolumn{8}{l}{\textbf{Cosmological Parameters}} \\
$\Omega_m$        & $0.290^{+0.040}_{-0.040}$ & $0.316^{+0.009}_{-0.009}$ & $0.311^{+0.007}_{-0.007}$ & $0.318^{+0.016}_{-0.016}$ & $0.303^{+0.006}_{-0.006}$ & $0.302^{+0.005}_{-0.005}$ & $0.292^{+0.006}_{-0.006}$ \\
$\sigma_8$        & $0.796^{+0.048}_{-0.048}$ & $0.813^{+0.008}_{-0.008}$ & $0.809^{+0.006}_{-0.006}$ & $0.825^{+0.011}_{-0.011}$ & $0.803^{+0.006}_{-0.006}$ & $0.803^{+0.005}_{-0.005}$ & $0.808^{+0.005}_{-0.005}$ \\
$H_0$ [km/s/Mpc]  & $69.5^{+9.4}_{-9.4}$      & $67.4^{+0.6}_{-0.6}$      & $67.6^{+0.5}_{-0.5}$      & $67.4^{+1.1}_{-1.1}$      & $68.3^{+0.5}_{-0.5}$      & $68.2^{+0.4}_{-0.4}$      & $69.2^{+0.5}_{-0.5}$ \\
$S8$             & $0.777^{+0.017}_{-0.017}$ & $0.834^{+0.016}_{-0.016}$ & $0.824^{+0.014}_{-0.014}$ & $0.850^{+0.031}_{-0.031}$ & $0.808^{+0.009}_{-0.010}$ & $0.806^{+0.009}_{-0.009}$ & $0.798^{+0.011}_{-0.011}$ \\
$\tau$            & ---                       & $0.0554^{+0.0091}_{-0.0093}$ & $0.00575^{+0.0059}_{-0.0058}$ & $0.0581^{+0.0060}_{-0.0060}$ & $0.0532^{+0.0089}_{-0.0090}$ & $0.0575^{+0.0060}_{-0.0059}$ & $0.0597^{+0.0057}_{-0.0057}$ \\
\hline
\multicolumn{8}{l}{\textbf{Selected Nuisance Parameters}} \\
$A_{IA}$          & $0.318^{+0.351}_{-0.351}$ & --- & --- & --- & $0.518^{+0.280}_{-0.278}$ & $0.503^{+0.280}_{-0.277}$ & $0.464^{+0.299}_{-0.297}$ \\
$b_{\mathrm{LOWZ}}$ & $1.833^{+0.128}_{-0.128}$ & --- & --- & --- & $1.840^{+0.039}_{-0.040}$ & $1.841^{+0.038}_{-0.039}$ & $1.828^{+0.039}_{-0.039}$ \\
$b_{\mathrm{CMASS}}$& $2.080^{+0.128}_{-0.129}$ & --- & --- & --- & $2.094^{+0.026}_{-0.026}$ & $2.096^{+0.024}_{-0.025}$ & $2.081^{+0.025}_{-0.025}$ \\
$\log T_{AGN}$    & $7.799^{+0.482}_{-0.482}$ & --- & --- & --- & $7.775^{+0.467}_{-0.485}$ & $7.809^{+0.483}_{-0.473}$ & $7.789^{+0.472}_{-0.480}$ \\
\hline
\end{tabular}
}
\end{table}
 
We present constraints on cosmological parameters within the baseline $\Lambda\mathrm{CDM}$ model from both individual CMB experiments and low-$z$ datasets separately as well as our baseline (\textit{Planck} PR4 + low-$z$) combination in Fig.~\ref{fig:baseline_single} and an overview of the constraints in Table~\ref{tab:lcdm_constraints}. While ACT+WMAP shows a mild $1.3\sigma$ preference for a higher $\sigma_8$ than \textit{Planck} PR4, all other parameters agree within $1\sigma$. Incorporating low-$z$ data with \textit{Planck} PR4 (our baseline scenario, orange contours in Fig.~\ref{fig:baseline_single}) shifts constraints along the degeneracy directions toward lower $S8$. 

Moving onto Fig.~\ref{fig:baseline_combined}, our analysis reveals broad consistency between the \textit{Planck} (PR3 and PR4) and ACT+WMAP CMB experiments, with the joint CMB+low-$z$ constraints showing agreement at better than $1\sigma$ for most $\Lambda\mathrm{CDM}$ parameters. When comparing our baseline \textit{Planck} PR4 + low-$z$ configuration with the ACT+WMAP + low-$z$ combination, we observe moderate tensions in $\Omega_m$ ($1.3\sigma$) and $H_0$ ($1.6\sigma$). Furthermore, the initial $1.3\sigma$ tension in $\sigma_8$ between ACT+WMAP and \textit{Planck} PR4 (before inclusion of low-$z$ data) is reduced to $0.7\sigma$ in the combined analysis. The observed parameter shifts can be understood by examining the degeneracy directions shown in Fig.~\ref{fig:baseline_single}. While both CMB experiments exhibit similar parameter degeneracies, \textit{Planck}'s tighter constraints limit the extent to which the low-$z$ data can shift its parameter estimates. In contrast, the comparatively weaker constraints from ACT+WMAP allow the low-$z$ data to guide the combined constraints along these degeneracy directions toward lower $S8$ values, naturally leading to a reduction in both $\sigma_8$ and $\Omega_m$, while simultaneously driving $H_0$ higher. We find that additionally including the DESI Y1 BAO data in the analysis (dashed contours in Fig.~\ref{fig:baseline_combined}) produces small shifts that further reduce the tension between ACT+WMAP and \textit{Planck} PR4 to $<1\sigma$ in all parameters (see Appendix~\ref{appendix:lcdm_plots} for the table of constraints in this scenario). Overall we find that the three CMB likelihoods considered in this analysis give consistent constraints in $\Lambda \mathrm{CDM}$ and hence provide a solid foundation for exploring how this consistency evolves when considering extensions to the baseline model. 

\subsubsection{Amplitude tension revisited}
We are now in a position to answer the question: to what extent can the $S8$ tension be described as an amplitude discrepancy? To do this we can compare the $S8$ inference from the free amplitude test in Section~\ref{subsec:internal_consistency} with the $S8$ found in the standard $\Lambda \mathrm{CDM}$ analysis presented above for our baseline data combination. When freeing amplitudes for the \textit{Planck} PR4 + low-$z$ combination we find $S8=0.820^{+0.017}_{-0.017}$ which closely matches the constraint from \textit{Planck} PR4 alone ($S8=0.824^{+0.014}_{-0.014}$). This data-driven assessment indicates that the $S8$ tension can be mostly attributed to an amplitude discrepancy of the LSS data and in particular the amplitude of the KiDS-1000 redshift bin 4 as explained in Section~\ref{subsec:internal_consistency}. 

\subsection{Dynamical dark energy \label{subsec:dynamical_de_results}} 

\begin{figure}[!htb]
\centering
\includegraphics[width=0.8\textwidth]{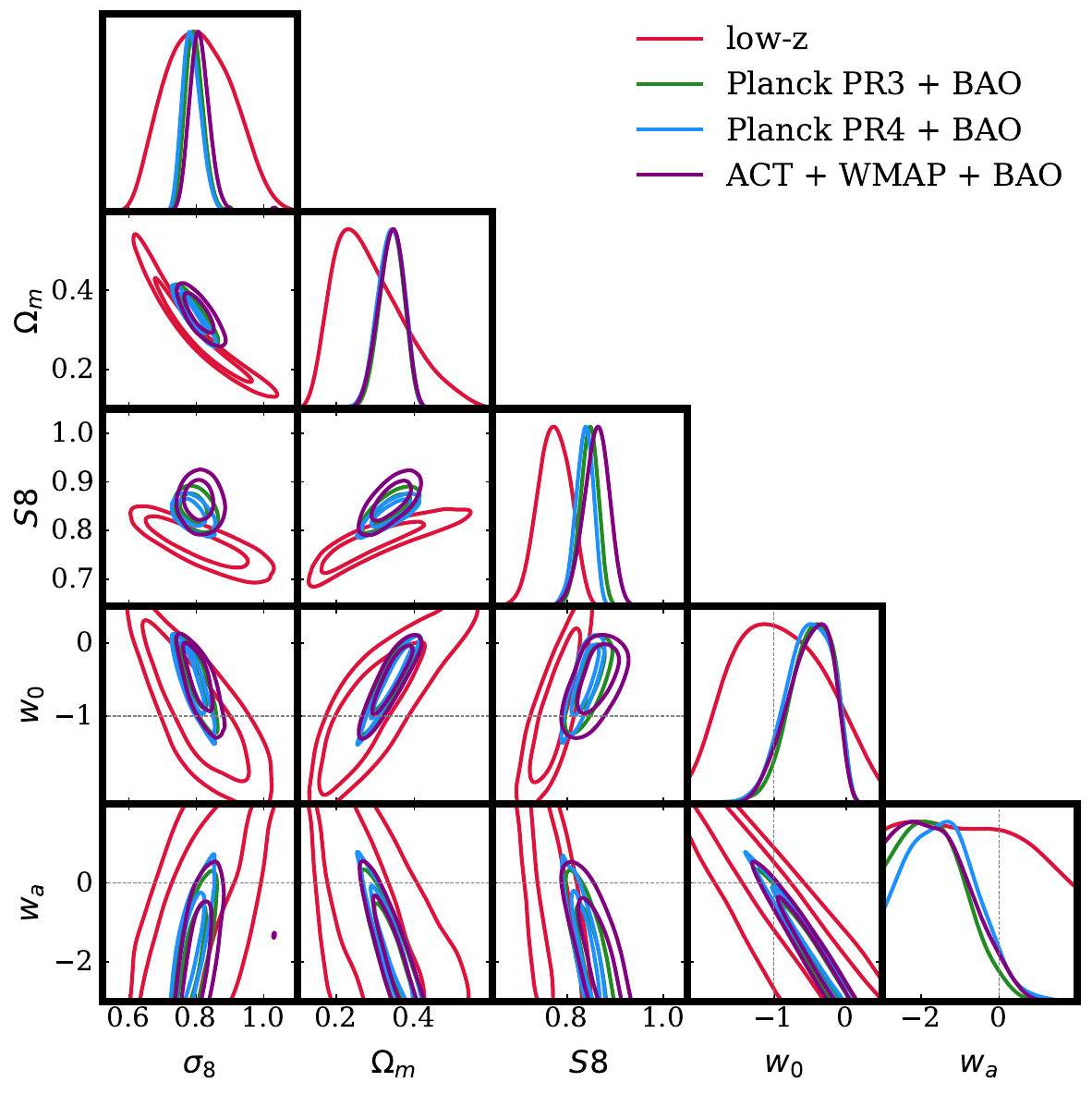}
\caption{The constraints in the $w_0w_a$ model considering the low-$z$ data and the various CMB primary likelihoods analyzed in this work combined with BAO information. The $\Lambda \mathrm{CDM}$ regime is shown in the gray marker points which represent a cosmological constant.\label{fig:w0wa_single}}
\end{figure}

\begin{figure}[!htb]
    \centering
    \includegraphics[width=0.75\textwidth]{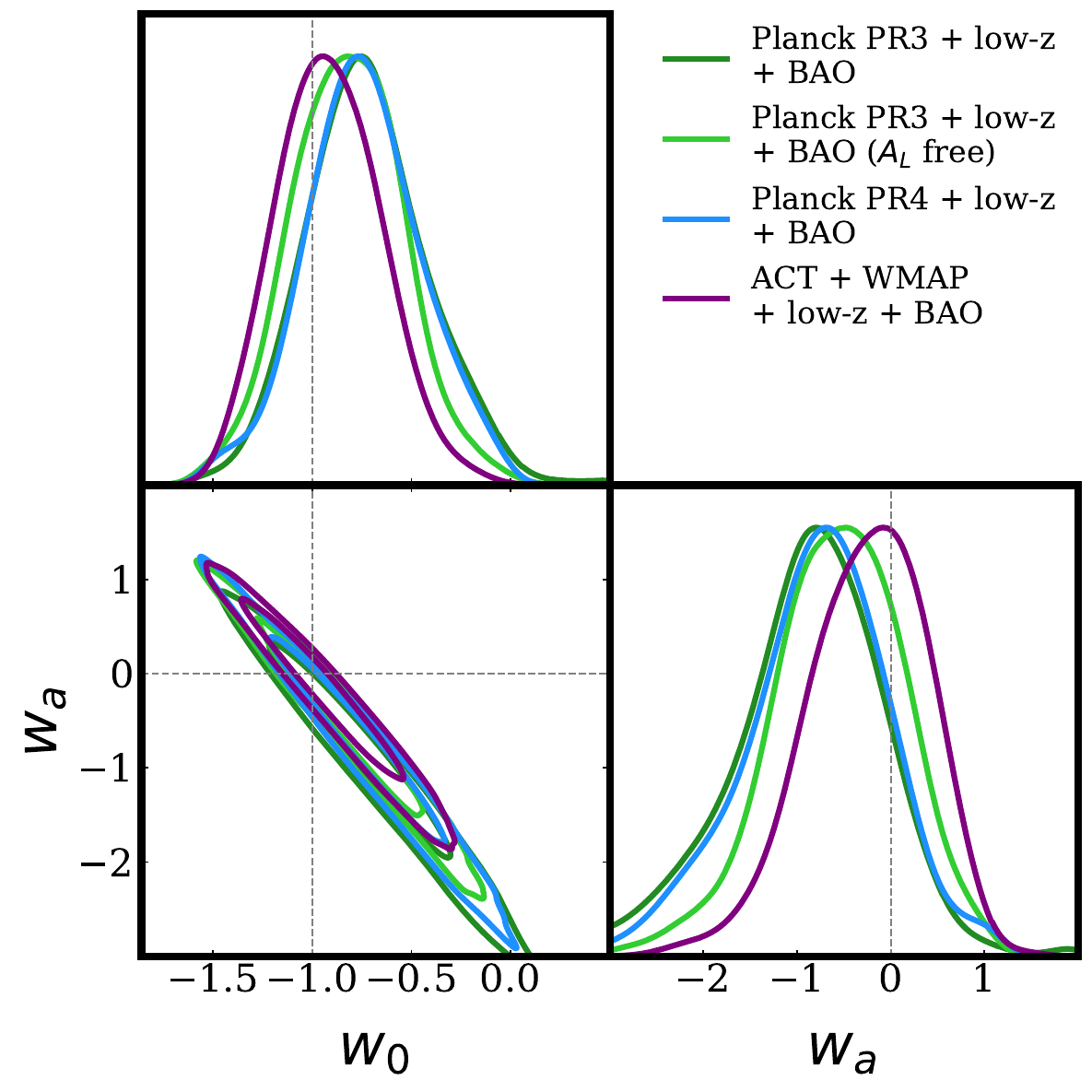}
    \caption{The constraints in the $w_0w_a$ model considering the combination of the CMB, BAO and low-$z$ datasets. The $\Lambda \mathrm{CDM}$ regime is shown in the gray marker points which represent a cosmological constant. \label{fig:w0wa_multi}}
\end{figure}

\begin{table}[ht]
\centering
\caption{Cosmological parameter constraints in the $w_0w_a\mathrm{CDM}$ model using single and combined datasets. 
\label{tab:w0wa_constraints}}
\resizebox{\textwidth}{!}{
\setlength{\tabcolsep}{1.5pt}
\renewcommand{\arraystretch}{1.0}
\begin{tabular}{lccccccc}
\hline
\hline
\textbf{Parameter} 
 & low-$z$ 
 & \makecell{\textit{Planck} \\ PR3 + BAO} 
 & \makecell{\textit{Planck} \\ PR4 + BAO} 
 & \makecell{ACT \\ + WMAP + BAO} 
 & \makecell{\textit{Planck} PR3 \\ +low-$z$ + BAO} 
 & \makecell{\textit{Planck} PR4 \\ +low-$z$ + BAO} 
 & \makecell{ACT+WMAP \\ +low-$z$ + BAO} \\
\hline
\multicolumn{8}{l}{\textbf{Cosmological Parameters}} \\
$\Omega_m$ 
  & $0.290^{+0.091}_{-0.10}$ 
  & $0.346^{+0.035}_{-0.028}$ 
  & $0.339^{+0.034}_{-0.034}$ 
  & $0.345^{+0.037}_{-0.030}$ 
  & $0.317^{+0.031}_{-0.032}$ 
  & $0.313^{+0.031}_{-0.031}$ 
  & $0.302^{+0.028}_{-0.028}$ \\[6pt]
$\sigma_8$ 
  & $0.810^{+0.101}_{-0.102}$ 
  & $0.795^{+0.027}_{-0.027}$ 
  & $0.788^{+0.028}_{-0.028}$ 
  & $0.812^{+0.030}_{-0.028}$ 
  & $0.792^{+0.028}_{-0.030}$ 
  & $0.795^{+0.028}_{-0.028}$ 
  & $0.803^{+0.027}_{-0.028}$ \\[6pt]
$H_0\,[\mathrm{km/s/Mpc}]$ 
  & $71.8^{+14.4}_{-15.1}$ 
  & $64.7^{+2.9}_{-3.1}$ 
  & $64.9^{+3.2}_{-3.2}$ 
  & $65.1^{+3.0}_{-3.2}$ 
  & $67.1^{+3.4}_{-3.2}$ 
  & $67.2^{+3.3}_{-3.3}$ 
  & $68.4^{+3.1}_{-3.2}$ \\[6pt]
$S8$ 
  & $0.772^{+0.033}_{-0.034}$ 
  & $0.850^{+0.023}_{-0.015}$ 
  & $0.835^{+0.018}_{-0.018}$ 
  & $0.868^{+0.032}_{-0.018}$ 
  & $0.812^{+0.012}_{-0.016}$ 
  & $0.810^{+0.014}_{-0.014}$ 
  & $0.804^{+0.015}_{-0.015}$ \\
\hline 
\multicolumn{8}{l}{\textbf{Dynamical dark energy parameters}} \\
$w_0$ 
  & $-0.936^{+0.669}_{-0.692}$ 
  & $-0.485^{+0.301}_{-0.313}$ 
  & $-0.524^{+0.325}_{-0.337}$ 
  & $-0.524^{+0.315}_{-0.336}$ 
  & $-0.724^{+0.302}_{-0.308}$ 
  & $-0.791^{+0.301}_{-0.290}$ 
  & $-0.911^{+0.269}_{-0.268}$ \\[6pt]
$w_a$ 
  & $-0.633^{+1.624}_{-1.623}$ 
  & $-1.685^{+0.862}_{-0.834}$ 
  & $-1.467^{+0.933}_{-0.890}$ 
  & $-1.628^{+0.939}_{-0.926}$ 
  & $-0.861^{+0.778}_{-0.753}$ 
  & $-0.659^{+0.716}_{-0.746}$ 
  & $-0.255^{+0.647}_{-0.657}$ \\
\hline
FoM in $w_0,w_a$ subspace 
 & $2.11$ 
 & $18.8$
 & $18.1$
 & $14.7$
 & $24.1$
 & $27.1$
 & $31.5$ \\
\hline 
 
\end{tabular}
}
\end{table}
We consider a dynamical dark energy model described by the Chevallier--Polarski--Linder (CPL) parameterization~\cite{Chevallier:2000qy, Linder:2002et, DESI:2024kob}:
\begin{equation}
\label{eqt:w0wa}
   w(a) = w_0 + w_a \bigl(1-a\bigr),
\end{equation}
and investigate the associated constraints using our multiprobe framework. Figure~\ref{fig:w0wa_single} shows the results when the cosmic microwave background (CMB) data are combined with DESI Y1 BAO and for the low-$z$ data separately. We recover the previously reported DESI Y1 BAO hint of dynamical dark energy~\cite{DESI:2024mwx} under all three CMB likelihoods tested, although we note a slightly reduced preference for dynamical dark energy when using either the ACT+WMAP or \textit{Planck} PR4 likelihoods compared to the \textit{Planck} PR3 likelihood. In particular, the \textit{Planck} PR3 result continues to rule out a cosmological constant ($w_0=-1, w_a=0$) at $>2\sigma$, whereas ACT+WMAP and \textit{Planck} PR4 both allow consistency with $\Lambda \mathrm{CDM}$ at just within the $2\sigma$ contours. The small difference between the PR4 and PR3 inferences is potentially related to PR3's stronger preference for $A_L>1$~\cite{RoyChoudhury:2024wri}. In Fig.~\ref{fig:w0wa_single} we also see that the projected large-scale structure measurements can provide non-negligible constraining power on dynamical dark energy with low-$z$ data alone sufficient to place a (weak) constraint on $w_0$, with a central value just $0.1\sigma$ away from the $\Lambda \mathrm{CDM}$ value of $w_0=-1$.  

When all three data sets (CMB, DESI Y1 BAO, and low-$z$) are combined (see Fig.~\ref{fig:w0wa_multi}), there is a marked improvement in constraining power in the $(w_0, w_a)$ parameter space.  A convenient way to characterize this improvement is via the Figure of Merit (FoM) which we compute as: 
\begin{equation} \label{equation:fom}
\text{FoM} \;=\; \frac{1}{\sqrt{\bigl|\hat{\Sigma}\bigl(w_0, w_a\bigr)\bigr|}},
\end{equation}
where we compute the parameter covariance $\Sigma$ from our MCMC chains (see values in the bottom row of Table~\ref{tab:w0wa_constraints}). In our baseline scenario, we observe about a $50\%$ increase in the FoM relative to the CMB+BAO constraints alone. Moreover, for the combined data sets, the constraints on $(w_0, w_a)$ are consistent with a cosmological constant within $\sim 1\sigma$ for each of the CMB likelihoods considered. We find that the ACT+WMAP data prefer a cosmology that is highly compatible with $\Lambda \mathrm{CDM}$ with $w_0$ $0.4\sigma$ higher and $w_a$ $0.4\sigma$ lower than $w_0=-1, w_a=0$, whilst our baseline combination with \textit{Planck} PR4 exhibits a shift in the same directions of $0.7\sigma $ and $1\sigma$ respectively. The combination with \textit{Planck} PR3 shows a mildly stronger $\sim1\sigma$ deviation from a cosmological constant. We investigate the impact of allowing $A_L$ to vary in this scenario, finding a constraint on the excess lensing parameter of $A_L=1.17^{+0.06}_{-0.06}$, representing a $3\sigma$ deviation from the expected value. We find that once this parameter is allowed to vary the constraints move closer to the cosmological constant scenario (light green contour in Fig.~\ref{fig:w0wa_multi}) allowing this within $1\sigma$ (see also Ref.~\cite{Chan-GyungPark:2025cri} who find a similar impact of $A_L$ in this scenario). We checked that varying $A_L$ does not produce any noticeable shifts in the constraints when considering the other CMB likelihoods (\textit{Planck} PR4 and ACT + WMAP) in this framework. 

The origins of the shift toward a cosmological constant when including the low-$z$ data can be understood by examining the degeneracy directions in Fig.~\ref{fig:w0wa_single}. We see that moving along the $w_0,w_a$ degeneracy direction away from $\Lambda \mathrm{CDM}$ causes an increase in $\Omega_m$, a decrease in $\sigma_8$ but an overall increase in $S8$ as the increase in $\Omega_m$ outweighs the slight decrease in $\sigma_8$ (recall $S8 = \sigma_8\sqrt{\Omega_m/0.3}$). The low-$z$ data prefer a lower $\Omega_m$ and $S8$ than the CMB data and hence the addition of this data pulls the CMB + DESI Y1 BAO constraint along the degeneracy directions towards $\Lambda \mathrm{CDM}$.

\begin{figure}[!htb]
    \centering
    \includegraphics[width=0.8\textwidth]{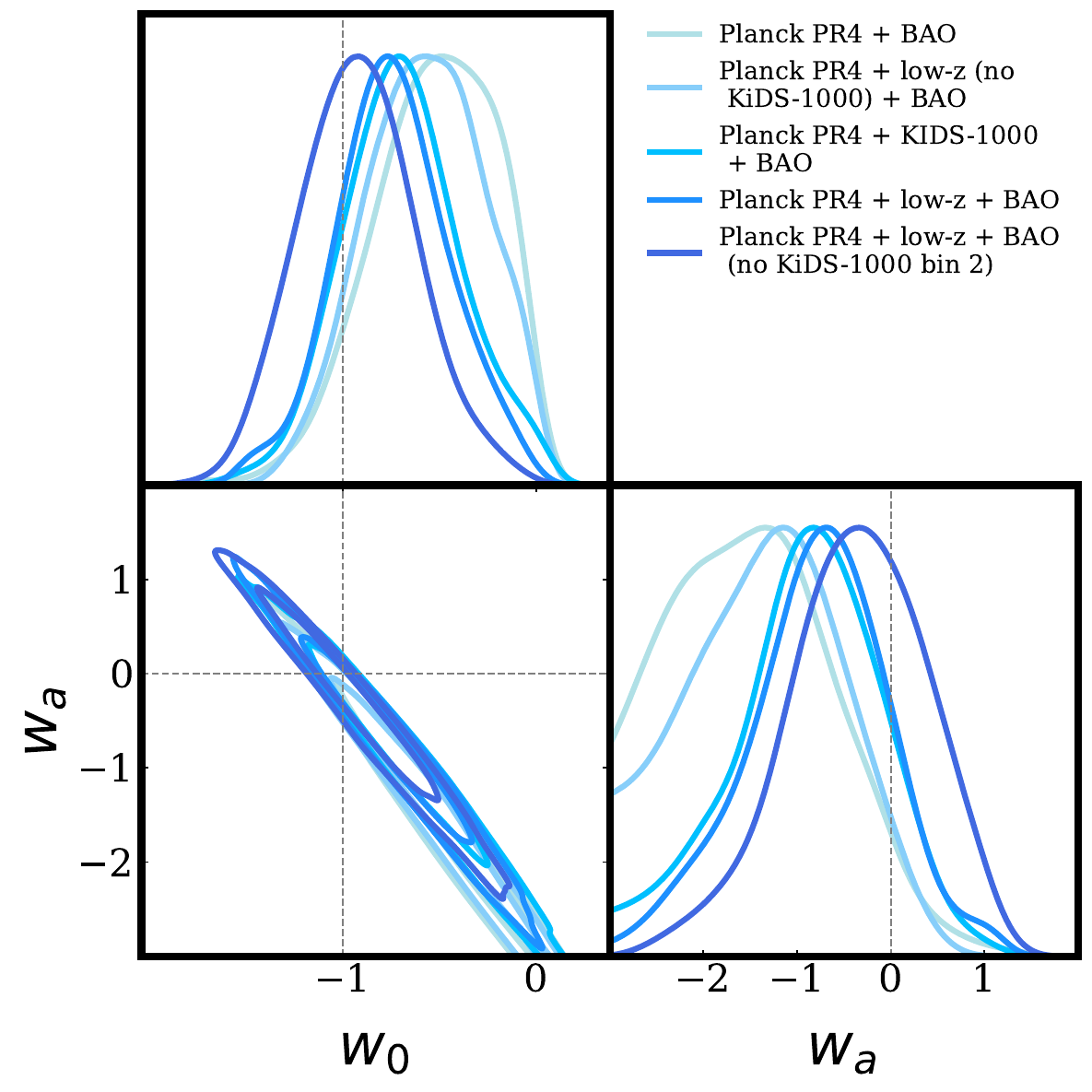}
    \caption{Constraints in the $w_0w_a$ plane when combining the \textit{Planck} PR4 likelihood with various other datasets, demonstrating where the preference for the observed shift toward $w_0=-1, w_a=0$ is coming from within the low-$z$ data vector. The $\Lambda \mathrm{CDM}$ regime is shown in the gray marker points which represent a cosmological constant. \label{fig:w0wa_info}}
\end{figure}

Figure~\ref{fig:w0wa_info} illustrates how different subsets of the low-$z$ data vector, analyzed in combination with our baseline \textit{Planck} PR4 likelihood, contribute to pulling the constraints toward $w_0=-1, w_a=0$. We find that the KiDS-1000 weak lensing data set is particularly influential, although the other components of the low-$z$ data vector also contribute to improving the overall constraining power and give a small shift in the central values (see the second lightest blue contour in Fig.~\ref{fig:w0wa_info}). Given the sensitivity of the dynamical dark energy constraint to the weak lensing dataset we accordingly see a bigger shift in contours when removing the KiDS-1000 redshift bin 2 that we identified as anomalous in Section~\ref{subsec:internal_consistency} compared to the $\Lambda \mathrm{CDM}$ case. The darkest blue contour in Fig.~\ref{fig:w0wa_info} shows that removing this bin from the analysis further drives the constraints towards $\Lambda \mathrm{CDM}$, this can be understood as the data in this bin prefer a high lensing amplitude than the rest of the KiDS dataset (cf. the finding that $m_{\gamma2}>0$ at $1.8\sigma$ when considering the low-$z$ data alone in Section~\ref{subsec:internal_consistency}), hence removing this bin slightly lowers the inferred $S8$ and $\Omega_m$ from the low-$z$ data vector which in turn causes a stronger pull towards $\Lambda \mathrm{CDM}$ when the data are combined. See Appendix~\ref{appendix:kids_bin_2} for further discussion of the influence of removing this bin on the constraints in the $w_0w_a\mathrm{CDM}$ model, including the full parameter constraints in this scenario for each of the three CMB likelihoods. Overall, these findings suggest that current projected large-scale structure measurements, and in particular weak lensing data, strongly limit the extent to which dynamical dark energy models can differ from a cosmological constant. 

\subsection{Neutrino masses}

\subsubsection{$\nu \Lambda \mathrm{CDM}$ model}

\begin{figure}[!htb]
\centering
\includegraphics[width=0.75\textwidth]{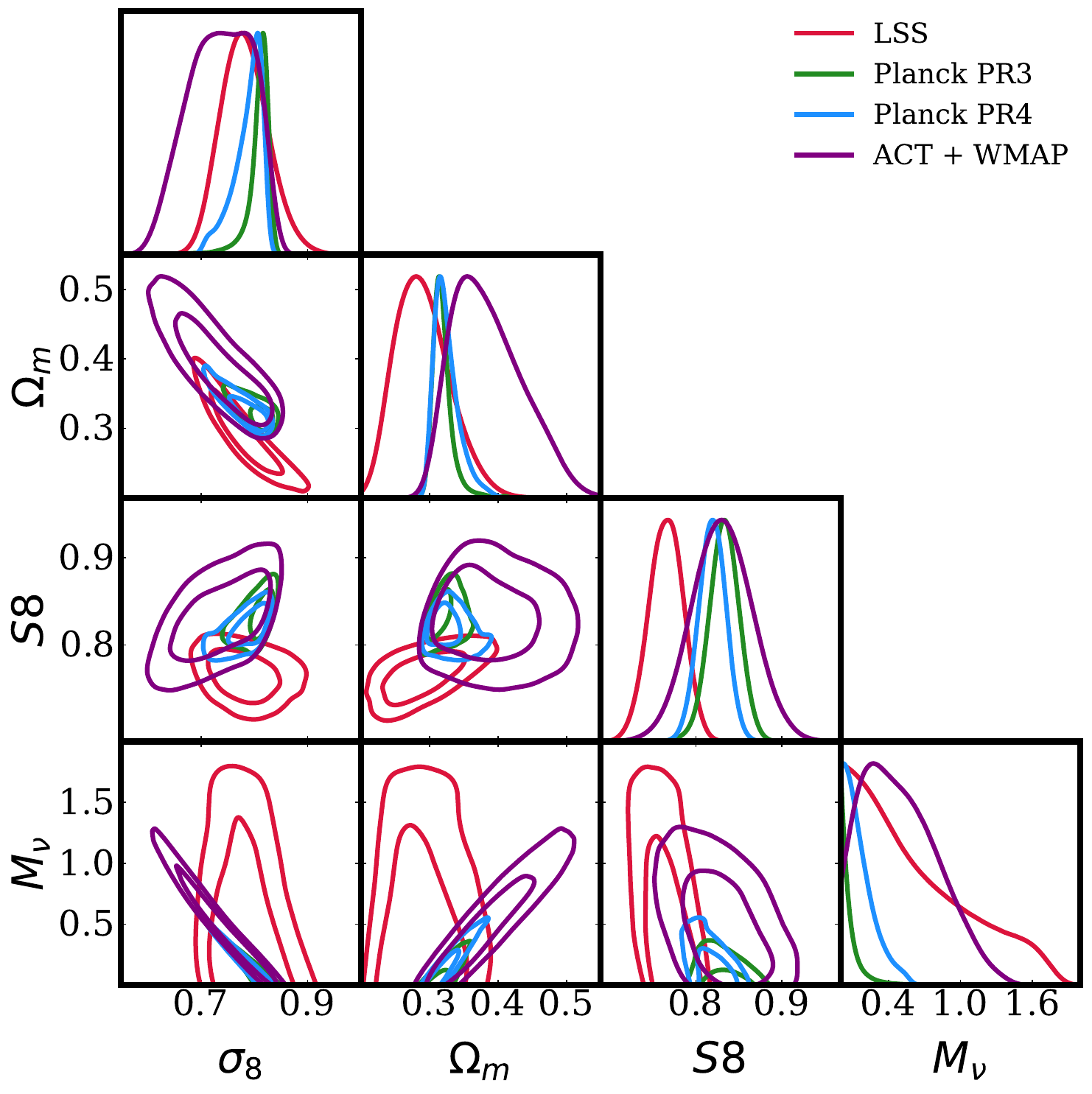}
\caption{The constraints in the $\nu \Lambda \mathrm{CDM}$ model considering the low-$z$ data and the various CMB primary likelihoods analyzed in this work.\label{fig:baseline_single_mnu}}
\end{figure}
\begin{figure}[!htb]
\centering
\includegraphics[width=0.8\textwidth]{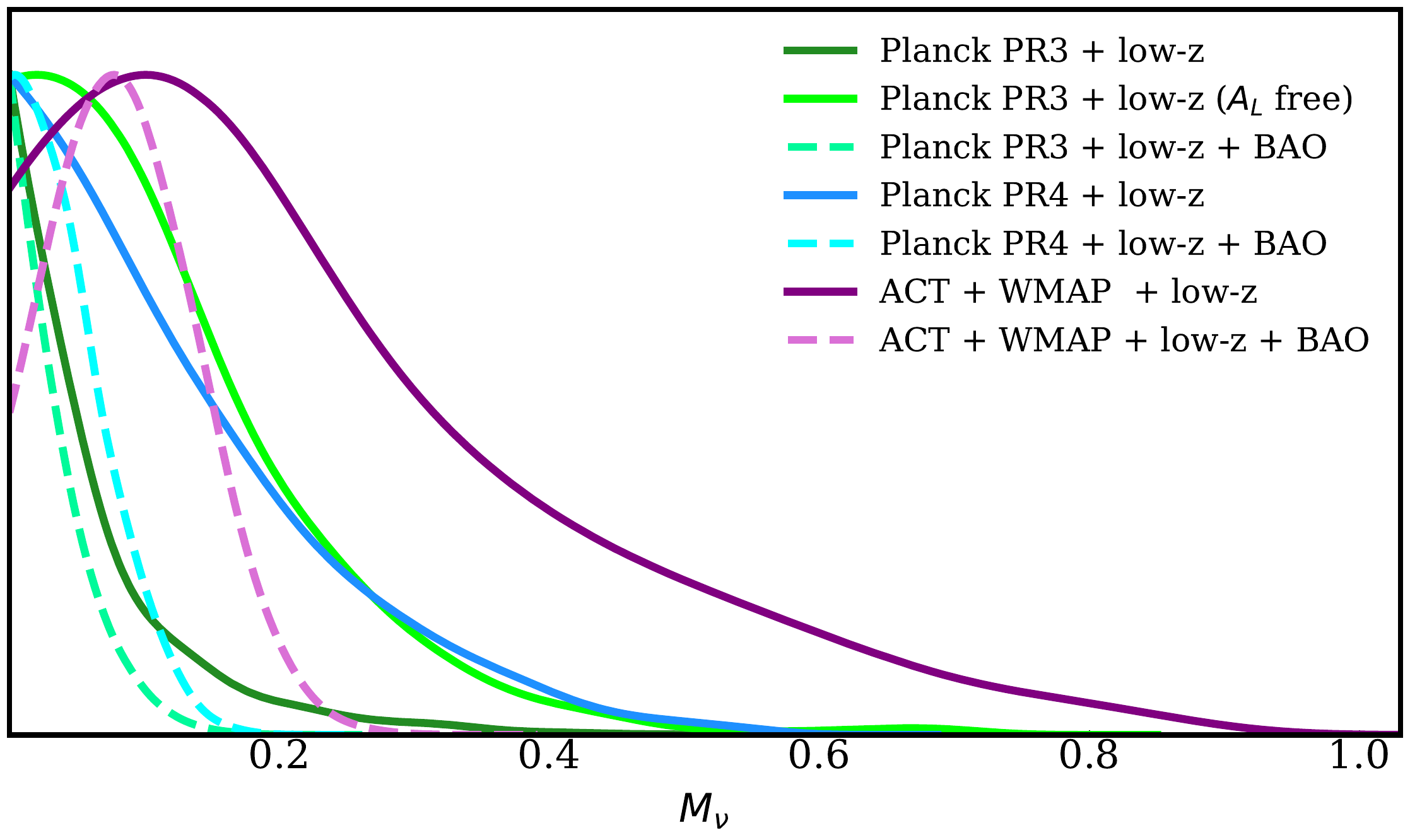}
\caption{The 1D neutrino mass constraints in the $\nu \Lambda \mathrm{CDM}$ model considering the combination of low-$z$ data with CMB likelihoods and DESI Y1 BAO measurements.\label{fig:multi_1d-constraints}}
\end{figure}

\begin{table}[!htb]
\centering
\caption{95\% CL upper limits on the sum of neutrino masses, $M_\nu$, in the $\nu\Lambda \mathrm{CDM}$ model from various datasets. 
We group the datasets by whether they include only CMB, CMB+low-$z$, CMB+BAO, or CMB+low-$z$+BAO. 
``Freed $A_L$?'' indicates whether the CMB lensing amplitude $A_L$ is allowed to vary (``Yes'') or set to $1$ (``No''). 
All CMB datasets here exclude the CMB lensing auto-correlation, which is instead incorporated into the low-$z$ dataset.\label{tab:nuLCDM}}
\begin{tabular}{lcc}
\toprule
\multirow{2}{*}{\textbf{Dataset}}  & \multirow{2}{*}{\textbf{Freed $A_L$?}} & \multirow{2}{*}{$\boldsymbol{M_\nu}$ \textbf{(eV)}}\\
& & \\
\midrule
\addlinespace[5pt]
\multicolumn{3}{l}{\textbf{CMB only}} \\
\addlinespace[5pt]
\quad low-$z$ only            & No  & $<1.55$ \\
\quad \textit{Planck} PR3     & No  & $<0.25$ \\
\quad \textit{Planck} PR4     & No  & $<0.46$ \\
\quad ACT DR4 + WMAP          & No  & $<1.07$ \\
\midrule

\multicolumn{3}{l}{\textbf{CMB + low-$z$}} \\
\addlinespace[5pt]
\quad \textit{Planck} PR3 + low-$z$        & No  & $<0.22$ \\
\quad \textit{Planck} PR3 + low-$z$        & Yes & $<0.32$ \\
\quad \textit{Planck} PR4 + low-$z$        & No  & $<0.34$ \\
\quad \textit{Planck} PR4 + low-$z$        & Yes & $<0.35$ \\
\quad ACT DR4 + WMAP + low-$z$             & No  & $<0.66$ \\
\midrule

\multicolumn{3}{l}{\textbf{CMB + BAO}} \\
\addlinespace[5pt]
\quad \textit{Planck} PR3 + BAO   & No  & $<0.09$ \\
\quad \textit{Planck} PR3 + BAO   & Yes & $<0.20$ \\
\quad \textit{Planck} PR4 + BAO   & No  & $<0.11$ \\
\quad \textit{Planck} PR4 + BAO   & Yes & $<0.17$ \\
\quad ACT DR4 + WMAP + BAO        & No  & $<0.21$ \\
\midrule

\multicolumn{3}{l}{\textbf{CMB + low-$z$ + BAO}} \\
\addlinespace[5pt]
\quad \textit{Planck} PR3 + low-$z$ + BAO   & No  & $<0.09$ \\
\quad \textit{Planck} PR3 + low-$z$ + BAO   & Yes & $<0.13$ \\
\quad \textit{Planck} PR4 + low-$z$ + BAO   & No  & $<0.12$ \\
\quad \textit{Planck} PR4 + low-$z$ + BAO   & Yes & $<0.13$ \\
\quad ACT DR4 + WMAP + low-$z$ + BAO        & No  & $<0.18$ \\
\bottomrule
\end{tabular}
\end{table}

We present constraints on the sum of neutrino masses, $M_{\nu}$, from various uncombined likelihoods in Fig.~\ref{fig:baseline_single_mnu}. The ACT+WMAP combination yields $M_\nu < 1.07\,\mathrm{eV}$, slightly more constraining than the $M_\nu < 1.2\,\mathrm{eV} $ reported in Ref.~\cite{ACT:2020gnv} due to our stronger $\tau$ prior but remaining weaker than constraints from either \textit{Planck} likelihood. This reflects at least in part the lower statistical power of this dataset due to the lack of $C_\ell^{TE}$ data below $\ell=24$ and $C_\ell^{EE}$ data below $\ell=350$ (in contrast to \textit{Planck}'s full coverage from $\ell=2$ to $350$) and related to this the reliance on an external $\tau$ prior that does not fully capture parameter correlations. While all three likelihoods share similar degeneracy directions with $M_\nu$, ACT+WMAP's broader limits translate to wider uncertainties in the other cosmological parameters as well. We also find that the low-$z$ data alone can place a weak constraint on the neutrino mass of $M_\nu < 1.551\mathrm{eV}$.

Figure~\ref{fig:multi_1d-constraints} illustrates the one-dimensional constraints on $M_{\nu}$ when combining CMB data with low-$z$ data and optionally DESI Y1 BAO measurements (dashed contours). We find $M_{\nu}<0.34\,\mathrm{eV}$ upon combining \textit{Planck} PR4 with the low-$z$ data compared to $M_{\nu}<0.46\,\mathrm{eV}$ for the \textit{Planck} PR4 data alone representing a $26\%$ decrease in the 95\% CL upper limit (see Table~\ref{tab:nuLCDM} for an overview of the neutrino mass constraints). As illustrated in Fig.~\ref{fig:baseline_single_mnu}, the low-$z$ data constrain $\Omega_m$ in a manner that weakly correlates with $M_\nu$, helping to lift the degeneracies seen in CMB-only analyses and improve constraining power (see also Ref.~\cite{Loverde:2024nfi}). 

Including DESI Y1 BAO data further sharpens this limit due to additional constraining power on the late-time expansion history, yielding \(M_\nu < 0.122\,\mathrm{eV}\). This result is competitive with current bounds, although it is somewhat weaker than the \(0.072\,\mathrm{eV}\) reported by the DESI collaboration which used a combination of \textit{Planck} PR3 (including CMB lensing) and BAO~\cite{DESI:2024mwx}. This widening partly reflects our use of the \hillipop\ likelihood, which is known to relax neutrino mass constraints relative to the \textit{Planck} PR3 likelihood used by the DESI team \cite{Escudero:2024uea, Allali:2024aiv, Naredo-Tuero:2024sgf}. We checked that we recover a constraint compatible with the DESI results when using the same combination of data (finding $M_\nu < 0.73\mathrm{eV}$ at 95\% CL).

As for the CMB alone case, we find that the constraints from combinations of data with ACT+WMAP are significantly weaker than with the \textit{Planck} likelihoods. We additionally find a posterior that peaks at $M_\nu>0$, in contrast to the constraints with \textit{Planck} that peak at $M_\nu=0$ (see Ref.~\cite{DiValentino:2021imh} who also find neutrino mass posteriors that peak above $M_\nu=0$ when considering ACT+WMAP data combinations). See Appendix~\ref{appendix:nulcdm_plots} for an overview of the full parameter constraints when combining with BAO in this model.

An important consideration is how robust the neutrino mass constraints are to potential systematic effects in the data. In Paper~I, we showed that a consistent combination of CMB and large-scale structure (LSS) observations can mitigate the impact of $A_{\mathrm{lens}}$-like systematics. Here, we find that allowing $A_{L}$ to vary causes only a small shift in the neutrino mass bound for our baseline combination of \textit{Planck} PR4 low-$z$ data and DESI Y1 BAO, moving the upper limit from $M_\nu < 0.12\,\mathrm{eV}$ to $M_\nu < 0.13\,\mathrm{eV}$. This modest change arises partly because the \textit{Planck} PR4 likelihood exhibits only a mild ($0.75\sigma$) preference for $A_L > 1$~\cite{Tristram:2023haj}, but is also aided by the constraining power of the low-$z$ data, which helps preserve robust results. Indeed, when we analyze \textit{Planck} PR4 + DESI Y1 BAO without the addional low-$z$ data, marginalizing over $A_{L}$ broadens the limit more substantially, from $M_\nu < 0.11\,\mathrm{eV}$ to $M_\nu < 0.17\,\mathrm{eV}$ (see Table~\ref{tab:nuLCDM}). In contrast, the \textit{Planck} PR3 data exhibit a stronger preference for $A_L > 1$, so even combining PR3 with low-$z$ data cannot fully diminish the impact of this systematic. We find that when allowing this parameter to vary the constraints from \textit{Planck} PR3 become almost identical to those of PR4 when combining with low-$z$ data (see the light green solid line in Fig.~\ref{fig:multi_1d-constraints} compared to the light blue solid line), suggesting the difference in neutrino mass constraint from these likelihoods can be attributed to the impact of the excess lensing residual in \textit{Planck} PR3. We also checked the impact of varying $A_L$ on the ACT+WMAP + low-$z$ + BAO neutrino mass constraints, finding that this produces almost no shift in the posterior, consistent with the fact that the ACT team finds no preference for a deviation from $A_L=1$~\cite{ACT:2020gnv}. Finally, we verified that removing the second KiDS-1000 redshift bin has a negligible impact on the constraints and conclusions presented in this section.

Overall, our baseline (\textit{Planck}~PR4 + low-$z$ + DESI Y1 BAO) provides a robust, holistic bound on $M_\nu$ unaffected by $A_{\mathrm{lens}}$-like systematics. This same combination but swapping \textit{Planck} PR4 with ACT+WMAP data gives a neutrino mass posterior that, in contrast to the \textit{Planck} posteriors, peaks at $M_\nu>0$. Upcoming ACT CMB primary data (ACT DR6) will likely determine whether this is due to a genuine preference for a higher inferred neutrino mass from ACT+WMAP compared to \textit{Planck} or simply reflects the lower statistical power of these data.

\subsubsection{$\nu w_0w_a$ model} 
\begin{table}[!htb]
\centering
\caption{Neutrino mass constraints in the $\nu w_0w_a\mathrm{CDM}$ model. Quoted values are 95\% CL upper limits if shown as `$<$', or mean $\pm 1\sigma$ otherwise.\label{tab:nuw0wa}}
\begin{tabular}{lcc}
\toprule
\textbf{Dataset}  & \textbf{Freed $A_L$?} & $\boldsymbol{M_\nu}$ \textbf{(eV)} \\
\midrule
\textit{Planck} PR3 + low-$z$ + BAO   & No  & $<0.26$ \\
\textit{Planck} PR3 + low-$z$ + BAO   & Yes & $0.16^{+0.09}_{-0.10}$ \\
\textit{Planck} PR4 + low-$z$ + BAO   & No  & $0.16^{+0.09}_{-0.09}$ \\
ACT DR4 + WMAP + low-$z$ + BAO        & No  & $0.26^{+0.10}_{-0.11}$ \\
\bottomrule
\end{tabular}
\end{table}

\begin{figure}
\centering
\includegraphics[width=0.6\textwidth]{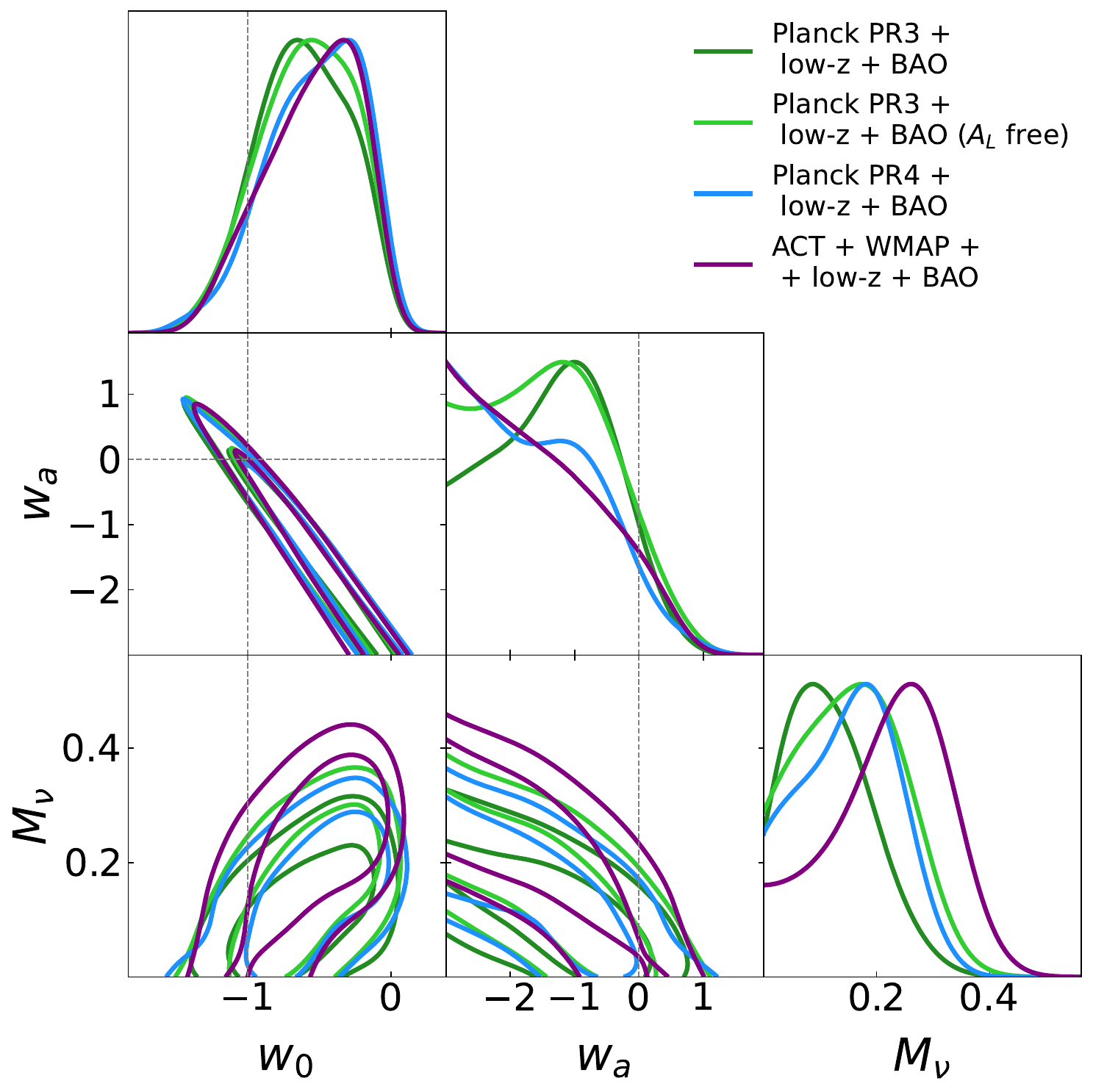}
\caption{$\nu w_0w_a$ model constraints for the three CMB likelihoods in the pipeline combined with low-$z$ and DESI Y1 BAO data.\label{fig:mnu_w0wa}}
\end{figure}

In this section, we analyze neutrino mass constraints under the Chevallier--Polarski--Linder (CPL) parameterization of the dark energy equation of state. While previous studies have explored neutrino mass within this extended parameter space using DESI Y1 BAO in combination with \textit{Planck}, ACT, and Type~Ia supernova data \cite{Du:2024pai, Reboucas:2024smm, Escudero:2024uea, RoyChoudhury:2024wri, Elbers:2024sha}, we provide the first analysis incorporating our $9\times2$pt low-$z$ data vector.

Figure~\ref{fig:mnu_w0wa} and Table~\ref{tab:nuw0wa} show the constraints obtained by combining three CMB likelihoods with the low-$z$ dataset and DESI Y1 BAO data. As expected, the ability to constrain neutrino mass diminishes in this extended parameter space, since the background effects of neutrino mass are partially degenerate with evolving dark energy at low redshifts~\cite{Allison:2015qca}. Although in the extended scenario with weakened constraining power we cannot fully close the $w_0$--$w_a$ contour, the $\Lambda\mathrm{CDM}$ point ($w_0=-1,\,w_a=0$) remains very close to the $1\sigma$ contours for each of the CMB experiments considered. See Appendix~\ref{appendix:nuw0wa_plots} for the full parameter constraints in this model. 

Consistent with earlier work \cite{Elbers:2024sha, RoyChoudhury:2024wri}, we find a preference for larger neutrino masses in dynamical dark energy models. In particular, the \textit{Planck} PR4 and ACT+WMAP combinations yield constraints of $M_\nu = 0.160^{+0.087}_{-0.095}\,\mathrm{eV}$ and $M_\nu = 0.255^{+0.101}_{-0.112}\,\mathrm{eV}$, respectively. Consistent with the $\nu\Lambda\mathrm{CDM}$ results, we see that the ACT+WMAP combination allows for a higher neutrino mass compared to the \textit{Planck} likelihoods, albeit with a slightly larger error bar. For \textit{Planck} PR3 with a variable $A_L$, the result $M_\nu = 0.160^{+0.094}_{-0.099}\,\mathrm{eV}$ closely matches the \textit{Planck} PR4 constraint, and we see a $\sim3\sigma$ preference for $A_L>1$ ($A_L = 1.18 \pm 0.06$). Varying $A_L$ has a negligible impact on constraints from the other two CMB likelihoods. 

The results presented in this section are relevant to the ongoing debate in the literature regarding the compatibility of cosmological neutrino mass constraints with limits from neutrino oscillations, which require $M_\nu \ge 0.06\,\mathrm{eV}$ under normal ordering. In particular, it has been shown that the \textit{Planck} PR3 + DESI BAO combination in a $\nu\Lambda\mathrm{CDM}$ context favors (unphysical) negative neutrino mass~\cite{Craig:2024tky, Green:2024xbb}, possibly linked to the preference for $A_L>1$~\cite{Elbers:2024sha}. In contrast, the constraints presented here for \textit{Planck} PR4 and ACT+WMAP, which remain robust when $A_L$ is varied yield a posterior that peaks at $M_\nu>0$ at $2.4\sigma$ for ACT+WMAP and $1.6\sigma$ for \textit{Planck} PR4, fully compatible with oscillation experiments. This demonstrates the sensitivity of neutrino mass constraints to the assumed cosmological model, which should be considered when discussing incompatibility with oscillation data. Nevertheless, these findings must be interpreted with caution. Neutrino mass limits in dynamical dark energy scenarios depend on the assumed dynamical dark energy model and parameterization \cite{Elbers:2024sha, Reboucas:2024smm}. Furthermore, it should be highlighted that, as also pointed out in Ref.~\cite{Reboucas:2024smm}, the degeneracies in Fig.~\ref{fig:mnu_w0wa} are such that models with higher neutrino mass represent greater departures from the cosmological constant scenario. 

Overall, these results provide a foundation for further investigation required to disentangle neutrino mass and dynamical dark energy constraints and test the consistency of these constraints with neutrino oscillation data. It will be interesting to see how the hint of a neutrino mass constraint from the ACT+WMAP data in this extended scenario evolves with upcoming higher-resolution ACT DR6 data. 

\subsection{Model comparison\label{subsec:model_comparison}}

\begin{table}[ht]
    \centering
    \caption{Comparison of \(\Delta \mathrm{AIC}\) compared to the baseline $\Lambda \mathrm{CDM}$ model for different cosmological models under various datasets. The first block shows results \emph{without} BAO, and 
      the second block shows results \emph{with} BAO.\label{tab:model_comparison}}
    \label{tab:deltaAIC}
    
    \begin{adjustbox}{max width=0.9\linewidth,fbox} 
    \renewcommand{\arraystretch}{1.2} 
    \setlength{\tabcolsep}{6pt}       
    
    \begin{tabular}{@{}lccc@{}}
    \multicolumn{4}{c}{\textbf{No BAO}} \\
    \hline
       & \textit{Planck} PR3 + low-$z$ 
       & \textit{Planck} PR4 + low-$z$ 
       & ACT + WMAP + low-$z$ \\
    \hline
    \(\nu\Lambda \mathrm{CDM}\) & +2.22 & +3.12 & +1.95 \\
    \(w_0 w_a \mathrm{CDM}\)                 & -0.95 & +2.48 & +3.40 \\
    \(\nu w_0 w_a \mathrm{CDM}\)             & +1.50 & +4.95 & +4.90 \\
    \hline
    \multicolumn{4}{c}{\textbf{With BAO}} \\
    \hline
       & \textit{Planck} PR3 + low-$z$ + BAO
       & \textit{Planck} PR4 + low-$z$ + BAO
       & ACT + WMAP + low-$z$ + BAO \\
    \hline
    \(\nu\Lambda \mathrm{CDM}\) & -1.14 & +0.24 & 2.33  \\
    \(w_0 w_a \mathrm{CDM}\)                 & -1.39 & +1.13 & 5.73 \\
    \(\nu w_0 w_a \mathrm{CDM}\)             & +0.06 & +2.99 & 4.87 \\
    \end{tabular}
    \end{adjustbox}
\end{table}

Finally, we perform model comparison to determine if there is a statistical preference for any of the extensions we have explored in this study. Following~\cite{1100705}, we can compute the Akaike information criterion (AIC), which represents an unbiased estimator of the expected Kullback discrepancy; a measure of the similarity of the true underlying model and the candidate model~\cite{aic_review}. It accounts for a differing number of free parameters, penalizing a higher number of free parameters that represent a more complex model, which does not lead to a sufficient improvement in fit. For a given model it is given by:
\begin{eqnarray}
\text{AIC} = 2k + \min(\chi^2)
\end{eqnarray}
where $k$ is the number of parameters and $\min(\chi^2)$ denotes the lowest $\chi^2$ found in our MCMC chains. A model with a lower AIC is statistically preferred.

We summarize these findings in Table~\ref{tab:model_comparison}. For our baseline dataset combination, $\Lambda \mathrm{CDM}$ achieves the lowest AIC; however, the extended models do not exceed the usual $\Delta\mathrm{AIC} > 10$ threshold for strong rejection~\cite{burnham_aic}\footnote{This threshold corresponds to an evidence ratio (calculated by $\exp{(-\frac{1}{2}\Delta})$) of $\sim 150$ in favor of the model with the smallest AIC~\cite{burnham_aic}.}. The only instances where adding parameters slightly lowers the AIC relative to $\Lambda\mathrm{CDM}$ occur when using the \textit{Planck} PR3 likelihood with low-$z$ data (in dynamical dark energy or $\nu\Lambda\mathrm{CDM}$ when including BAO), though even there the improvements are insufficient to make a strong statistical claim. This mild preference may relate to the $A_{\mathrm{lens}}$ tension in the PR3 likelihood, as others have noted that this tension can favor dynamical dark energy~\cite{Chan-GyungPark:2025cri, Elbers:2024sha}. Similarly, we demonstrated that the presence for $A_L>1$ in PR3 can artificially prefer models with lower neutrino masses which may explain the improved fit in $\nu\Lambda\mathrm{CDM}$. We note that our MCMC samples may not always include the exact best-fit point for each model. A dedicated minimization routine could improve the AIC estimates, but we leave this refinement to future work. Overall, we find no strong statistical basis to discard the model extensions explored here, although there is a modest AIC preference for $\Lambda\mathrm{CDM}$ in our baseline scenario. Hence, following Occam's razor, we see no compelling statistical motivation to replace $\Lambda\mathrm{CDM}$ with any of these extensions.

\section{Conclusions}
\label{sec:conclusion} 

In this work, we simultaneously analyzed several cosmological datasets using our multiprobe framework developed in Paper I~\cite{Reeves:2023kjx}, which includes a simulation-based covariance and a rapid \texttt{JAX} emulator-based inference pipeline. We used three different CMB datasets (\textit{Planck} PR3, \textit{Planck} PR4 and ACT + WMAP), DESI Y1 BAO data and a $9\times2$pt low-$z$ combination of KiDS-1000 weak lensing, BOSS DR12 galaxy clustering and \textit{Planck} PR3 lensing/ISW including all possible cross-correlations. We derived constraints in the baseline $\Lambda \mathrm{CDM}$ model and, motivated by recent DESI results, dynamical dark energy and neutrino mass extensions.  

Within the context of the $\Lambda \mathrm{CDM}$ model, we find a mild ($< 2 \sigma$) tension between the low-$z$ and CMB datasets in the full parameter space which is partially alleviated when allowing beyond-$\Lambda \mathrm{CDM}$ extensions. This mild discrepancy aligns with what is commonly referred to as the ``$S8$ tension'' widely reported in the literature. By introducing an independent amplitude parameter for each spectrum in our pipeline, we identified the second-highest redshift bin of KiDS-1000 (bin 4) as carrying the largest share of this tension. Other amplitude parameters remain consistent with their expected values for BOSS DR12, \textit{Planck} CMB lensing, \textit{Planck} PR4, and ACT + WMAP. However, we identify evidence for a mild discrepancy in the KiDS-1000 second-lowest redshift bin (bin 2) which prefers an anomalously high amplitude. Whilst the data in this bin has almost no impact on our cosmological constraints in $\Lambda \mathrm{CDM}$ and $\nu\Lambda \mathrm{CDM}$, we show this can produce mild though non-negligible shifts when considering dynamical dark energy models. It will be interesting to see if the mild statistical preference for the anomalously high amplitude remains in the forthcoming KiDS-1000 legacy data analysis~\cite{Kids_legacy}. We find more significant evidence of a discrepancy in the \textit{Planck} PR3 primary power spectrum likelihood. This can be interpreted as a recasting of the $A_{\mathrm{lens}}$ anomaly which can impact neutrino mass and dynamical dark energy inferences. Based on these findings, and the reliance of the ACT+WMAP data on an external $\tau$-prior, we define our baseline setup as \textit{Planck} PR4 combined with the low-$z$ dataset and optionally the DESI Y1 BAO data. 

We present a novel $9 \times 2$pt low-$z$ measurement of $S8=0.777^{+0.17}_{-0.17}$ in $\Lambda \mathrm{CDM}$. When combining CMB data and low-$z$ data in the $\Lambda\mathrm{CDM}$ framework we observe a high level of agreement between ACT and \textit{Planck} and demonstrate that mild shifts in $H_0$ and $\Omega_m$ can be traced back to the comparatively weaker constraint from ACT+WMAP allowing for greater influence from the low-$z$ data to shift parameter constraints along degeneracy directions. 

Moving beyond $\Lambda \mathrm{CDM}$, we explore a dynamical dark energy model using the CPL parameterization. We find that adding low-$z$ data significantly improves constraints on the CPL parameters, achieving a 50\% improvement in the figure of merit (FoM) for the $(w_0, w_a)$ plane in our baseline configuration. Notably, including low-$z$ data removes the $\sim 2\sigma$ preference for evolving dark energy present in CMB+BAO alone, with the combined constraints now accommodating a cosmological constant ($w_0 = -1$, $w_a = 0$) within $1\sigma$ for all CMB datasets considered. We identify the KiDS-1000 weak lensing dataset as the primary driver of this shift towards $\Lambda \mathrm{CDM}$, though other components of the low-$z$ data vector also contribute. Furthermore, removing KiDS-1000 redshift bin 2 slightly strengthens the preference for a cosmological constant. These results suggest that strong deviations from a cosmological constant—particularly along the CMB+BAO degeneracy directions towards a decaying dark energy equation of state (i.e., $w_0 > -1$, $w_a < 0$)—are disfavored by current projected LSS data. These findings highlight the importance of current and future LSS data in constraining dark energy models. Upcoming weak lensing surveys, such as Euclid~\cite{EUCLID:2011zbd}, LSST~\cite{LSSTScience:2009jmu}, and the Roman Space Telescope~\cite{Akeson:2019biv}, will further enhance this capability.

Next, we turn our attention to constraints on the total neutrino mass. In the minimal extension, \(\nu\Lambda\mathrm{CDM}\), our baseline data combination---consisting of \textit{Planck} PR4, low-$z$, and DESI Y1 BAO---yields an upper limit of $M_\nu < 0.12 \mathrm{eV}$. This bound is robust to $A_{\mathrm{lens}}$-like systematics which is partially due to the consistent combination with the low-$z$ data vector~\cite{Reeves:2023kjx}. We reconfirmed previous reports of the $A_{\mathrm{lens}}$ anomaly in the \textit{Planck} PR3 likelihood producing anomalously tight neutrino mass constraints~\cite{Naredo-Tuero:2024sgf}. When this is accounted for, by varying $A_L$ in the analysis, we find constraints that are indistinguishable from the equivalent combinations with \textit{Planck} PR4 data. We find the constraints from combinations of data with ACT+WMAP to be both broader and shifted to slightly higher $M_\nu$ compared to combinations with \textit{Planck}. This small shift towards higher $M_\nu$ likely reflects the comparatively lower signal-to-noise ratio in ACT+WMAP data and the use of an external $\tau$-prior. Upcoming high-resolution ACT DR6 data will be crucial in clarifying this to determine if there is a genuine preference in the ACT+WMAP data for higher $M_\nu$. 

When allowing for a dynamical dark energy background, we observe that neutrino mass constraints both shift toward higher values and broaden significantly compared to $\nu \Lambda \mathrm{CDM}$. In our baseline configuration---combining \textit{Planck} PR4, low-$z$ and BAO measurements---we find a $1.8\sigma$ preference for a nonzero neutrino mass, $M_\nu = 0.16^{+0.09}_{-0.09} \mathrm{eV}$, while using ACT+WMAP instead of \textit{Planck} PR4 further strengthens this to $2.4\sigma$, yielding $M_\nu = 0.26^{+0.10}_{-0.11}$. Notably, these values are entirely consistent with the lower bound of $M_\nu>0.06\,\mathrm{eV}$ required by neutrino oscillation experiments, in contrast to the tighter upper limits found in $\nu \Lambda \mathrm{CDM}$ where the neutrino mass posterior has been shown to peak in the unphysical $M_\nu<0$ region~\cite{Green:2024xbb, Craig:2024tky}. This must be interpreted with caution, however, as neutrino mass constraints in a dynamical dark energy scenario depend on the specific parameterization and model used to describe dark energy~\cite{Elbers:2024sha, Reboucas:2024smm}. Nevertheless, these results highlight the importance of considering degeneracies between dark energy and neutrino physics, as the broader parameter space can accommodate larger neutrino masses. Future high-precision data will be critical to disentangle the interplay between evolving dark energy and neutrino mass. Finally, we show that there is consistency between the \textit{Planck} PR3 and PR4 inferences of the neutrino mass in both $\nu\Lambda \mathrm{CDM}$ and $\nu w_0 w_a \mathrm{CDM}$ once the $A_L$ parameter is also varied. 

We assessed the preference for the extended models compared to baseline $\Lambda \mathrm{CDM}$ using the AIC criterion. We find in our baseline configuration with \textit{Planck} PR4 that the $\Lambda \mathrm{CDM}$ model has the lowest AIC, though the differences in AIC are not sufficient to conclusively rule out any of the extensions we have explored in this work. Following Occam's razor we find no compelling evidence to replace the standard $\Lambda \mathrm{CDM}$ model with any of the extensions we analyzed. 

Our findings motivate several avenues for further investigation. Primarily, this work heightens the need to understand discrepancies between the high-redshift and low-redshift inferences of the $S8$ parameter as this plays a critical role when considering joint constraints on models beyond $\Lambda \mathrm{CDM}$. We find that the extensions considered in this work (both neutrino masses and dynamical dark energy) do not help to significantly alleviate this mild tension. In fact, the kind of decaying dark energy models (i.e. $w_0>-1, w_a<0$) favored by some combinations of CMB, BAO, and supernovae data lead to an \emph{increase} in the central value of the $S8$ posterior compared to $\Lambda \mathrm{CDM}$. Upcoming data from Euclid~\cite{EUCLID:2011zbd} and LSST~\cite{LSSTScience:2009jmu} as well as improvements in IA~\cite{Chen:2024vvk} and Baryonic~\cite{Hadzhiyska:2023wae} modeling will be necessary to investigate this further.  

On the data side, an immediate upgrade would be to replace the \textit{Planck} PR3 lensing maps with the higher-resolution ACT DR6~\cite{ACT:2023kun} or \textit{Planck} \texttt{NPIPE} lensing maps~\cite{Carron:2022eyg}. Furthermore, the inclusion of supernovae data in the framework, such as the Pantheon+ sample~\cite{Brout:2022vxf} or DESY5~\cite{DES:2024jxu}, would strengthen constraints on dynamical dark energy and neutrino masses. For example, Ref.~\cite{Loverde:2024nfi} find that the preference in current supernovae data for higher $\Omega_m$ than what is preferred by DESI Y1 BAO data allows for a higher neutrino mass when combined with CMB data. These additional data may also partially break the degeneracy between dynamical dark energy background effects and neutrino mass background effects, improving the neutrino mass bound in extended models. Our present constraints on dynamical dark energy and neutrino masses in an evolving dark energy scenario are also partly limited by our reliance on the CPL parameterization. In an upcoming study~\cite{Reeves:2025ghw}, we plan to adopt a more model-independent approach to reconstruct the expansion and growth history at low redshift from a combination of datasets.

Overall, this work demonstrates the power of combining multiple CMB and LSS datasets to (a) assess tensions between data, (b) identify and mitigate the impact of systematic effects, and (c) provide robust constraints on both the standard cosmological model and extensions. These key advantages will be increasingly important in the face of upcoming high-precision data.

\newpage
\appendix

\section{To what extent are we double counting BAO information?\label{appendix:bao_information}}

\begin{figure}
\centering
\includegraphics[scale=0.3]{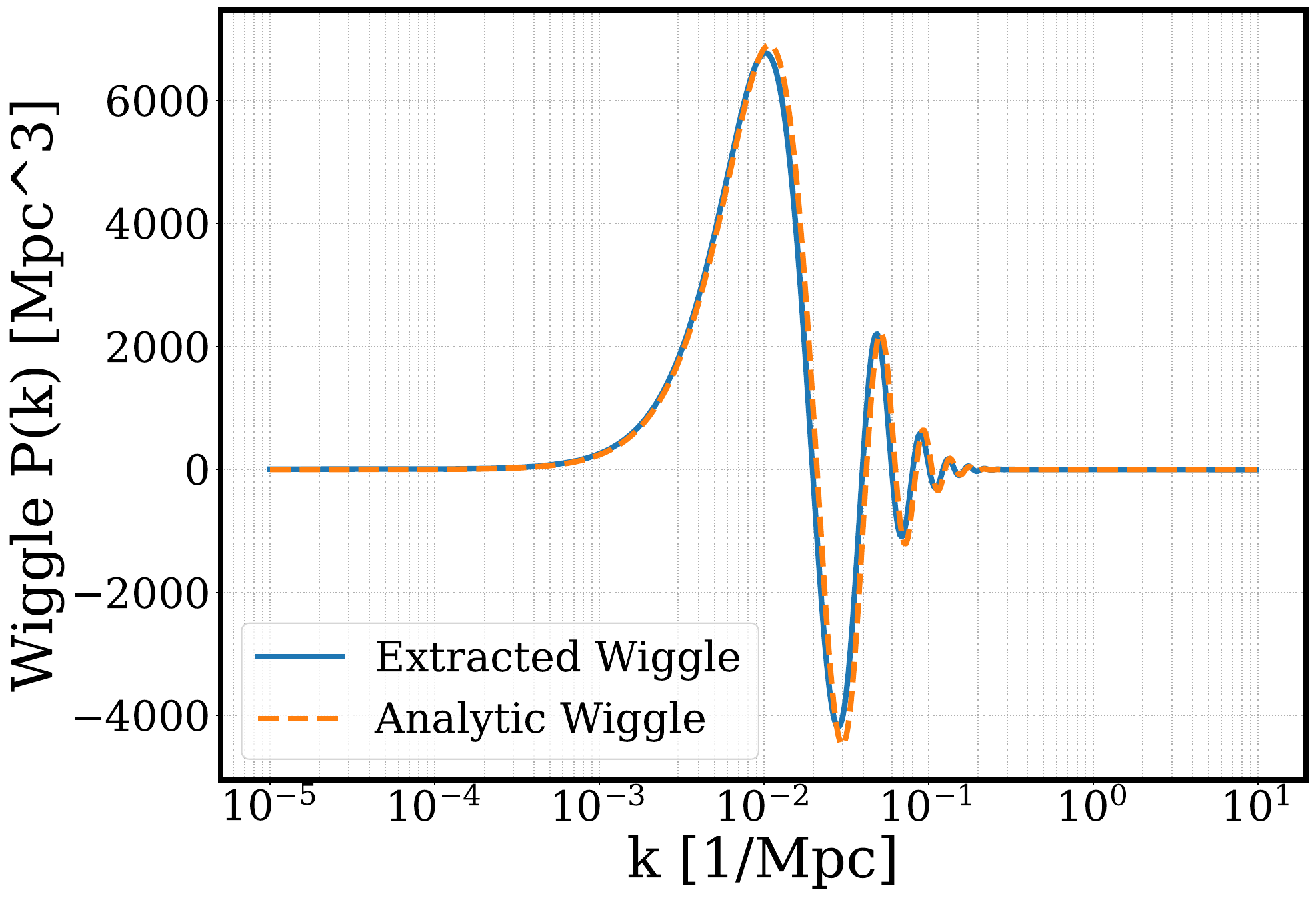}
\caption{The `wiggly' part of the power spectrum residual and the fit to it using the ansatz from Equation~\ref{eqt:wiggly_bao} at the \cosmogrid{} fiducial cosmology at $z=0$. \label{fig:wiggly_ps}}
\end{figure}

In this section, we validate our assumption that the DESI Y1 BAO measurements carry negligible cross-covariance with the low-$z$ probes included in our analysis. The intuition behind this assumption is that our low-$z$ data vector comprises projected quantities where the BAO information gets ``washed out'' by integrating over the line of sight. Furthermore, despite sharing a similar redshift range, the BOSS DR12 data we use shares only $\sim 30\%$ of the same objects with the BGS and LRG1 samples of the DESI DR1 data~\cite{DESI:2024aax}.  

To estimate the information content in our low-$z$ data vector coming from BAOs we first decompose the matter power spectrum into its ``smooth'' and ``wiggly'' components following Refs.~\cite{Philcox:2020vvt, Hamann:2010pw, Chudaykin:2020aoj}, using an approximate analytic for the ``wiggly'' component:
\begin{equation}\label{eqt:wiggly_bao}
    P_{w}(P_{nw}, k, \alpha) \approx \gamma\, P_{nw}(k)\,
    \sin\left(\frac{k \,\ell_{\mathrm{BAO}}}{\alpha}\right)\,
    \exp\!\bigl[-k^2\bigl(\Sigma_{\mathrm{NL}}^2 + \Sigma_{\mathrm{Silk}}^2\bigr)\bigr],
\end{equation}
where $P_{nw}$ is the smooth (no-wiggle) power spectrum, $\ell_{\mathrm{BAO}}\approx 105\,h^{-1}\mathrm{Mpc}$, $\Sigma_{\mathrm{NL}}\approx 3\,h^{-1}\mathrm{Mpc}$, $\Sigma_{\mathrm{Silk}}\approx 5\,h^{-1}\mathrm{Mpc}$, and $\alpha$ is the isotropic Alcock--Paczynski parameter. We obtain $P_{nw}$ by performing a ``wiggly--non-wiggly split'' via the discrete spectral analysis method of Refs.~\cite{Hamann:2010pw, Chudaykin:2020aoj}, which identifies and removes BAO-induced features in the Fourier transform of $\ln\bigl(k\,P(k)\bigr)$, replacing them by a smooth spline before inverse-transforming back to $k$-space. Figure~\ref{fig:wiggly_ps} shows that Eq.~\eqref{eqt:wiggly_bao} accurately reproduces the BAO residual in our fiducial cosmology, providing a suitable order-of-magnitude model of the BAO signal.

Next, to examine the sensitivity of our framework to the BAO scale, we construct a \texttt{JAX}-based theoretical calculation that combines the smooth component $P_{nw}(k,z)$ with its ``wiggly'' counterpart $P_w$ in each DESI Y1 BAO redshift bin. Assuming linear theory, which is accurate for an order of magnitude assessment of the cross-covariance, we write
\begin{equation}
    P_\mathrm{tot}(k,z) \;=\; P_{nw}(k, 0)\,D^2(z)\;+\;
    P_w\bigl[P_{nw}(k, 0)\,D^2(z),\,k,\,\alpha(z)\bigr],
\end{equation}
where $D(z)$ is the linear growth factor. We then project $P_\mathrm{tot}(k,z)$ into angular power spectra $C_\ell$ with a Limber integral using each probe's window function. In this way, we can systematically vary $\alpha(z)$ and assess the extent to which BAO wiggles affect our low-$z$ observables.

\begin{figure}
\centering
\includegraphics[scale=0.2]{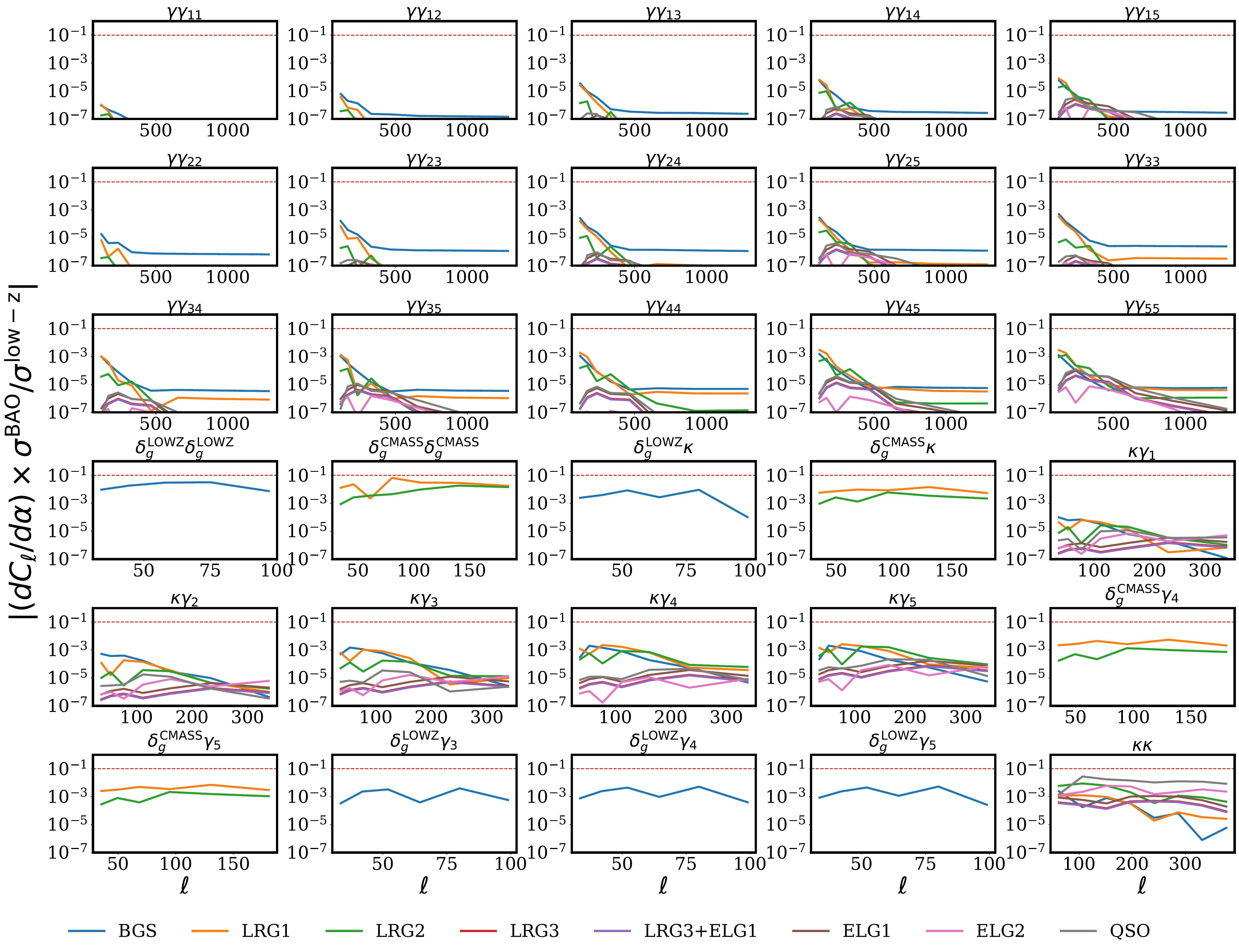}
\caption{The impact of a $1\sigma^{\mathrm{BAO}}$ shift on the low-$z$ data vector normalized by the low-$z$ error bars. As described in the text this is less than 10\% of an error bar (dashed red line) across the entire data vector demonstrating that the information content on the BAO scale from the low-$z$ data vector is negligible.  \label{fig:bao_info}}
\end{figure}

\begin{figure}
\centering
\includegraphics[width=\textwidth]{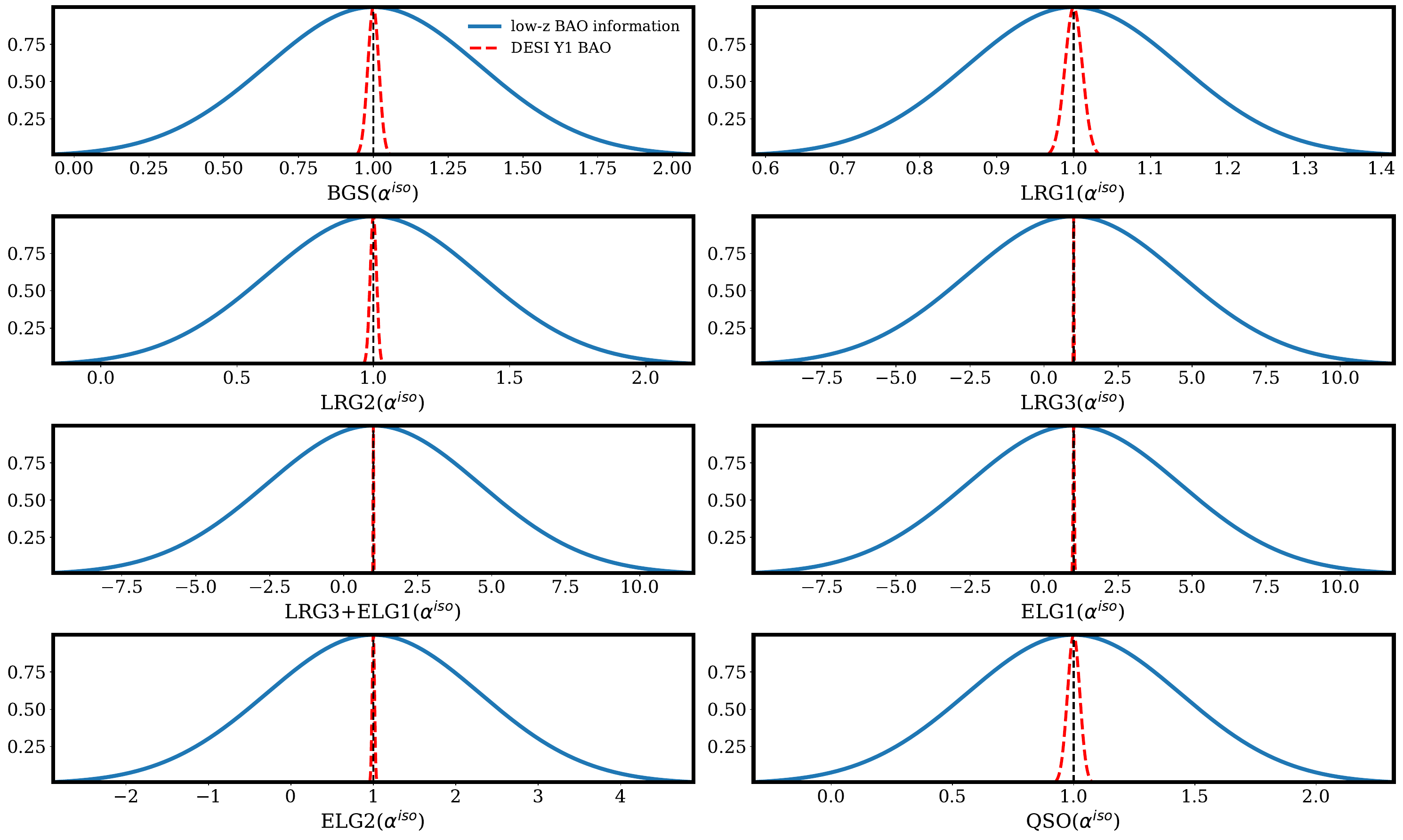}
\caption{1D Fisher forecast of the constraining power from the low-$z$ data vector on the $\alpha^{iso}$ BAO parameters compared to the measured DESI Y1 sensitivities at the reference value of $\alpha_{\mathrm{iso}}=1$ demonstrating that the BAO constraints are dominated by DESI Y1 BAO data. The densities are \emph{not} normalized for visualization purposes. \label{fig:bao_fisher}}
\end{figure}

We use \texttt{JAX}'s automatic differentiation to compute the derivative of our low-$z$ angular power spectra $C_\ell$ with respect to each $\alpha(z)$ parameter which we vary stepwise in the redshift bins associated with the DESI Y1 BAO measurements. We can then compute the $\Delta C_\ell$ associated with a $1\sigma$ shift in $\alpha(z)$ (where we take the $1\sigma$ uncertainties from the DESI Y1 BAO measurements presented in Table~15 of Ref.~\cite{DESI:2024uvr}), which we further normalize by the error bar associated with our low-$z$ data vector points. As illustrated in Fig.~\ref{fig:bao_info}, the largest effect across the low-$z$ data vector corresponds to a $\sim 0.1\,\sigma$ shift in our low-$z$ data vector for the CMASS autocorrelation angular power spectrum and the DESI LRG1 bin, confirming that any BAO signal in our low-$z$ dataset is marginal compared to the precision of the DESI Y1 BAO measurements.

To further demonstrate this we can also compute the Fisher information matrix: 
\begin{equation}
    \mathcal{F} \;=\; 
    \Bigl(\frac{\partial \vec{T}}{\partial \alpha_i}\Bigr)^T\,C^{-1}_{ij}\,
    \Bigl(\frac{\partial \vec{T}}{\partial \alpha_j}\Bigr),
\end{equation}
where $\vec{T}(\vec{\alpha})$ is the theory prediction for the low-$z$ data vector, which depends on the $\alpha^{\mathrm{iso}}$ parameters. The inverse of $\mathcal{F}$ (the ``precision matrix'') provides an estimate of how sensitively our low-$z$ data constrain the $\alpha$ parameters. We present the 1D Gaussian distributions representing this constraining power in Fig.~\ref{fig:bao_fisher}. Comparing the resulting 1D constraints with the DESI Y1 BAO uncertainties from Table~15 of Ref.~\cite{DESI:2024uvr}, we find that the low-$z$ data vector is at least four times weaker (in terms of the width of the distribution) compared to DESI Y1 BAO in constraining the $\alpha$ parameters, with the strongest constraint from low-$z$ on the parameter associated with the LRG1 sample which we found above is weakly constrained by the CMASS auto-correlation.

Overall, with this `back of the envelope' calculation we demonstrate that the BAO information content in our joint constraints comes essentially exclusively from the DESI Y1 BAO sample which completely dominates over the information in our low-$z$ data vector, justifying our assumption in ignoring the cross-covariance between BAO and our low-$z$ data vector.  

\section{Lensing MC normalization correction \label{appendix:mc_norm}}

\begin{figure}[htbp]
    \centering
    %
    \begin{subfigure}{0.45\textwidth}
        \centering
        \includegraphics[scale=0.23]{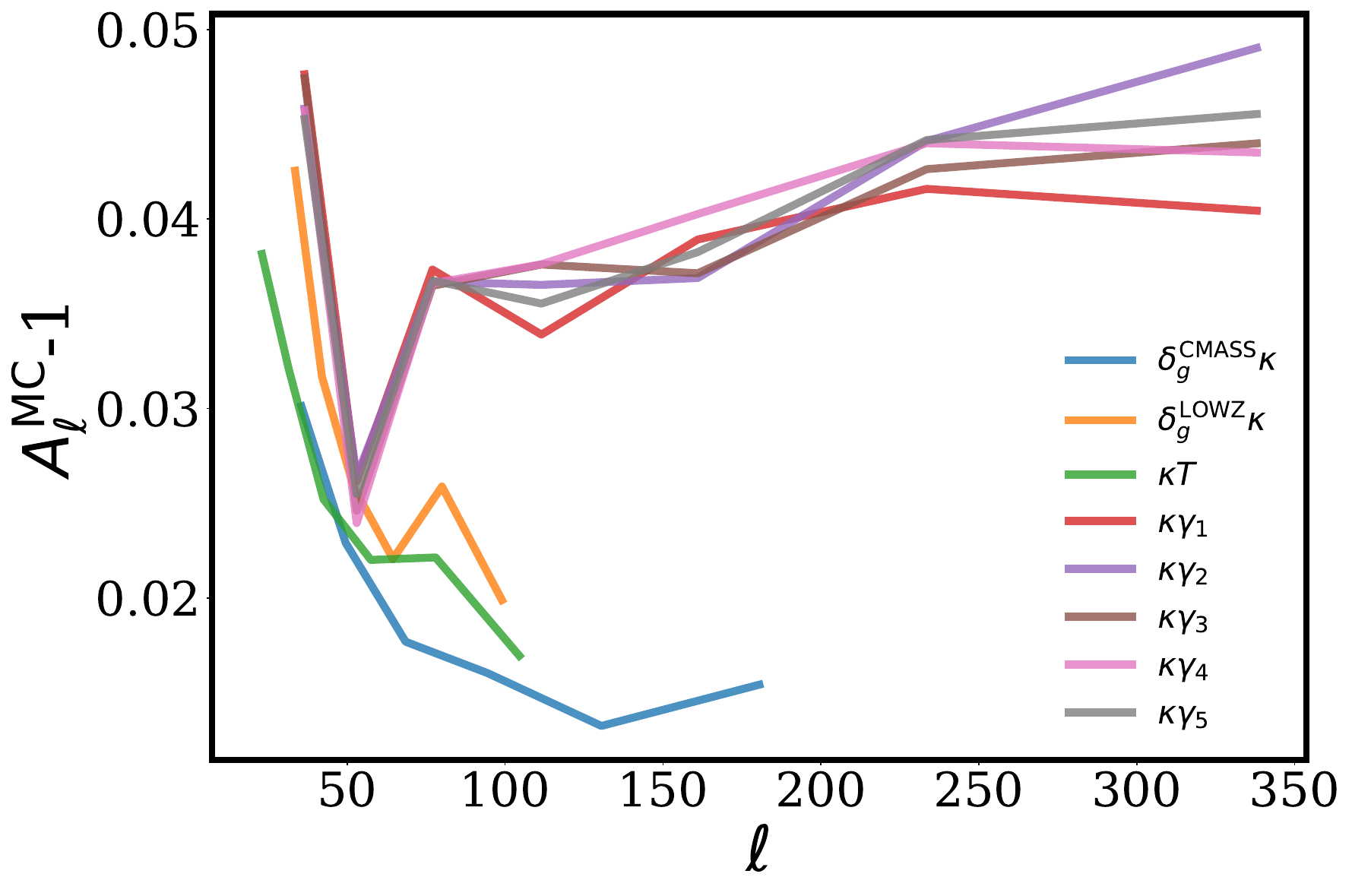}
        \caption{Value of the computed MC norm correction for the bandpowers of the CMB lensing cross-correlations.}
        \label{fig:mc_norm}
    \end{subfigure}
    \quad
    %
    \begin{subfigure}{0.45\textwidth}
        \centering
        \includegraphics[width=\textwidth]{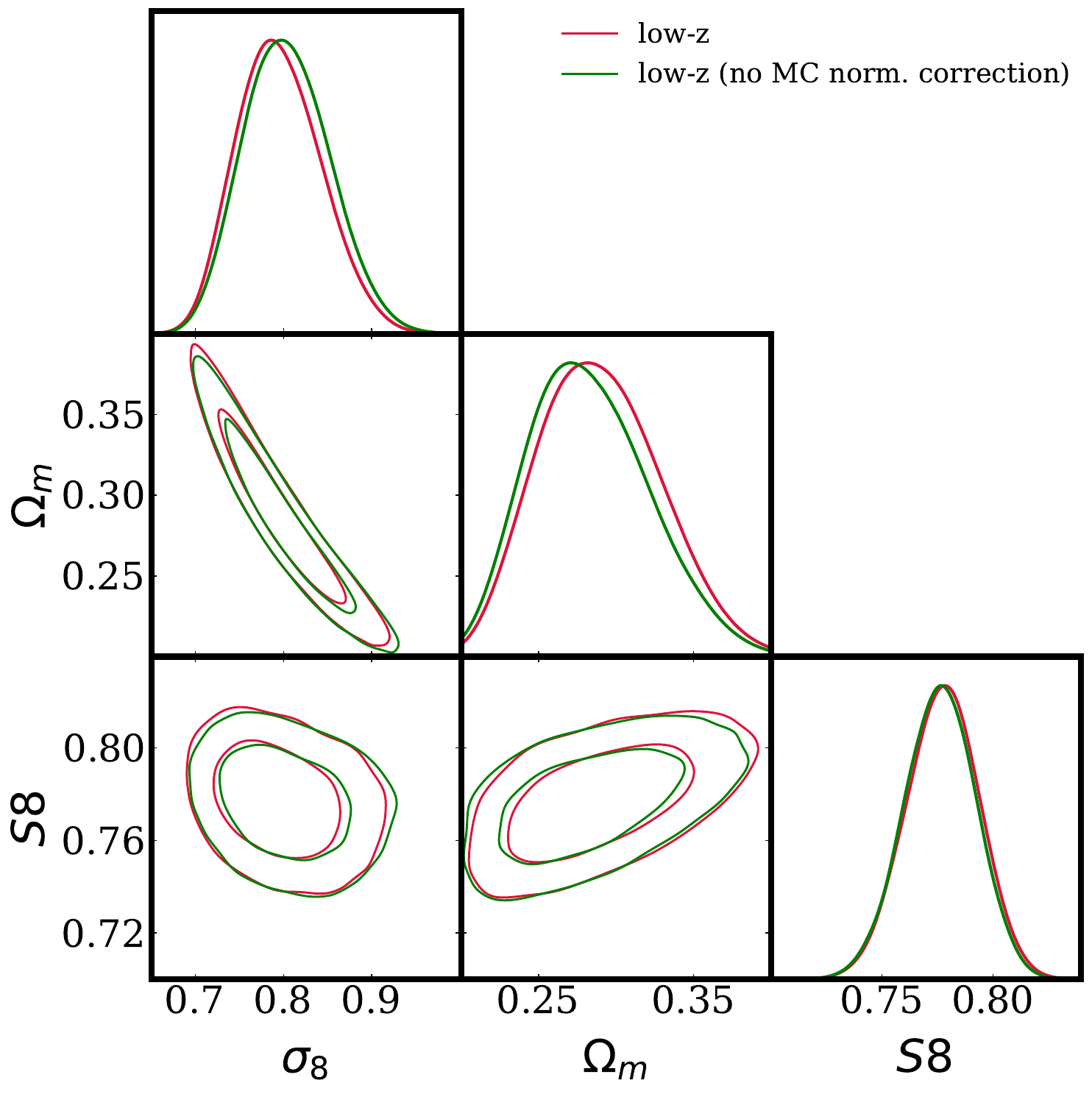}
        \caption{Impact of the MC norm correction on the inference in $\Lambda \mathrm{CDM}$ for the low-$z$ data vector.}
        \label{fig:impact_of_mc}
    \end{subfigure}
    \caption{The value of the computed MC norm correction (\subref{fig:mc_norm}) and its impact on inference (\subref{fig:impact_of_mc}).}
    \label{fig:combined}
\end{figure}

When performing CMB lensing reconstruction in the presence of a mask, one must account for an additional normalization correction. This correction is required to capture the mode couplings introduced by the anisotropic filtering and masking of the raw CMB Temperature and Polarization maps used in the reconstruction. As first noted in~\cite{ACT:2023oei}, the required normalization differs between CMB lensing cross-correlations and auto-correlations, and must therefore be recalculated for each new cropecifieds-correlation measurement. If neglected, this mis-normalization will bias the cosmological inference from CMB lensing cross-correlations typically giving a low bias in the inferred $S8$ value.

Following the prescription of~\cite{Sailer:2024coh}, we address this issue by multiplying our CMB lensing cross-correlation measurement by a Monte Carlo ``MC norm correction'' obtained from a suite of simulated CMB lensing reconstructions. Concretely, we define
\begin{equation}
\label{eq:MCcorr}
(\mathrm{MC\ norm\ correction})_{L} \;=\;
\frac{
  \displaystyle \sum_{\ell} W_{L\ell}
      \sum_{i=1}^{N_{\mathrm{sim}}}
      \sum_{m=-\ell}^{\ell}
      \bigl\{ M^\kappa \,\hat{\kappa}^{\,i}\bigr\}_{\ell m}
      \bigl\{ M^g \,\kappa^{\,i}\bigr\}^{*}_{\ell m}
}{
  \displaystyle \sum_{\ell} W_{L\ell}
      \sum_{i=1}^{N_{\mathrm{sim}}}
      \sum_{m=-\ell}^{\ell}
      \bigl\{ M^\kappa \,\kappa^{\,i}\bigr\}_{\ell m}
      \bigl\{ M^g \,\kappa^{\,i}\bigr\}^{*}_{\ell m}
}\,,
\end{equation}
where $M^\kappa$ ($M^g$) is the CMB lensing (galaxy) mask, and $\hat{\kappa}^{\,i}$ ($\kappa^{\,i}$) denotes the reconstructed (input) CMB lensing convergence from the $i^\text{th}$ simulation, and we use the convention
\[
\bigl\{AB\bigr\}_{\ell m}
\,\equiv\,
\int d^2\hat{n}\,Y_{\ell m}^*(\hat{n})\,A(\hat{n})\,B(\hat{n})\,,
\]
where $W_{L\ell}$ is the window function associated with the bandpower $L$. 

Figure~\ref{fig:mc_norm} shows the magnitude of this Monte Carlo correction for each of the CMB lensing cross-correlations computed in the pipeline. In turn, Fig.~\ref{fig:impact_of_mc} shows the effect of applying this correction on our cosmological parameter constraints in the baseline $\Lambda \mathrm{CDM}$ analysis, yielding a modest upward shift of $+0.05\sigma$ in $S8$ compared to the case where this correction is omitted. This negligible shift is smaller than has been observed in some previous analyses~\cite{ACT:2023oei, Sailer:2024coh}. This is mainly due to our use of the lower signal to noise \textit{Planck} PR3 maps, and partly due to the fact that the low-$z$ data vector also contains galaxy weak lensing auto-correlation data from KiDS-1000 that is unaffected by this correction. Our findings match with ~\cite{Xu:2023qmp} who also find this correction to be negligible in the context of a $6 \times 2$pt analysis including \textit{Planck} PR3 CMB lensing, DES weak lensing, and galaxy clustering. 

\section{Impact of KiDS-1000 redshift bin 2 \label{appendix:kids_bin_2}}
As we showed in Section~\ref{subsec:internal_consistency} the second redshift bin of the KiDS dataset has a discordantly high amplitude compared to the rest of the KiDS-1000 dataset which is detected at over $3 \sigma$ when combining with CMB data. This also contributes to worsening the goodness of fit metric for the low-$z$ dataset. This may be caused by a statitistical fluctuation or a systematic problem with the data in the bin and the KiDS team posits that this can be plausibly explained by high-z interlopers within the bin~\cite{KiDS:2020suj}. In this work, we decide to keep the data within this bin in our fiducial analysis and here we demonstrate the impact of removing this bin on our results.

\subsection{$\Lambda\mathrm{CDM}$}

\begin{figure}
\centering
\includegraphics[width=0.75\textwidth]{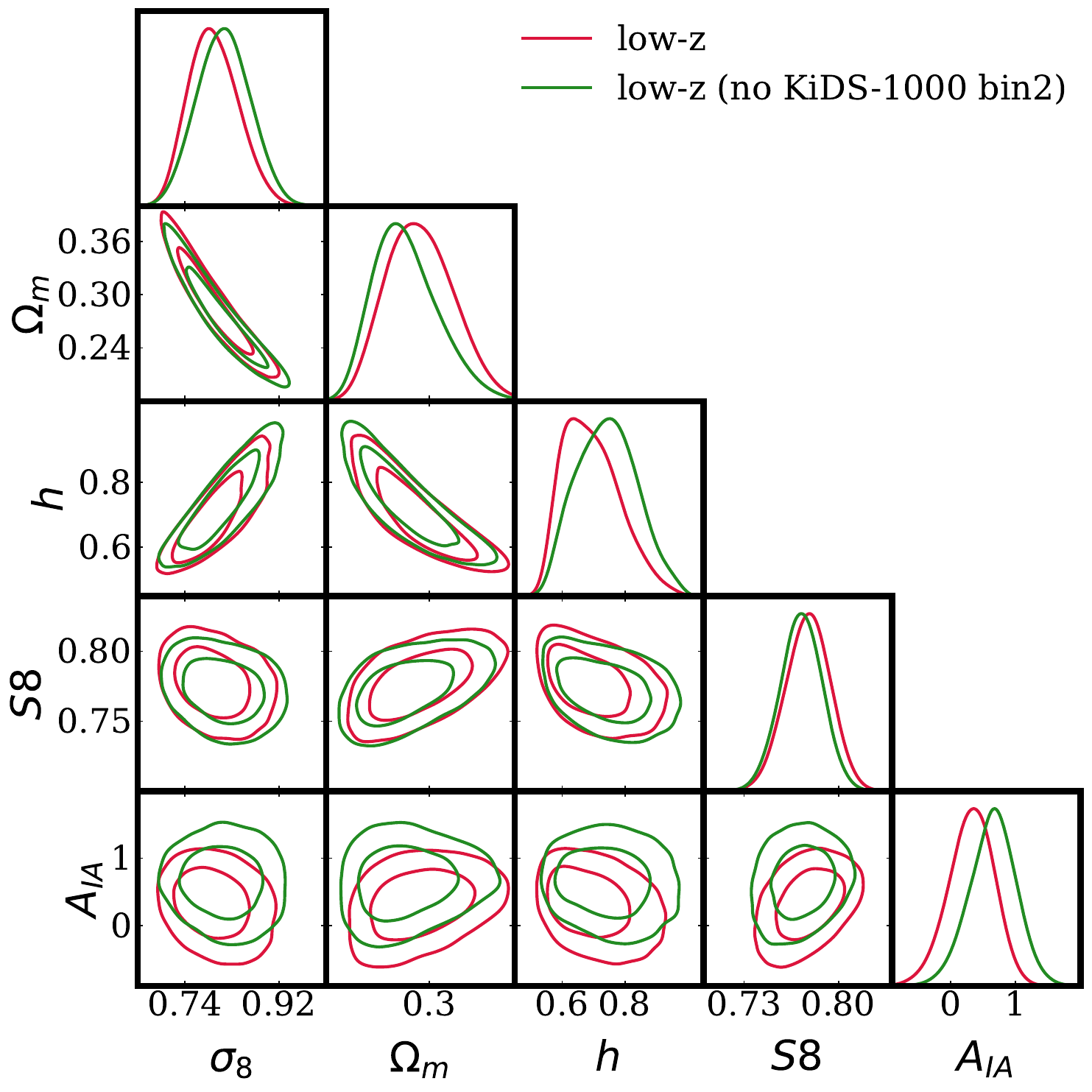}
\caption{Impact of removing redshift bin 2 on the low-$z$ constraints in $\Lambda \mathrm{CDM}$. \label{fig:red_bin_2}}
\end{figure}

In Fig.~\ref{fig:red_bin_2} we show the impact of removing KiDS-1000 redshift bin 2 from our low-$z$ data vector in the $\Lambda \mathrm{CDM}$ model. We find small shifts in the cosmological parameters. In particular, we find $\Delta S8=-0.2\sigma$ going from with bin 2 to without bin 2. The direction of this relatively small shift can be understood as we showed in Section~\ref{subsec:internal_consistency} that the data in this bin prefer a higher weak lensing power spectrum amplitude compared to the data in the other bins. This means removing the bin from the analysis pushes the $S8$ constraint towards lower values. We see a larger shift for the amplitude of the intrinsic alignment signal $A_{IA}$ ($+0.7 \sigma$), this is because KiDS-1000 redshift bin 2 is at relatively low redshift where the signal is dominated by the intrinsic alignment contribution and is therefore particularly sensitive to this parameter. We find that when combining the low-$z$ data with CMB there is a negligible shift in the cosmological parameters (within $0.1\sigma$ for all cosmological parameters), though the shift in $A_{IA}$ remains in the same direction with at a reduced significance of ($+0.3 \sigma$). 

\subsection{Dynamical dark energy}

\begin{figure}[!htb]
    \centering
    \includegraphics[width=0.75\textwidth]{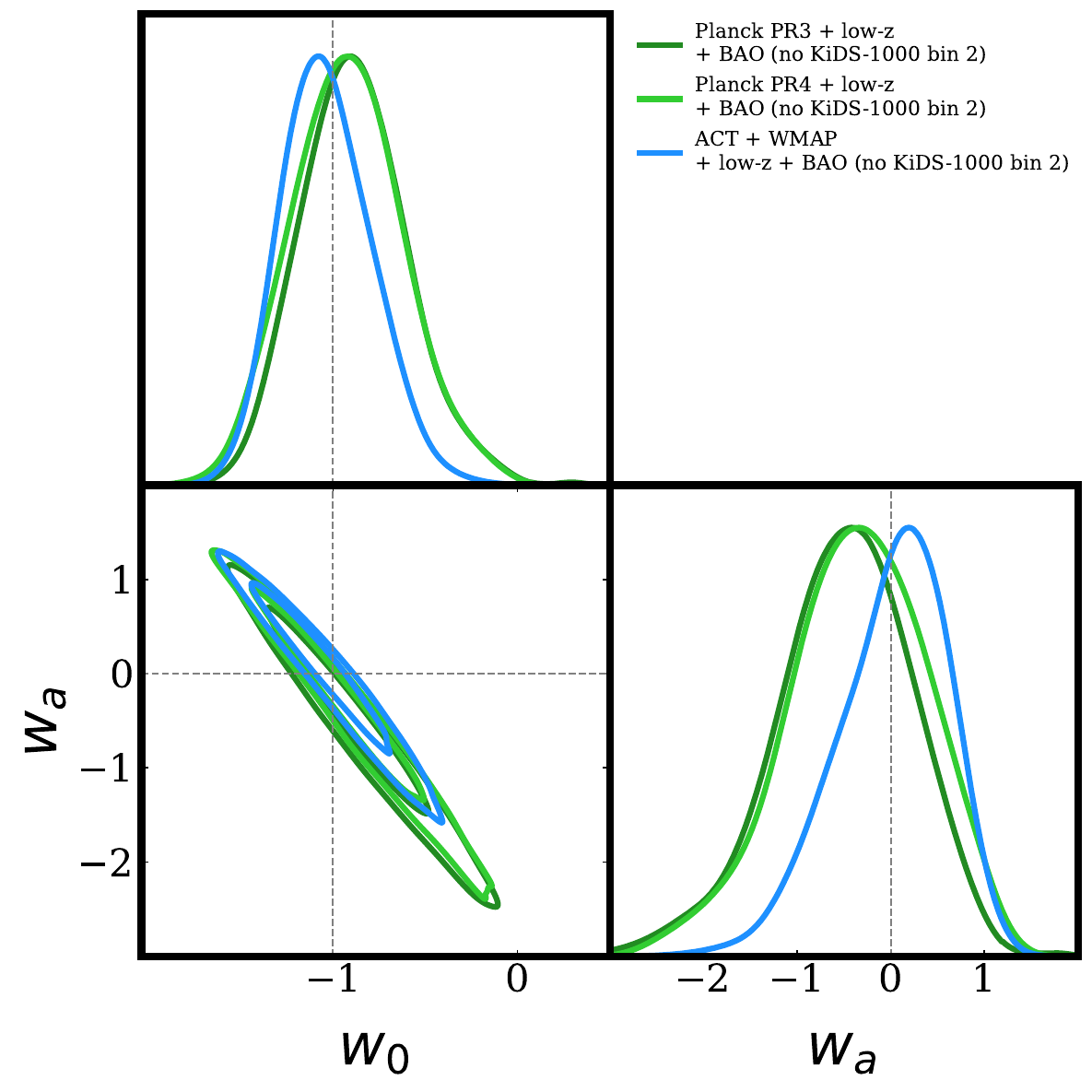}
    \caption{The constraints in the $w_0w_a$ model considering the combination of the CMB, BAO and low-$z$ datasets when removing the KiDS-1000 redshift bin 2 from the analysis.\label{fig:w0wa_multi_nozbin2}}
\end{figure}

\begin{table}[ht]
\centering
\caption{Cosmological parameter constraints in the $w_0w_a\mathrm{CDM}$ model when removing the second KiDS-1000 redshift bin from the analysis \label{tab:w0wa_constraints_nozbin2}}
\setlength{\tabcolsep}{6pt}
\renewcommand{\arraystretch}{1.1}
\begin{tabular}{lccc}
\hline
\hline
\textbf{Parameter} 
 & \makecell{\textit{Planck} PR3 \\ + low-$z$ + BAO \\ (no KiDS-1000 bin 2)} 
 & \makecell{\textit{Planck} PR4 \\ + low-$z$ + BAO \\ (no KiDS-1000 bin 2)} 
 & \makecell{ACT + WMAP \\ + low-$z$ + BAO \\ (no KiDS-1000 bin 2)} \\
\hline
\multicolumn{4}{l}{\textbf{Cosmological Parameters}} \\
\hline 
$\Omega_m$ 
  & $0.301^{+0.030}_{-0.029}$ 
  & $0.300^{+0.030}_{-0.030}$ 
  & $0.288^{+0.026}_{-0.026}$ \\[6pt]
$\sigma_8$ 
  & $0.808^{+0.028}_{-0.029}$ 
  & $0.807^{+0.029}_{-0.029}$ 
  & $0.816^{+0.027}_{-0.026}$ \\[6pt]
$H_0\,[\mathrm{km/s/Mpc}]$ 
  & $68.7^{+3.29}_{-3.40}$ 
  & $68.7^{+3.50}_{-3.46}$ 
  & $69.9^{+3.11}_{-3.10}$ \\[6pt]
$S8$ 
  & $0.807^{+0.0146}_{-0.0143}$ 
  & $0.804^{+0.0146}_{-0.0146}$ 
  & $0.798^{+0.0143}_{-0.0141}$ \\[6pt]
\multicolumn{4}{l}{\textbf{Dynamical dark energy parameters}} \\
\hline 
$w_0$ 
  & $-0.881^{+0.293}_{-0.283}$ 
  & $-0.919^{+0.309}_{-0.302}$ 
  & $-1.031^{+0.247}_{-0.253}$ \\[6pt]
$w_a$ 
  & $-0.499^{+0.712}_{-0.734}$ 
  & $-0.362^{+0.741}_{-0.775}$ 
  & $0.010^{+0.613}_{-0.590}$ \\
\hline
\end{tabular}
\end{table}

For the dynamical dark energy model we find a greater impact from removing the second redshift bin of KiDS-1000. As shown in Section~\ref{subsec:dynamical_de_results}, removing this bin generally drives the $w_0, w_a$ results further towards the $\Lambda \mathrm{CDM}$ values when considering the full combination of CMB + low-$z$ + DESI Y1 BAO data. In this section, we provide the full set of parameter constraints for each of the different CMB likelihoods without the KiDS-1000 redshift bin 2 in the analysis in Fig.~\ref{fig:w0wa_multi_nozbin2} and Table~\ref{tab:w0wa_constraints_nozbin2}. We see that the trend for the other two CMB experiments is the same as discussed in Section~\ref{subsec:dynamical_de_results}: the removal of the KiDS-1000 redshift bin 2 generally pushes the constraints further towards $\Lambda \mathrm{CDM}$.

\subsection{$\nu w_0 w_a \mathrm{CDM}$}

\begin{figure}[!htb]
    \centering
    \includegraphics[width=0.75\textwidth]{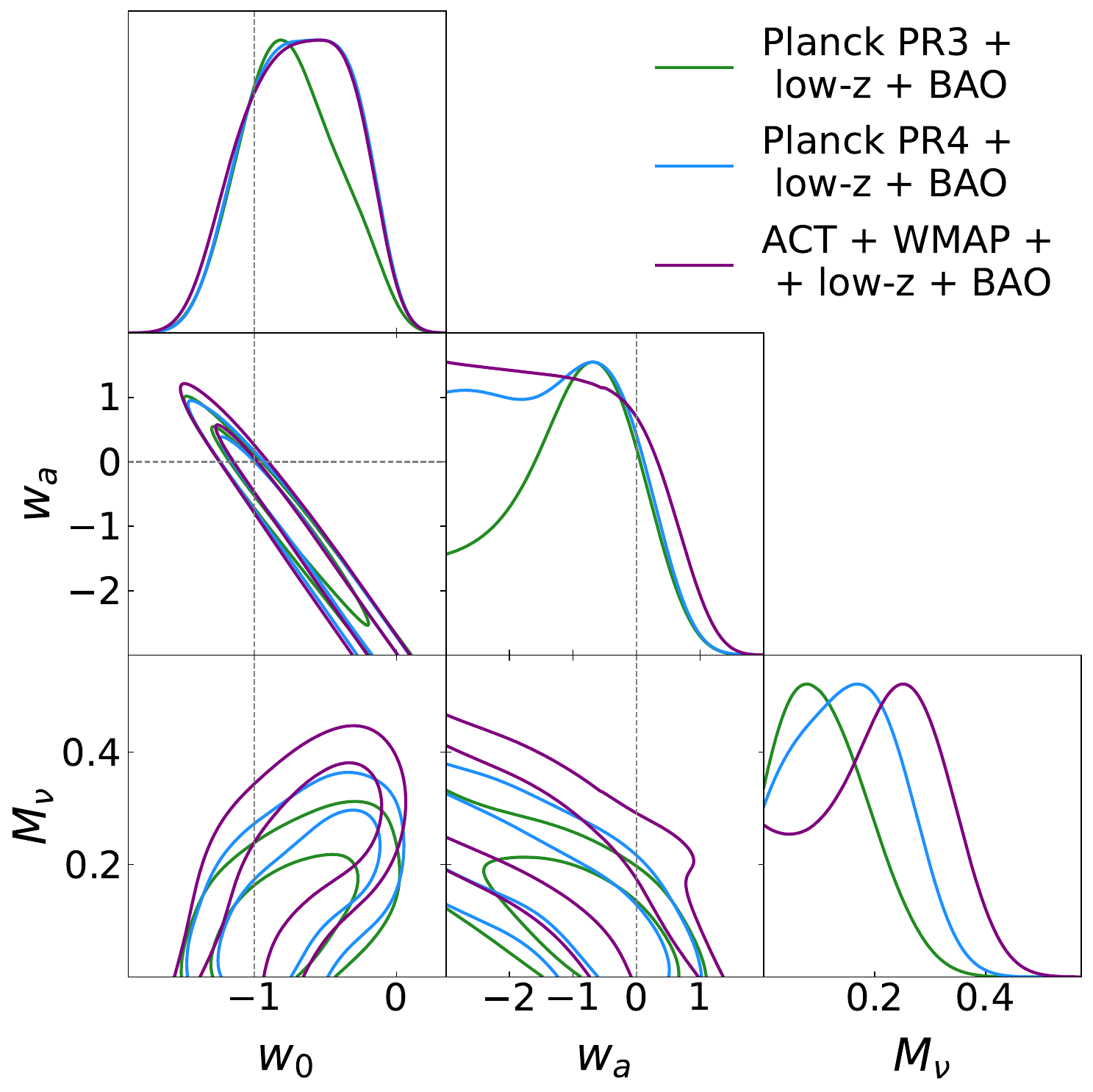}
    \caption{The constraints in the $\nu w_0w_a$ model considering the combination of the CMB, BAO and low-$z$ datasets when removing the KiDS-1000 redshift bin 2 from the analysis.\label{fig:nuw0wa_multi_nozbin2}}
\end{figure}

\begin{table}[ht]
\centering
\caption{Cosmological parameter constraints in the $\nu w_0w_a\mathrm{CDM}$ model when removing the second KiDS-1000 redshift bin from the analysis \label{tab:nuw0wa_constraints_nozbin2}}
\setlength{\tabcolsep}{6pt}
\renewcommand{\arraystretch}{1.1}
\begin{tabular}{lccc}
\hline
\hline
\textbf{Parameter} 
 & \makecell{\textit{Planck} PR3 \\ (Planck 2018) + LSS + BAO \\ (no KiDS-1000 bin 2)} 
 & \makecell{\textit{Planck} PR4 \\ (Hillipop) + LSS + BAO \\ (no KiDS-1000 bin 2)} 
 & \makecell{ACT + WMAP \\ + LSS + BAO \\ (no KiDS-1000 bin 2)} \\
\hline
\multicolumn{4}{l}{\textbf{Cosmological Parameters}} \\
\hline 
$\Omega_m$ 
  & $0.314^{+0.035}_{-0.036}$ 
  & $0.323^{+0.038}_{-0.038}$ 
  & $0.320^{+0.040}_{-0.041}$ \\[6pt]
$\sigma_8$ 
  & $0.791^{+0.036}_{-0.035}$ 
  & $0.779^{+0.037}_{-0.037}$ 
  & $0.771^{+0.036}_{-0.046}$ \\[6pt]
$H_0\,[\mathrm{km/s/Mpc}]$ 
  & $67.5^{+3.73}_{-3.65}$ 
  & $66.6^{+3.76}_{-3.84}$ 
  & $67.3^{+4.36}_{-3.65}$ \\[6pt]
$S8$ 
  & $0.807^{+0.0141}_{-0.0140}$ 
  & $0.806^{+0.0141}_{-0.0142}$ 
  & $0.793^{+0.0089}_{-0.0213}$ \\[6pt]
\multicolumn{4}{l}{\textbf{Extension parameters}} \\
\hline 
$M_\nu\,[\mathrm{eV}]$        
 & $<0.26$ 
 & $0.159^{+0.100}_{-0.094}$ 
 & $0.22^{+0.125}_{-0.110}$ \\
$w_0$ 
  & $-0.742^{+0.342}_{-0.367}$ 
  & $-0.664^{+0.389}_{-0.394}$ 
  & $-0.683^{+0.405}_{-0.403}$ \\[6pt]
$w_a$ 
  & $-0.977^{+1.054}_{-0.951}$ 
  & $-1.232^{+1.201}_{-1.131}$ 
  & $-1.160^{+1.276}_{-1.287}$ \\
\hline
\end{tabular}
\end{table}

In the $\nu w_0 w_a \mathrm{CDM}$ model, we find small shifts when excluding the KiDS-1000 second redshift bin which again pushes the constraints in the $w_0w_a$ plane towards $\Lambda \mathrm{CDM}$ (see Table~\ref{tab:nuw0wa_constraints_nozbin2} and Fig.~\ref{fig:nuw0wa_multi_nozbin2}). The neutrino mass constraints are less affected and we still find a small preference for $M_\nu>0$ for the baseline combination with \textit{Planck} PR4 ($M_\nu =0.159^{+0.100}_{-0.094}$, representing a $1.7\sigma$ constraint) and for the combination with ACT + WMAP ($M_\nu = 0.22^{+0.125}_{-0.110}$, representing a $2.0\sigma$ constraint).

\section{Extra parameter constraint tables}

\subsection{$\Lambda\mathrm{CDM}$ \label{appendix:lcdm_plots}}


\begin{table}[ht]
    \centering
    \caption{Cosmological and selected nuisance parameter constraints for the $\Lambda\mathrm{CDM}$ model using the combination of CMB, low-$z$, and BAO data.\label{tab:baoconstraints}}
    \resizebox{\textwidth}{!}{
    \setlength{\tabcolsep}{4pt}
    \renewcommand{\arraystretch}{1.1}
    \begin{tabular}{lccc}
    \toprule
    \textbf{Parameter} & \textit{Planck} PR3 + low-$z$ + BAO & \textit{Planck} PR4 + low-$z$ + BAO & ACT+WMAP + low-$z$ + BAO \\
    \midrule
    \multicolumn{4}{l}{\textbf{Cosmological Parameters}} \\
    $\sigma_{8}$   & $0.803 \pm 0.006$ & $0.803 \pm 0.005$ & $0.808 \pm 0.005$ \\
    $H_0\,[\mathrm{km/s/Mpc}]$ & $68.5 \pm 0.4$ & $68.4 \pm 0.4$ & $69.0 \pm 0.5$ \\
    $\Omega_{m}$    & $0.300 \pm 0.005$ & $0.300 \pm 0.004$ & $0.294 \pm 0.005$ \\
    $S_8$            & $0.803 \pm 0.009$ & $0.802 \pm 0.008$ & $0.799 \pm 0.009$ \\
    \midrule
    \multicolumn{4}{l}{\textbf{Selected Nuisance Parameters}} \\
    $b_{\mathrm{LOWZ}}$ & $1.839 \pm 0.039$ & $1.841 \pm 0.038$ & $1.828 \pm 0.039$ \\
    $b_{\mathrm{CMASS}}$ & $2.092 \pm 0.027$ & $2.093 \pm 0.026$ & $2.081 \pm 0.025$ \\
    $A_{IA}$             & $0.505 \pm 0.279$ & $0.503 \pm 0.290$ & $0.502 \pm 0.303$ \\
    $\log T_{AGN}$       & $7.822 \pm 0.479$ & $7.799 \pm 0.473$ & $7.792 \pm 0.477$ \\
    \bottomrule
    \end{tabular}
    }
\end{table}

In Table~\ref{tab:baoconstraints} we show the constraints in $\Lambda \mathrm{CDM}$ when including the DESI Y1 BAO information. We see that with the inclusion of the DESI Y1 BAO data, the constraints from each of the different CMB experiments become compatible at the $< 1\sigma$ level and the addition of this data marginally improves on constraining power compared to the combination of low-$z$ and CMB.

\subsection{$\nu \Lambda\mathrm{CDM}$ \label{appendix:nulcdm_plots}}
\begin{table}[ht]
\centering
\caption{Cosmological and selected nuisance parameter constraints for the $\nu \Lambda\mathrm{CDM}$ model for the combination of CMB, low-$z$, and BAO data.\label{tab:baoconstraints_nuLCDM}}
\resizebox{\textwidth}{!}{
\setlength{\tabcolsep}{2.5pt}
\renewcommand{\arraystretch}{1.0}
\begin{tabular}{lccc}
\toprule
\textbf{Parameter} & \textit{Planck} PR3 + low-$z$ + BAO & \textit{Planck} PR4 + low-$z$ + BAO & ACT+WMAP + low-$z$ + BAO \\
\midrule
\multicolumn{4}{l}{\textbf{Cosmological Parameters}} \\
$H_0\,[\mathrm{km/s/Mpc}]$      
 & $68.7^{+0.4}_{-0.4}$ 
 & $68.5^{+0.4}_{-0.4}$ 
 & $68.8^{+0.6}_{-0.6}$ \\

$\Omega_{m}$        
 & $0.2975^{+0.0051}_{-0.0051}$ 
 & $0.2988^{+0.0053}_{-0.0053}$ 
 & $0.2962^{+0.0064}_{-0.0065}$ \\

$S_8$               
 & $0.804^{+0.009}_{-0.009}$ 
 & $0.803^{+0.008}_{-0.008}$ 
 & $0.797^{+0.011}_{-0.011}$ \\

$\sigma_{8}$        
 & $0.808^{+0.008}_{-0.008}$ 
 & $0.805^{+0.008}_{-0.008}$ 
 & $0.802^{+0.012}_{-0.012}$ \\
\midrule
\multicolumn{4}{l}{\textbf{Extension Parameters}} \\
$M_\nu\,[\mathrm{eV}]$        
 & $<0.12$ 
 & $<0.09$ 
 & $<0.18$ \\
\bottomrule
\end{tabular}
}

\end{table}

In Table~\ref{tab:baoconstraints_nuLCDM}, we present the full parameter constraints in the $\nu\Lambda\mathrm{CDM}$ model for the three CMB likelihoods considered in this work combined with the low-$z$ data and BAO.

\subsection{$\nu w_0w_a\mathrm{CDM}$ \label{appendix:nuw0wa_plots}}
\begin{table}[ht]
\centering
\caption{Cosmological parameter constraints for the $\nu w_0w_a \mathrm{CDM}$ model for the combination of CMB, low-$z$, and BAO data.\label{tab:baoconstraints_nuw0wa}}
\resizebox{\textwidth}{!}{
\setlength{\tabcolsep}{2.5pt}
\renewcommand{\arraystretch}{1.0}
\begin{tabular}{lccc}
\toprule
\textbf{Parameter} & \textit{Planck} PR3 + low-$z$ + BAO & \textit{Planck} PR4 + low-$z$ + BAO & ACT+WMAP + low-$z$ + BAO \\
\midrule
\multicolumn{4}{l}{\textbf{Cosmological Parameters}} \\
$H_0\,[\mathrm{km/s/Mpc}]$    
 & $66.0^{+3.4}_{-3.4}$ 
 & $64.9^{+3.3}_{-3.4}$ 
 & $65.2^{+3.3}_{-3.4}$ \\

$\Omega_m$                  
 & $0.329^{+0.035}_{-0.036}$ 
 & $0.339^{+0.036}_{-0.036}$ 
 & $0.337^{+0.037}_{-0.036}$ \\

$S_8$  
 & $0.813^{+0.014}_{-0.014}$ 
 & $0.812^{+0.014}_{-0.014}$ 
 & $0.806^{+0.015}_{-0.015}$ \\

$\sigma_{8}$               
 & $0.779^{+0.034}_{-0.033}$ 
 & $0.767^{+0.033}_{-0.033}$ 
 & $0.762^{+0.034}_{-0.036}$ \\
\midrule
\multicolumn{4}{l}{\textbf{Extension Parameters}} \\
$M_\nu\,[\mathrm{eV}]$        
 & $<0.26$ 
 & $0.156^{+0.094}_{-0.085}$ 
 & $0.255^{+0.112}_{-0.101}$ \\

$w_0$                         
 & $-0.605^{+0.338}_{-0.358}$ 
 & $-0.525^{+0.354}_{-0.354}$ 
 & $-0.541^{+0.368}_{-0.350}$ \\

$w_a$                         
 & $-1.312^{+1.048}_{-0.949}$ 
 & $-1.541^{+1.070}_{-1.045}$ 
 & $-1.537^{+1.080}_{-1.133}$ \\
\bottomrule
\end{tabular}
}

\end{table}

In Table~\ref{tab:baoconstraints_nuw0wa}, we present the full parameter constraints in the $\nu w_0w_a \mathrm{CDM}$ model for the three CMB likelihoods considered in this work combined with the low-$z$ data and BAO. 

\acknowledgments
AR thanks Antón Baleato Lizancos for a careful reading of the manuscript and very helpful feedback. Further thanks to Pierre Zhang for insightful conversations surrounding the BAO content of LSS data, Marco Bonici for helpful comments surrounding building $w_0w_a\mathrm{CDM}$ emulators, Tilman Tröster and Marika Asgari for further insight into KiDS-1000 redshift bin 2, Gerrit Farren and Noah Sailer for helpful conversations around the lensing MC normalization correction and Erik Rosenberg for helpful discussion surrounding \textit{Planck} PR4. Thanks to Azucena Garvía Bosshard for a careful reading of the manuscript. Finally thanks to Simone Ferraro and Martin White for insightful discussions and feedback on the results presented here. 

We used functionalities provided by \texttt{JAX}~\cite{jax2018github}, \texttt{numpy}~\cite{harris2020array}, \texttt{scipy}~\cite{2020SciPy-NMeth}, and
\texttt{matplotlib}~\cite{Hunter:2007} for this work. Job arrays were submitted using esub-epipe~\cite{Zurcher:2022clh, Zurcher:2020dvu}. We used \texttt{emcee}~\cite{Foreman-Mackey:2012any} to run our MCMC chains. We used the \texttt{GetDist} package for our contour plot visualizations~\cite{Lewis:2019xzd}. Some of the results in this paper have been derived using the \texttt{healpy} and \healpix packages.

\newpage{}
\bibliographystyle{jcap}
\bibliography{xcorr.bib}
\end{document}